\documentclass[acmsmall]{acmart}

\setcopyright{cc}
\setcctype{by}
\acmDOI{10.1145/3871791}
\acmYear{2027}
\acmJournal{PACMPL}
\acmVolume{11}
\acmNumber{POPL}
\acmArticle{89}
\acmMonth{1}
\acmSubmissionID{popl27main-p911-p}
\received{2026-07-09}
\received[accepted]{2026-10-05}

\usepackage{booktabs}
\usepackage{pifont}
\usepackage{listings}
\usepackage{quantikz}        
\usepackage{agda-style}      
\usepackage{agda}
\renewcommand{\AgdaCodeStyle}{\small}
\usetikzlibrary{backgrounds,arrows,shapes,shapes.geometric,shapes.misc}
\pgfdeclarelayer{edgelayer}
\pgfdeclarelayer{nodelayer}
\pgfsetlayers{background,edgelayer,nodelayer,main}
\tikzstyle{none}=[inner sep=0mm]
\tikzstyle{tikzfig}=[baseline=-0.25em,scale=0.5]

\definecolor{normGreen}{RGB}{214, 243, 221}
\definecolor{normRed}{RGB}{249, 196, 196}

\tikzstyle{gate}=[shape=rectangle, text height=1.5ex, text depth=0.25ex, yshift=0.5mm, fill=white, draw=black, minimum height=7mm, yshift=-0.5mm, minimum width=7mm, font={\small}, tikzit category=circuit]
\tikzstyle{small gate}=[shape=rectangle, text height=1.5ex, text depth=0.25ex, yshift=0.5mm, fill=white, draw=black, minimum height=5mm, yshift=-0.5mm, minimum width=7mm, font={\small}, tikzit category=circuit]
\tikzstyle{big gate}=[shape=rectangle, text height=1.5ex, text depth=0.25ex, yshift=0.5mm, fill=white, draw=black, minimum height=13mm, yshift=-0.5mm, minimum width=5mm, font={\small}, tikzit category=circuit]
\tikzstyle{box}=[shape=rectangle, text height=1.5ex, text depth=0.25ex, yshift=0.5mm, fill=white, draw=black, minimum height=6mm, yshift=-1mm, minimum width=5mm, font={\footnotesize}]
\tikzstyle{medium box}=[shape=rectangle, text height=1.5ex, text depth=0.25ex, yshift=0.5mm, fill=white, draw=black, minimum height=13mm, yshift=-0.5mm, minimum width=5mm, font={\footnotesize}]
\tikzstyle{large box}=[shape=rectangle, text height=1.5ex, text depth=0.25ex, yshift=0.5mm, fill=white, draw=black, minimum height=14mm, yshift=-0.5mm, minimum width=5mm, font={\footnotesize}]
\tikzstyle{giant box}=[shape=rectangle, text height=1.5ex, text depth=0.25ex, yshift=0.5mm, fill=white, draw=black, minimum height=23mm, yshift=-0.5mm, minimum width=5mm, font={\footnotesize}]
\tikzstyle{Z dot}=[inner sep=0.1mm, font={\tiny\boldmath}, minimum size=2mm, shape=circle, draw=black, fill=zxgreen, tikzit fill={rgb,255: red,216; green,248; blue,216}]
\tikzstyle{Z phase dot}=[minimum size=5mm, font={\footnotesize\boldmath}, shape=rectangle, rounded corners=2mm, inner sep=0.2mm, outer sep=-2mm, scale=0.8, tikzit shape=circle, draw=black, fill=zxgreen, tikzit draw=blue, tikzit fill={rgb,255: red,216; green,248; blue,216}]
\tikzstyle{Z bold dot}=[inner sep=0mm, minimum size=2mm, shape=circle, draw=black, fill=zxgreen, tikzit fill={rgb,255: red,216; green,248; blue,216}, line width=1.2pt, tikzit category=zx]
\tikzstyle{Z bold phase dot}=[minimum size=5mm, font={\footnotesize\boldmath}, shape=rectangle, rounded corners=2mm, inner sep=0.2mm, outer sep=-2mm, scale=0.8, tikzit shape=circle, draw=black, fill=zxgreen, tikzit draw=blue, tikzit fill={rgb,255: red,216; green,248; blue,216}, line width=1.2pt]
\tikzstyle{Z tiny dot}=[inner sep=0.2mm, minimum size=1mm, shape=circle, draw=black, fill=zxgreen, tikzit fill={rgb,255: red,216; green,248; blue,216}, font={\tiny\boldmath}]
\tikzstyle{light Z dot}=[inner sep=0.1mm, font={\tiny\boldmath}, minimum size=2mm, shape=circle, draw={gray!80!white}, fill=zxgreen, tikzit fill={rgb,255: red,216; green,248; blue,216}]
\tikzstyle{highlight Z dot}=[inner sep=0.1mm, font={\tiny\boldmath}, very thick, minimum size=2mm, shape=circle, draw=blue, fill=zxgreen, tikzit fill={rgb,255: red,216; green,248; blue,216}]
\tikzstyle{X dot}=[Z dot, shape=circle, draw=black, fill=zxred, tikzit fill={rgb,255: red,221; green,165; blue,165}]
\tikzstyle{X phase dot}=[Z phase dot, tikzit shape=circle, tikzit draw=blue, fill=zxred, font={\footnotesize\boldmath}, tikzit fill={rgb,255: red,221; green,165; blue,165}]
\tikzstyle{X bold dot}=[inner sep=0mm, minimum size=2mm, shape=circle, draw=black, fill=zxred, tikzit fill={rgb,255: red,221; green,165; blue,165}, line width=1.2pt, tikzit category=zx]
\tikzstyle{X bold phase dot}=[Z phase dot, tikzit shape=circle, tikzit draw=blue, fill=zxred, font={\footnotesize\boldmath}, tikzit fill={rgb,255: red,221; green,165; blue,165}, line width=1.2pt]
\tikzstyle{X tiny dot}=[inner sep=0mm, minimum size=1mm, shape=circle, draw=black, fill=zxred, tikzit fill={rgb,255: red,221; green,165; blue,165}]
\tikzstyle{light X dot}=[Z dot, shape=circle, draw={gray!80!white}, fill=zxred, tikzit fill={rgb,255: red,221; green,165; blue,165}]
\tikzstyle{hadamard}=[fill=zxhad, draw=black, shape=rectangle, inner sep=0.6mm, minimum height=1.5mm, minimum width=1.5mm, tikzit fill=yellow, font={\footnotesize\boldmath}]
\tikzstyle{bold hadamard}=[fill=zxhad, draw=black, shape=rectangle, inner sep=0.6mm, minimum height=1.5mm, minimum width=1.5mm, line width=1.2pt, tikzit fill=yellow, font={\footnotesize\boldmath}]
\tikzstyle{paulibox}=[fill=zxpaulibox, draw=black, shape=rectangle, inner sep=0.6mm, minimum height=5mm, minimum width=5mm, font={\footnotesize}, text height=1.5ex, text depth=0.25ex, tikzit fill={rgb,255: red,138; green,159; blue,198}]
\tikzstyle{big paulibox}=[fill=zxpaulibox, draw=black, shape=rectangle, inner sep=0.6mm, minimum height=15mm, minimum width=5mm, font={\footnotesize}, text height=1.5ex, text depth=0.25ex, tikzit fill={rgb,255: red,138; green,159; blue,198}]
\tikzstyle{vertex}=[inner sep=0.2mm, minimum size=1mm, shape=circle, draw=black, fill=black]
\tikzstyle{vertex set}=[inner sep=0.2mm, minimum size=1mm, shape=circle, draw=black, fill=white, font={\footnotesize\boldmath}]
\tikzstyle{small black dot}=[fill=black, draw=black, shape=circle, inner sep=0pt, minimum width=1.2mm, tikzit category=circuit]
\tikzstyle{cnot ctrl}=[fill=black, draw=black, shape=circle, inner sep=0pt, minimum width=1.2mm, tikzit category=circuit]
\tikzstyle{cnot targ}=[fill=white, draw=white, shape=circle, tikzit category=circuit, label={center:$\bigoplus$}, inner sep=0pt, minimum width=2.1mm, tikzit fill={rgb,255: red,102; green,204; blue,255}, tikzit draw=black]
\tikzstyle{left state}=[fill=white, draw=black, shape=regular polygon, regular polygon sides=3, regular polygon rotate=-30, scale=0.7, inner sep=0.5pt, minimum size=1cm]
\tikzstyle{ket}=[fill=white, draw=black, shape=regular polygon, regular polygon sides=3, regular polygon rotate=-30, scale=0.7, inner sep=0.5pt, minimum size=1cm]
\tikzstyle{right state}=[fill=white, draw=black, shape=regular polygon, regular polygon sides=3, regular polygon rotate=30, scale=0.7, inner sep=0.5pt, minimum size=1cm]
\tikzstyle{bra}=[fill=white, draw=black, shape=regular polygon, regular polygon sides=3, regular polygon rotate=30, scale=0.7, inner sep=0.5pt, minimum size=1cm]
\tikzstyle{wide left state}=[fill=white, draw=black, minimum height=8mm, minimum width=1.5cm, shape=isosceles triangle, isosceles triangle stretches, shape border rotate=180, scale=0.7, inner sep=1pt]
\tikzstyle{wide right state}=[fill=white, draw=black, minimum height=8mm, minimum width=1.5cm, shape=isosceles triangle, isosceles triangle stretches, scale=0.7, inner sep=1pt]
\tikzstyle{scalar}=[shape=rectangle, text height=1.5ex, text depth=0.25ex, yshift=0.5mm, fill=white, draw=black, minimum height=5mm, yshift=-0.5mm, minimum width=5mm, font={\small}]
\tikzstyle{pointer edge part}=[very thick, gray]
\tikzstyle{white dot}=[Z dot, tikzit fill={rgb,255: red,216; green,248; blue,216}]
\tikzstyle{white phase dot}=[Z phase dot, tikzit fill={rgb,255: red,216; green,248; blue,216}, tikzit draw=blue]
\tikzstyle{gray dot}=[X dot, tikzit fill={rgb,255: red,221; green,165; blue,165}]
\tikzstyle{gray phase dot}=[X phase dot, tikzit fill={rgb,255: red,221; green,165; blue,165}, tikzit draw=blue]
\tikzstyle{small hadamard}=[hadamard, tikzit shape=rectangle, tikzit fill=yellow]
\tikzstyle{clabel}=[fill=white, draw=none, shape=rectangle, tikzit fill={rgb,255: red,56; green,255; blue,242}, font={\footnotesize}, inner sep=1pt]
\tikzstyle{diag label}=[fill=white, draw=none, shape=circle, tikzit fill={rgb,255: red,56; green,255; blue,242}, inner sep=1pt]
\tikzstyle{empty diagram}=[draw={gray!40!white}, dashed, shape=rectangle, minimum width=1cm, minimum height=1cm]
\tikzstyle{empty diagram small}=[draw={gray!50!white}, dashed, shape=rectangle, minimum width=0.6cm, minimum height=0.5cm]
\tikzstyle{growbox}=[fill=zxpaulibox, draw=black, shape=Ebox, tikzit shape=rectangle, minimum height=6mm]
\tikzstyle{rmat}=[draw, signal, fill={zx_grey}, signal to=east, signal from=west, inner sep=1pt, minimum height=6pt, tikzit fill={rgb,255: red,191; green,191; blue,191}, tikzit shape=rectangle, very thick]
\tikzstyle{lmat}=[draw, signal, fill={zx_grey}, signal to=west, signal from=east, inner sep=1pt, minimum height=6pt, tikzit fill={rgb,255: red,164; green,162; blue,191}, tikzit shape=rectangle, very thick]
\tikzstyle{umat}=[draw, signal, fill={zx_grey}, signal to=north, signal from=south, inner sep=1pt, minimum width=6pt, tikzit fill={rgb,255: red,187; green,159; blue,191}, tikzit shape=rectangle, very thick]
\tikzstyle{dmat}=[draw, signal, fill={zx_grey}, signal to=south, signal from=north, inner sep=1pt, minimum width=6pt, tikzit fill={rgb,255: red,222; green,176; blue,219}, tikzit shape=rectangle, very thick]
\tikzstyle{A box}=[gate, fill=normRed]
\tikzstyle{B box}=[medium box, fill=normRed]
\tikzstyle{C box}=[small gate, fill=normRed]
\tikzstyle{D box}=[medium box, fill=normGreen]
\tikzstyle{E box}=[gate, fill=normGreen]
\tikzstyle{F box}=[small gate, fill=normGreen]
\tikzstyle{vertex number}=[fill=white, draw=black, shape=circle, inner sep=.5pt]
\tikzstyle{vertex number 2}=[fill=white, draw=IMSBlue, shape=circle, inner sep=.5pt]

\tikzstyle{hadamard edge}=[-, dashed, dash pattern=on 2pt off 0.5pt, thick, draw={rgb,255: red,68; green,136; blue,255}]
\tikzstyle{hadamard edge alt}=[-, dashed, dash pattern=on 2pt off 0.5pt, thick, draw={rgb,255: red,150; green,200; blue,255}]
\tikzstyle{box edge}=[-, dashed, dash pattern=on 2pt off 0.5pt, thick, draw={rgb,255: red,203; green,192; blue,225}]
\tikzstyle{brace edge}=[-, tikzit draw=blue, decorate, decoration={brace,amplitude=1mm,raise=-1mm}]
\tikzstyle{diredge}=[->]
\tikzstyle{double edge}=[-, double, shorten <=-1mm, shorten >=-1mm, double distance=2pt]
\tikzstyle{gray edge}=[-, {gray!80!white}]
\tikzstyle{pointer edge}=[->, very thick, gray]
\tikzstyle{boldedge}=[-, line width=1.6pt, shorten <=-0.17mm, shorten >=-0.17mm]
\tikzstyle{highlight edge}=[-, line width=1.6pt, shorten <=-0.17mm, shorten >=-0.17mm, draw=blue]
\tikzstyle{simple}=[-]
\tikzstyle{dashed edge}=[-, dashed, dash pattern=on 2pt off 0.5pt, draw=black]
\tikzstyle{light dashed edge}=[-, dashed, dash pattern=on 2pt off 0.5pt, draw={rgb,255: red,191; green,191; blue,191}]
\tikzstyle{Z normal}=[-, tikzit fill={rgb,255: red,221; green,165; blue,165}, fill=normRed]
\tikzstyle{X normal}=[-, tikzit fill={rgb,255: red,216; green,248; blue,216}, fill=normGreen]
\tikzstyle{normal}=[-, tikzit fill={rgb,255: red,216; green,248; blue,216}, fill=normYellow]
\tikzstyle{light-fill}=[-, fill={rgb,255: red,230; green,230; blue,230}, tikzit fill={rgb,255: red,191; green,191; blue,191}, draw=none, tikzit draw={rgb,255: red,191; green,191; blue,191}]

\newcommand{\ufig}[2][1]{$\vcenter{\hbox{\scalebox{#1}{{\tikzstyle{every picture}=[tikzfig]\input{figures/#2.tikz}}}}}$}

\newcommand{\aid}[1]{\mbox{\textsf{\small #1}}}
\newcommand{\aidop}[1]{\ifmmode\mathbin{\aid{#1}}\else\aid{#1}\fi}
\newcommand{\aidrel}[1]{\ifmmode\mathrel{\aid{#1}}\else\aid{#1}\fi}

\begin{document}

\title{Normal Forms and Complete Relations for Circuits in Agda}

\author{Xiaoning Bian}
\orcid{0009-0002-1042-669X}
\affiliation{%
  \institution{Tsinghua University}
  \city{Beijing}
  \country{China}}
\email{bian@mail.tsinghua.edu.cn}

\author{Yuan Feng}
\correspondingauthor
\orcid{0000-0002-3097-3896}
\affiliation{%
  \institution{Tsinghua University}
  \city{Beijing}
  \country{China}}
\email{yuan\_feng@tsinghua.edu.cn}

\begin{abstract}
A set of equations between circuits is \emph{complete} if it derives every
true equation.  In a normal-form proof of completeness, much of the work
lies in showing how to reduce a normal form after adjoining one more gate.
Recent completeness proofs come with appendices hundreds
of pages long, produced and checked by hand.  This paper presents an Agda
library that mechanizes the mathematics these proofs rest on: monoid and group
presentations by generators and relations, normal forms, and completeness of
finite relation sets, specialized to wire-indexed circuits.  The library
represents presented monoids as inductively generated congruences on free
monoids of words, represents normal forms as the standard-library function
bundles (injections, right inverses, bijections) out of the word setoid, and
derives completeness from soundness and a \emph{unique normal form} witness by
a single generic lemma.  Normal forms are built by a constructive,
monoid-level version of the Reidemeister--Schreier method: a tower of
verified coset tables adds one digit to the normal form per level, as in a
mixed-radix number system.  The library also provides presentation-level
constructions (direct, semidirect, and amalgamated free products, and group
extensions) with their presentation theorems, and the corresponding semantic
constructions, which are intended for the Agda standard library.  We validate
the design on a catalogue of verified presentations:
symmetric groups as circuits (sound and complete for both an endofunction and
a permutation semantics), cyclic groups, wreath products $\mathbb{Z}_N \wr S_n$
(at $N = 2$, the signed permutations), Pauli groups modulo phases in all prime
dimensions, and, as amalgamated free products with verified alternating normal
forms, the presentations of the single-qubit and single-qutrit Clifford+$T$
monoids and of $U_3(\mathbb{Z}[\tfrac12,i])$.  All headline results are
collected in a single index module; the verified development is safe, axiom-free
Agda.  We compare in detail
with formalizations of group presentations, normal forms, and quantum circuits
in Agda, Lean, Rocq, and Isabelle.

\end{abstract}



\maketitle

\section{Introduction}\label{sec:intro}

Circuit design relies on equational reasoning.  Optimizing a quantum
circuit, verifying a compiler pass, and normalizing a diagram are all
justified by a finite set of local rewrite rules that preserve the
semantics of the circuit.  Two questions arise for any such rule set.
\emph{Soundness}, that every rule preserves the semantics, is usually easy:
it is a finite collection of small algebraic identities.
\emph{Completeness}, that every true semantic equation can be derived from
the rules, is much harder.

Completeness proofs are large computations.  The standard strategy defines
a \emph{normal form} for circuits, proves that every circuit can be rewritten
to normal form using the rules, and proves that distinct normal forms denote
distinct semantic objects.  The middle step accounts for most of the work:
for every way a generator can meet a normal form, one must
show that the combination reduces to a normal form again.  The case
analysis is mechanical but very large, as the sizes of published proofs
show.  Selinger's presentation of the $n$-qubit Clifford group already
required a substantial rewrite system and normal form
analysis~\cite{selinger2015clifford}.  Its recent qutrit
analogue by Li, Mosca, Ross, van~de~Wetering, and Zhao, the first
completeness result for a gate set of quantum circuits in an odd prime
dimension, runs to more than fifty pages, with a further supplement of
167 pages of derivations~\cite{li2025qutrit}.  The first complete equational theory for full
quantum circuits, by Cl\'ement, Heurtel, Mansfield, Perdrix, and
Valiron~\cite{clement2023complete}, and its minimal
refinement~\cite{clement2024minimal}, rest on intricate normal-form
manipulations; and Cl\'ement's complete equational theory for
Real-Clifford+$CH$ circuits consists of a thirteen-page paper attached to an
appendix of 264 pages of equational
derivations~\cite{clement2026realcliffordch}.  Tens of pages of mechanical
calculation likewise support the
rig-categorical completeness theorems for quantum programming
languages~\cite{carette2024squareroots,fang2026hadamard} and the
recent uniform simplification of the rule sets of six near-Clifford circuit
fragments~\cite{blake2026simpler}.  Such proofs are long, regular, and
computational, and are therefore good candidates for machine checking.

This paper reports on an Agda library for this purpose: a
reusable formal framework for \emph{presentations of monoids and
groups by generators and relations}, for \emph{normal forms}, and for
\emph{completeness} of finite relation sets, with circuits (words of
wire-indexed gates) as the motivating instance. The library is written in
safe, axiom-free Agda~\cite{norell2007thesis,bove2009agda} on top of the
Agda standard library~\cite{daggitt2025stdlib,agdastdlib}.
A presented monoid is a setoid on words rather than a quotient type, so
the library is written in plain Agda, not cubical Agda.  The library
computes with syntax, and its central devices (sections of normal forms,
coset representatives) are maps into the syntax, which need
concrete representatives as values; \S\ref{sec:setoids} discusses the
trade-off.

\subsection{Contributions}\label{sec:contributions}

\begin{enumerate}
\item[(0)] \textbf{A library for circuit normal forms and complete relation
  sets.}  An Agda library in which a relation set is an inductive
  family on words, the presented monoid is the inductively generated
  congruence closure (\S\ref{sec:presentations}), normal-form witnesses are the
  standard library's function bundles out of the word setoid, and a single
  generic lemma, \aid{by-normalization}, turns soundness plus a \emph{unique
  normal form} into completeness (\S\ref{sec:nfproperty}).  A verified,
  monoid-level Reidemeister--Schreier engine constructs normal forms from
  verified coset tables, one per level
  (\S\ref{sec:engine}).

\item[(1)] \textbf{Inductively defined circuits, for easy normalization and
  completeness proofs.}  A generic circuit layer represents $n$-wire circuits
  as words over wire-indexed generators, and extends any gate-specific
  relation set with the structural rules (shift congruence and
  far-commutation) shared by every circuit presentation
  (\S\ref{sec:circuits}).  The indexing discipline is chosen so that
  coset tables can be written by direct pattern matching and, where their
  obligations reduce to computation, checked by Agda's evaluator.

\item[(2)] \textbf{A catalogue of verified presentations}
  (\S\ref{sec:examples}).  Seven nontrivial example families, several
  of them formalizing presentations from the research literature:
  cyclic groups of every positive order (and $\mathbb{N}$ at order
  $0$); symmetric groups $S_n$ presented as circuits, proved sound and
  complete for a loose endofunction semantics and presented for the
  group of permutations of $\mathsf{Fin}\;n$; wreath products
  $\mathbb{Z}_N \wr S_n$, the groups of \emph{signed permutations} and
  their generalizations; Pauli groups modulo phases,
  $(\mathbb{Z}_p \times \mathbb{Z}_p)^n$ for every prime $p$,
  assembled purely compositionally; and, as amalgamated free products
  with verified alternating normal forms, the presentations of the
  single-qubit and single-qutrit Clifford+$T$ monoids and of the
  monoid $U_3(\mathbb{Z}[\tfrac12,i])$.

\item[(3)] \textbf{Monoid-level amalgamation and
    Reidemeister--Schreier.}  We formalize coset tables, the
  Reidemeister--Schreier subgroup presentation, and amalgamated free
  products with their alternating normal form at the monoid level, where
  inverses are not primitive (\S\ref{sec:engine},
  \S\ref{sec:constructions}).  The method follows Bian and Selinger's
  syntactic verifications~\cite{bian2023cliffordt2,bian2023cliffordcs3,
    bian2023thesis}; here each construction is a reusable module with its
  hypotheses packaged as a record.

\item[(4)] \textbf{Presentation theorems for the product
    constructions.}  The library builds relation sets for direct,
  $n$-fold direct, semidirect, and amalgamated free products from
  those of the factors, and lifts normal forms through each
  construction.  For each, we also prove the presentation theorem: if
  the factors' relation sets present groups, the constructed relation
  set presents the corresponding product of those groups.  A further
  theorem does the same for an arbitrary group extension, given the
  normalization and realization data of \S\ref{sec:constructions}.

\item[(5)] \textbf{New constructions and theorems for the standard
  library.}  The semantic
  side of the above required algebraic constructions absent from the Agda
  standard library: semidirect products of monoids and groups via a bundled
  \aid{Action} record; amalgamated free products of monoids as a pushout
  presented by words over the disjoint union of carriers; group extensions
  and split extensions as short exact sequences; monoid epimorphisms; the
  fact that a monoid homomorphism between groups is a group homomorphism;
  and the group of permutations of $\mathsf{Fin}\;n$ bundled as a
  \aid{Group} (\S\ref{sec:forstdlib}).
\end{enumerate}

All headline results in this paper are collected in a single index
module \texttt{MainTheorems.agda}. Each typechecks with the flags
\texttt{-{}-cubical-compatible} and \texttt{-{}-safe}: there are no
postulates, no unfinished holes, and no appeals to classical axioms
anywhere in the development.  Therefore, the development is
constructive; in particular, normal forms and their sections
compute, so a normal form with decidable equality gives a decision
procedure for the word problem of its theory.  Wherever a statement in
this paper is not formalized, the text says so explicitly, and
\S\ref{sec:threats} discusses the main such gap.

\subsection{Paper Outline}

Section~\ref{sec:permutations} walks through the permutation case
study.  Section~\ref{sec:preliminaries} recalls the group theory
behind the design: normal forms from subgroup chains and transversals,
and their mixed-radix reading.  Section~\ref{sec:design} presents the
library layer by layer: words and presentations, the zoo of
normal-form witnesses, inductively defined circuits, the
Reidemeister--Schreier engine, and the product constructions with
their presentation theorems.  Section~\ref{sec:examples} is the
catalogue of verified results. Section~\ref{sec:related} surveys
related work and compares with other formalizations.
Section~\ref{sec:discussion} discusses design decisions, statistics,
and lessons learned, and Section~\ref{sec:conclusion} concludes.

\begin{code}[hide]%
\>[0]\AgdaKeyword{module}\AgdaSpace{}%
\AgdaModule{paper.sections.permutations}\AgdaSpace{}%
\AgdaKeyword{where}\<%
\\
\\[\AgdaEmptyExtraSkip]%
\>[0]\AgdaKeyword{open}\AgdaSpace{}%
\AgdaKeyword{import}\AgdaSpace{}%
\AgdaModule{Data.Nat}\AgdaSpace{}%
\AgdaKeyword{using}\AgdaSpace{}%
\AgdaSymbol{(}\AgdaDatatype{ℕ}\AgdaSpace{}%
\AgdaSymbol{;}\AgdaSpace{}%
\AgdaInductiveConstructor{suc}\AgdaSymbol{)}\<%
\\
\>[0]\AgdaKeyword{open}\AgdaSpace{}%
\AgdaKeyword{import}\AgdaSpace{}%
\AgdaModule{Data.Product}\AgdaSpace{}%
\AgdaKeyword{using}\AgdaSpace{}%
\AgdaSymbol{(}\AgdaOperator{\AgdaFunction{\AgdaUnderscore{}×\AgdaUnderscore{}}}\AgdaSpace{}%
\AgdaSymbol{;}\AgdaSpace{}%
\AgdaOperator{\AgdaInductiveConstructor{\AgdaUnderscore{},\AgdaUnderscore{}}}\AgdaSpace{}%
\AgdaSymbol{;}\AgdaSpace{}%
\AgdaField{proj₁}\AgdaSpace{}%
\AgdaSymbol{;}\AgdaSpace{}%
\AgdaField{proj₂}\AgdaSymbol{)}\<%
\\
\>[0]\AgdaKeyword{open}\AgdaSpace{}%
\AgdaKeyword{import}\AgdaSpace{}%
\AgdaModule{Data.Unit}\AgdaSpace{}%
\AgdaKeyword{using}\AgdaSpace{}%
\AgdaSymbol{(}\AgdaRecord{⊤}\AgdaSymbol{)}\<%
\\
\>[0]\AgdaKeyword{open}\AgdaSpace{}%
\AgdaKeyword{import}\AgdaSpace{}%
\AgdaModule{Level}\AgdaSpace{}%
\AgdaKeyword{using}\AgdaSpace{}%
\AgdaSymbol{(}\AgdaFunction{0ℓ}\AgdaSymbol{)}\<%
\\
\>[0]\AgdaKeyword{open}\AgdaSpace{}%
\AgdaKeyword{import}\AgdaSpace{}%
\AgdaModule{Relation.Binary}\AgdaSpace{}%
\AgdaKeyword{using}\AgdaSpace{}%
\AgdaSymbol{(}\AgdaFunction{Rel}\AgdaSymbol{)}\<%
\\
\\[\AgdaEmptyExtraSkip]%
\>[0]\AgdaKeyword{open}\AgdaSpace{}%
\AgdaKeyword{import}\AgdaSpace{}%
\AgdaModule{Notations}\AgdaSpace{}%
\AgdaKeyword{using}\AgdaSpace{}%
\AgdaSymbol{(}\AgdaInductiveConstructor{₁₊}\AgdaSpace{}%
\AgdaSymbol{;}\AgdaSpace{}%
\AgdaInductiveConstructor{₂₊}\AgdaSpace{}%
\AgdaSymbol{;}\AgdaSpace{}%
\AgdaInductiveConstructor{₃₊}\AgdaSymbol{)}\<%
\\
\>[0]\AgdaKeyword{open}\AgdaSpace{}%
\AgdaKeyword{import}\AgdaSpace{}%
\AgdaModule{Word.Base}\AgdaSpace{}%
\AgdaKeyword{using}\AgdaSpace{}%
\AgdaSymbol{(}\AgdaDatatype{Word}\AgdaSpace{}%
\AgdaSymbol{;}\AgdaSpace{}%
\AgdaFunction{WRel}\AgdaSpace{}%
\AgdaSymbol{;}\AgdaSpace{}%
\AgdaOperator{\AgdaInductiveConstructor{[\AgdaUnderscore{}]ʷ}}\AgdaSpace{}%
\AgdaSymbol{;}\AgdaSpace{}%
\AgdaInductiveConstructor{ε}\AgdaSpace{}%
\AgdaSymbol{;}\AgdaSpace{}%
\AgdaOperator{\AgdaInductiveConstructor{\AgdaUnderscore{}•\AgdaUnderscore{}}}\AgdaSpace{}%
\AgdaSymbol{;}\AgdaSpace{}%
\AgdaFunction{wmap}\AgdaSymbol{)}\<%
\\
\>[0]\AgdaKeyword{open}\AgdaSpace{}%
\AgdaKeyword{import}\AgdaSpace{}%
\AgdaModule{Presentation.Definitions}\AgdaSpace{}%
\AgdaKeyword{using}\AgdaSpace{}%
\AgdaSymbol{(}\AgdaOperator{\AgdaRecord{\AgdaUnderscore{}IsPresentationOf\AgdaUnderscore{}}}\AgdaSymbol{)}\<%
\\
\>[0]\AgdaKeyword{open}\AgdaSpace{}%
\AgdaKeyword{import}\AgdaSpace{}%
\AgdaModule{Examples.Groups.Symmetric.Tight.Semantics}\AgdaSpace{}%
\AgdaKeyword{using}\AgdaSpace{}%
\AgdaSymbol{(}\AgdaFunction{Permutation′-group}\AgdaSymbol{)}\<%
\\
\>[0]\AgdaKeyword{import}\AgdaSpace{}%
\AgdaModule{Examples.Groups.Symmetric.Theorems}\AgdaSpace{}%
\AgdaSymbol{as}\AgdaSpace{}%
\AgdaModule{SymThm}\<%
\\
\\[\AgdaEmptyExtraSkip]%
\>[0]\AgdaKeyword{postulate}\<%
\\
\>[0][@{}l@{\AgdaIndent{0}}]%
\>[2]\AgdaPostulate{proof-elided}\AgdaSpace{}%
\AgdaSymbol{:}\AgdaSpace{}%
\AgdaSymbol{∀}\AgdaSpace{}%
\AgdaSymbol{\{}\AgdaBound{ℓ}\AgdaSymbol{\}}\AgdaSpace{}%
\AgdaSymbol{\{}\AgdaBound{A}\AgdaSpace{}%
\AgdaSymbol{:}\AgdaSpace{}%
\AgdaPrimitive{Set}\AgdaSpace{}%
\AgdaBound{ℓ}\AgdaSymbol{\}}\AgdaSpace{}%
\AgdaSymbol{→}\AgdaSpace{}%
\AgdaBound{A}\<%
\\
\\[\AgdaEmptyExtraSkip]%
\>[0]\AgdaKeyword{private}\AgdaSpace{}%
\AgdaKeyword{variable}\<%
\\
\>[0][@{}l@{\AgdaIndent{0}}]%
\>[2]\AgdaGeneralizable{n}\AgdaSpace{}%
\AgdaSymbol{:}\AgdaSpace{}%
\AgdaDatatype{ℕ}\<%
\end{code}
\section{A Worked Example: Permutation Circuits}\label{sec:permutations}

We begin with the library's smallest complete case study: circuits built
from a single two-wire gate, the swap $\sigma$, which present the symmetric
groups.  The walkthrough covers, in order, the steps that the later
tower-based case studies repeat: gates, relations, cosets, the tower,
semantics, uniqueness, completeness, and the presentation theorem.

\subsection{Gates, Generators, and Circuits}

A gate set is a family $\aid{Gate} : \mathbb{N} \to \mathsf{Set}$ giving,
for each arity, the gates that act on that many adjacent wires.  For
permutations there is exactly one gate, on two wires.  The generic circuit
layer (\S\ref{sec:circuits}) uses the one- and two-wire gates of such a
family to form wire-indexed \emph{generators}: a gate placed at the bottom of
$n$ wires, or a
generator shifted up by one wire.

\begin{code}%
\>[0]\AgdaKeyword{data}\AgdaSpace{}%
\AgdaDatatype{Gate}\AgdaSpace{}%
\AgdaSymbol{:}\AgdaSpace{}%
\AgdaDatatype{ℕ}\AgdaSpace{}%
\AgdaSymbol{→}\AgdaSpace{}%
\AgdaPrimitive{Set}\AgdaSpace{}%
\AgdaKeyword{where}\<%
\\
\>[0][@{}l@{\AgdaIndent{0}}]%
\>[2]\AgdaInductiveConstructor{σ-gate}\AgdaSpace{}%
\AgdaSymbol{:}\AgdaSpace{}%
\AgdaDatatype{Gate}\AgdaSpace{}%
\AgdaNumber{2}\<%
\\
\>[0]\AgdaKeyword{data}\AgdaSpace{}%
\AgdaDatatype{Gen}\AgdaSpace{}%
\AgdaSymbol{:}\AgdaSpace{}%
\AgdaDatatype{ℕ}\AgdaSpace{}%
\AgdaSymbol{→}\AgdaSpace{}%
\AgdaPrimitive{Set}\AgdaSpace{}%
\AgdaKeyword{where}\<%
\\
\>[0][@{}l@{\AgdaIndent{0}}]%
\>[2]\AgdaInductiveConstructor{gate₁}\AgdaSpace{}%
\AgdaSymbol{:}\AgdaSpace{}%
\AgdaDatatype{Gate}\AgdaSpace{}%
\AgdaNumber{1}\AgdaSpace{}%
\AgdaSymbol{→}\AgdaSpace{}%
\AgdaDatatype{Gen}\AgdaSpace{}%
\AgdaSymbol{(}\AgdaInductiveConstructor{₁₊}\AgdaSpace{}%
\AgdaGeneralizable{n}\AgdaSymbol{)}%
\>[34]\AgdaComment{--\ a\ 1-ary\ gate\ on\ the\ bottom\ wire}\<%
\\
\>[2]\AgdaInductiveConstructor{gate₂}\AgdaSpace{}%
\AgdaSymbol{:}\AgdaSpace{}%
\AgdaDatatype{Gate}\AgdaSpace{}%
\AgdaNumber{2}\AgdaSpace{}%
\AgdaSymbol{→}\AgdaSpace{}%
\AgdaDatatype{Gen}\AgdaSpace{}%
\AgdaSymbol{(}\AgdaInductiveConstructor{₂₊}\AgdaSpace{}%
\AgdaGeneralizable{n}\AgdaSymbol{)}%
\>[34]\AgdaComment{--\ a\ 2-ary\ gate\ on\ the\ bottom\ two\ wires}\<%
\\
\>[2]\AgdaOperator{\AgdaInductiveConstructor{\AgdaUnderscore{}↥}}%
\>[7]\AgdaSymbol{:}\AgdaSpace{}%
\AgdaDatatype{Gen}\AgdaSpace{}%
\AgdaGeneralizable{n}\AgdaSpace{}%
\AgdaSymbol{→}\AgdaSpace{}%
\AgdaDatatype{Gen}\AgdaSpace{}%
\AgdaSymbol{(}\AgdaInductiveConstructor{₁₊}\AgdaSpace{}%
\AgdaGeneralizable{n}\AgdaSymbol{)}%
\>[34]\AgdaComment{--\ the\ same\ generator,\ one\ wire\ higher}\<%
\end{code}
\begin{code}[hide]%
\>[0]\AgdaFunction{Circuit}\AgdaSpace{}%
\AgdaSymbol{:}\AgdaSpace{}%
\AgdaDatatype{ℕ}\AgdaSpace{}%
\AgdaSymbol{→}\AgdaSpace{}%
\AgdaPrimitive{Set}\<%
\\
\>[0]\AgdaFunction{Circuit}\AgdaSpace{}%
\AgdaBound{n}\AgdaSpace{}%
\AgdaSymbol{=}\AgdaSpace{}%
\AgdaDatatype{Word}\AgdaSpace{}%
\AgdaSymbol{(}\AgdaDatatype{Gen}\AgdaSpace{}%
\AgdaBound{n}\AgdaSymbol{)}\<%
\\
\\[\AgdaEmptyExtraSkip]%
\>[0]\AgdaFunction{CRel}\AgdaSpace{}%
\AgdaSymbol{:}\AgdaSpace{}%
\AgdaDatatype{ℕ}\AgdaSpace{}%
\AgdaSymbol{→}\AgdaSpace{}%
\AgdaPrimitive{Set₁}\<%
\\
\>[0]\AgdaFunction{CRel}\AgdaSpace{}%
\AgdaBound{n}\AgdaSpace{}%
\AgdaSymbol{=}\AgdaSpace{}%
\AgdaFunction{Rel}\AgdaSpace{}%
\AgdaSymbol{(}\AgdaFunction{Circuit}\AgdaSpace{}%
\AgdaBound{n}\AgdaSymbol{)}\AgdaSpace{}%
\AgdaFunction{0ℓ}\<%
\\
\\[\AgdaEmptyExtraSkip]%
\>[0]\AgdaOperator{\AgdaFunction{\AgdaUnderscore{}↑}}\AgdaSpace{}%
\AgdaSymbol{:}\AgdaSpace{}%
\AgdaFunction{Circuit}\AgdaSpace{}%
\AgdaGeneralizable{n}\AgdaSpace{}%
\AgdaSymbol{→}\AgdaSpace{}%
\AgdaFunction{Circuit}\AgdaSpace{}%
\AgdaSymbol{(}\AgdaInductiveConstructor{₁₊}\AgdaSpace{}%
\AgdaGeneralizable{n}\AgdaSymbol{)}\<%
\\
\>[0]\AgdaOperator{\AgdaFunction{\AgdaUnderscore{}↑}}\AgdaSpace{}%
\AgdaSymbol{=}\AgdaSpace{}%
\AgdaFunction{wmap}\AgdaSpace{}%
\AgdaOperator{\AgdaInductiveConstructor{\AgdaUnderscore{}↥}}\<%
\\
\\[\AgdaEmptyExtraSkip]%
\>[0]\AgdaKeyword{pattern}\AgdaSpace{}%
\AgdaInductiveConstructor{σ-gen}\AgdaSpace{}%
\AgdaSymbol{=}\AgdaSpace{}%
\AgdaInductiveConstructor{gate₂}\AgdaSpace{}%
\AgdaInductiveConstructor{σ-gate}\<%
\\
\\[\AgdaEmptyExtraSkip]%
\>[0]\AgdaFunction{σ}\AgdaSpace{}%
\AgdaSymbol{:}\AgdaSpace{}%
\AgdaFunction{Circuit}\AgdaSpace{}%
\AgdaSymbol{(}\AgdaInductiveConstructor{₂₊}\AgdaSpace{}%
\AgdaGeneralizable{n}\AgdaSymbol{)}\<%
\\
\>[0]\AgdaFunction{σ}\AgdaSpace{}%
\AgdaSymbol{=}\AgdaSpace{}%
\AgdaOperator{\AgdaInductiveConstructor{[}}\AgdaSpace{}%
\AgdaInductiveConstructor{σ-gen}\AgdaSpace{}%
\AgdaOperator{\AgdaInductiveConstructor{]ʷ}}\<%
\\
\\[\AgdaEmptyExtraSkip]%
\>[0]\AgdaKeyword{infix}\AgdaSpace{}%
\AgdaNumber{4}\AgdaSpace{}%
\AgdaOperator{\AgdaDatatype{\AgdaUnderscore{}SRel,\AgdaUnderscore{}===\AgdaUnderscore{}}}\AgdaSpace{}%
\AgdaOperator{\AgdaDatatype{\AgdaUnderscore{}VRel,\AgdaUnderscore{}===\AgdaUnderscore{}}}\AgdaSpace{}%
\AgdaOperator{\AgdaPostulate{\AgdaUnderscore{}≈\AgdaUnderscore{}}}\<%
\\
\>[0]\AgdaKeyword{infixr}\AgdaSpace{}%
\AgdaNumber{10}\AgdaSpace{}%
\AgdaOperator{\AgdaInductiveConstructor{σ•\AgdaUnderscore{}}}\<%
\\
\\[\AgdaEmptyExtraSkip]%
\>[0]\AgdaKeyword{private}\AgdaSpace{}%
\AgdaKeyword{variable}\<%
\\
\>[0][@{}l@{\AgdaIndent{0}}]%
\>[2]\AgdaGeneralizable{w}\AgdaSpace{}%
\AgdaGeneralizable{v}\AgdaSpace{}%
\AgdaSymbol{:}\AgdaSpace{}%
\AgdaFunction{Circuit}\AgdaSpace{}%
\AgdaGeneralizable{n}\<%
\\
\\[\AgdaEmptyExtraSkip]%
\>[0]\AgdaKeyword{postulate}\<%
\\
\>[0][@{}l@{\AgdaIndent{0}}]%
\>[2]\AgdaOperator{\AgdaPostulate{\AgdaUnderscore{}≈\AgdaUnderscore{}}}\AgdaSpace{}%
\AgdaSymbol{:}\AgdaSpace{}%
\AgdaFunction{Circuit}\AgdaSpace{}%
\AgdaGeneralizable{n}\AgdaSpace{}%
\AgdaSymbol{→}\AgdaSpace{}%
\AgdaFunction{Circuit}\AgdaSpace{}%
\AgdaGeneralizable{n}\AgdaSpace{}%
\AgdaSymbol{→}\AgdaSpace{}%
\AgdaPrimitive{Set}\<%
\end{code}

Here $\aid{₁₊}\;n$ and $\aid{₂₊}\;n$ are the library's patterns for
$1 + n$ and $2 + n$, and $n$ is an implicit argument, so
$\aid{gate₂}\;\aid{σ-gate}$ is a generator at any width from two upward.
An $n$-wire \emph{circuit} is a word of generators, $\aid{Circuit}\;n =
\aid{Word}\;(\aid{Gen}\;n)$ (\S\ref{sec:words}): a single generator
$[\,g\,]$\aid{ʷ}, the empty circuit \aid{ε} ($n$ bare wires, the
identity circuit at that width), or a sequential composition
$c \aidop{•} c'$.  We write $\sigma$ for the one-letter circuit
$[\,\aid{gate₂}\;\aid{σ-gate}\;]$\aid{ʷ} and $c\;{\uparrow}$ for $c$ shifted
up one wire ($\aid{{\_}↑} = \aid{wmap}\;\aid{{\_}↥}$).  Read
diagrammatically, with wires drawn bottom-up and composition left to
right, $\sigma$ is a crossing of the bottom two wires, $\sigma\;{\uparrow}$
the same crossing one wire higher, and $\sigma \aidop{•} \sigma\;{\uparrow}
\aidop{•} \sigma$ the three-wire braid on the left of the
\aid{yang-baxter} equation drawn beside its axiom below.

\subsection{Relations: Two Axioms and the Structural Rules}

The group-specific relations are an inductive family indexed by the wire
count (a $\aid{WRel}\;X$ is a binary relation on words over $X$).  Each
axiom is stated once, with its gates at the bottom of the
wires and the width left open (the implicit $n$), so that one constructor
covers every width at which the axiom fits; the structural rule
\aid{cong↑} below moves it to any height.  (Agda names may be mixfix:
each underscore in \aid{{\_}SRel,{\_}==={\_}} marks an argument
position, so the family is applied as $n\;\aid{SRel,}\;w \aidrel{===}
v$, read ``$w = v$ is an axiom on $n$ wires''.)

\begin{center}
\begin{minipage}[c]{0.58\linewidth}
\begingroup\setlength{\mathindent}{0pt}
\begin{code}%
\>[0]\AgdaKeyword{data}\AgdaSpace{}%
\AgdaOperator{\AgdaDatatype{\AgdaUnderscore{}SRel,\AgdaUnderscore{}===\AgdaUnderscore{}}}\AgdaSpace{}%
\AgdaSymbol{:}\AgdaSpace{}%
\AgdaSymbol{(}\AgdaBound{n}\AgdaSpace{}%
\AgdaSymbol{:}\AgdaSpace{}%
\AgdaDatatype{ℕ}\AgdaSymbol{)}\AgdaSpace{}%
\AgdaSymbol{→}\AgdaSpace{}%
\AgdaFunction{WRel}\AgdaSpace{}%
\AgdaSymbol{(}\AgdaDatatype{Gen}\AgdaSpace{}%
\AgdaBound{n}\AgdaSymbol{)}\AgdaSpace{}%
\AgdaKeyword{where}\<%
\\
\>[0][@{}l@{\AgdaIndent{0}}]%
\>[2]\AgdaInductiveConstructor{order}%
\>[14]\AgdaSymbol{:}\AgdaSpace{}%
\AgdaSymbol{(}\AgdaInductiveConstructor{₂₊}\AgdaSpace{}%
\AgdaGeneralizable{n}\AgdaSymbol{)}\AgdaSpace{}%
\AgdaOperator{\AgdaDatatype{SRel,}}\AgdaSpace{}%
\AgdaFunction{σ}\AgdaSpace{}%
\AgdaOperator{\AgdaInductiveConstructor{•}}\AgdaSpace{}%
\AgdaFunction{σ}\AgdaSpace{}%
\AgdaOperator{\AgdaDatatype{===}}\AgdaSpace{}%
\AgdaInductiveConstructor{ε}\<%
\\
\>[2]\AgdaInductiveConstructor{yang-baxter}\AgdaSpace{}%
\AgdaSymbol{:}\AgdaSpace{}%
\AgdaSymbol{(}\AgdaInductiveConstructor{₃₊}\AgdaSpace{}%
\AgdaGeneralizable{n}\AgdaSymbol{)}\AgdaSpace{}%
\AgdaOperator{\AgdaDatatype{SRel,}}\AgdaSpace{}%
\AgdaFunction{σ}\AgdaSpace{}%
\AgdaOperator{\AgdaInductiveConstructor{•}}\AgdaSpace{}%
\AgdaFunction{σ}\AgdaSpace{}%
\AgdaOperator{\AgdaFunction{↑}}\AgdaSpace{}%
\AgdaOperator{\AgdaInductiveConstructor{•}}\AgdaSpace{}%
\AgdaFunction{σ}\AgdaSpace{}%
\AgdaOperator{\AgdaDatatype{===}}\AgdaSpace{}%
\AgdaFunction{σ}\AgdaSpace{}%
\AgdaOperator{\AgdaFunction{↑}}\AgdaSpace{}%
\AgdaOperator{\AgdaInductiveConstructor{•}}\AgdaSpace{}%
\AgdaFunction{σ}\AgdaSpace{}%
\AgdaOperator{\AgdaInductiveConstructor{•}}\AgdaSpace{}%
\AgdaFunction{σ}\AgdaSpace{}%
\AgdaOperator{\AgdaFunction{↑}}\<%
\end{code}
\endgroup
\end{minipage}
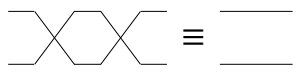
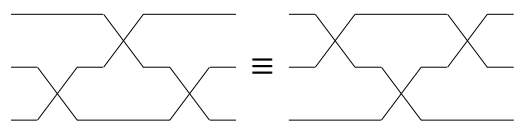
\hspace{2.5em}%
\begin{minipage}[c]{0.28\linewidth}\centering
  \ufig[0.35]{sn-order}\\[4pt]
  \ufig[0.35]{sn-yb}
\end{minipage}
\end{center}

\noindent These two axioms are two thirds of the Coxeter presentation of
$S_n$: $\sigma^2 = \varepsilon$ and the braid (Yang--Baxter) relation.
The third Coxeter relation, that swaps on disjoint wires commute, is not
specific to permutations: it is one of the relations every circuit
theory needs, whatever its gates.  By a
\emph{circuit theory} we mean a gate set with gate-specific relations,
presenting for each width $n$ a monoid of $n$-wire circuits under
sequential composition.  The structural rules are motivated by the
monoidal interpretation of circuits: the shift $c \mapsto c\;{\uparrow}$ is
``tensoring with an identity wire'', so equations must be preserved by
it, and the interchange law says that gates on disjoint wires commute.
With compatible tensor and symmetry laws, these are fixed-width
consequences of a symmetric monoidal category (a PROP).
Pen-and-paper presentations leave these rules implicit in the phrase ``a
congruence compatible with composition and tensor product''; the circuit
layer adds them generically: \aid{Lift-Relation} (\S\ref{sec:circuits}) extends any
gate-specific family to the full relation set \aid{{\_}VRel,{\_}==={\_}}
(here $\aid{CRel}\;n$ abbreviates $\aid{WRel}\;(\aid{Gen}\;n)$, the
binary relations on $n$-wire circuits):

\begin{code}%
\>[0]\AgdaKeyword{data}\AgdaSpace{}%
\AgdaOperator{\AgdaDatatype{\AgdaUnderscore{}VRel,\AgdaUnderscore{}===\AgdaUnderscore{}}}\AgdaSpace{}%
\AgdaSymbol{:}\AgdaSpace{}%
\AgdaSymbol{(}\AgdaBound{n}\AgdaSpace{}%
\AgdaSymbol{:}\AgdaSpace{}%
\AgdaDatatype{ℕ}\AgdaSymbol{)}\AgdaSpace{}%
\AgdaSymbol{→}\AgdaSpace{}%
\AgdaFunction{CRel}\AgdaSpace{}%
\AgdaBound{n}\AgdaSpace{}%
\AgdaKeyword{where}\<%
\\
\>[0][@{}l@{\AgdaIndent{0}}]%
\>[2]\AgdaInductiveConstructor{srel}%
\>[8]\AgdaSymbol{:}\AgdaSpace{}%
\AgdaGeneralizable{n}\AgdaSpace{}%
\AgdaOperator{\AgdaDatatype{SRel,}}\AgdaSpace{}%
\AgdaGeneralizable{w}\AgdaSpace{}%
\AgdaOperator{\AgdaDatatype{===}}\AgdaSpace{}%
\AgdaGeneralizable{v}\AgdaSpace{}%
\AgdaSymbol{→}\AgdaSpace{}%
\AgdaGeneralizable{n}\AgdaSpace{}%
\AgdaOperator{\AgdaDatatype{VRel,}}\AgdaSpace{}%
\AgdaGeneralizable{w}\AgdaSpace{}%
\AgdaOperator{\AgdaDatatype{===}}\AgdaSpace{}%
\AgdaGeneralizable{v}\<%
\\
\>[2]\AgdaInductiveConstructor{cong↑}\AgdaSpace{}%
\AgdaSymbol{:}\AgdaSpace{}%
\AgdaGeneralizable{n}\AgdaSpace{}%
\AgdaOperator{\AgdaDatatype{VRel,}}\AgdaSpace{}%
\AgdaGeneralizable{w}\AgdaSpace{}%
\AgdaOperator{\AgdaDatatype{===}}\AgdaSpace{}%
\AgdaGeneralizable{v}\AgdaSpace{}%
\AgdaSymbol{→}\AgdaSpace{}%
\AgdaSymbol{(}\AgdaInductiveConstructor{₁₊}\AgdaSpace{}%
\AgdaGeneralizable{n}\AgdaSymbol{)}\AgdaSpace{}%
\AgdaOperator{\AgdaDatatype{VRel,}}\AgdaSpace{}%
\AgdaGeneralizable{w}\AgdaSpace{}%
\AgdaOperator{\AgdaFunction{↑}}\AgdaSpace{}%
\AgdaOperator{\AgdaDatatype{===}}\AgdaSpace{}%
\AgdaGeneralizable{v}\AgdaSpace{}%
\AgdaOperator{\AgdaFunction{↑}}\<%
\\
\>[2]\AgdaInductiveConstructor{comm₁}\AgdaSpace{}%
\AgdaSymbol{:}\AgdaSpace{}%
\AgdaSymbol{(}\AgdaBound{h}\AgdaSpace{}%
\AgdaSymbol{:}\AgdaSpace{}%
\AgdaDatatype{Gate}\AgdaSpace{}%
\AgdaNumber{1}\AgdaSymbol{)}\AgdaSpace{}%
\AgdaSymbol{(}\AgdaBound{g}\AgdaSpace{}%
\AgdaSymbol{:}\AgdaSpace{}%
\AgdaDatatype{Gen}\AgdaSpace{}%
\AgdaGeneralizable{n}\AgdaSymbol{)}\AgdaSpace{}%
\AgdaSymbol{→}\AgdaSpace{}%
\AgdaSymbol{(}\AgdaInductiveConstructor{₁₊}\AgdaSpace{}%
\AgdaGeneralizable{n}\AgdaSymbol{)}\AgdaSpace{}%
\AgdaOperator{\AgdaDatatype{VRel,}}\AgdaSpace{}%
\AgdaOperator{\AgdaInductiveConstructor{[}}\AgdaSpace{}%
\AgdaBound{g}\AgdaSpace{}%
\AgdaOperator{\AgdaInductiveConstructor{↥}}\AgdaSpace{}%
\AgdaOperator{\AgdaInductiveConstructor{]ʷ}}\AgdaSpace{}%
\AgdaOperator{\AgdaInductiveConstructor{•}}\AgdaSpace{}%
\AgdaOperator{\AgdaInductiveConstructor{[}}\AgdaSpace{}%
\AgdaInductiveConstructor{gate₁}\AgdaSpace{}%
\AgdaBound{h}\AgdaSpace{}%
\AgdaOperator{\AgdaInductiveConstructor{]ʷ}}\AgdaSpace{}%
\AgdaOperator{\AgdaDatatype{===}}\AgdaSpace{}%
\AgdaOperator{\AgdaInductiveConstructor{[}}\AgdaSpace{}%
\AgdaInductiveConstructor{gate₁}\AgdaSpace{}%
\AgdaBound{h}\AgdaSpace{}%
\AgdaOperator{\AgdaInductiveConstructor{]ʷ}}\AgdaSpace{}%
\AgdaOperator{\AgdaInductiveConstructor{•}}\AgdaSpace{}%
\AgdaOperator{\AgdaInductiveConstructor{[}}\AgdaSpace{}%
\AgdaBound{g}\AgdaSpace{}%
\AgdaOperator{\AgdaInductiveConstructor{↥}}\AgdaSpace{}%
\AgdaOperator{\AgdaInductiveConstructor{]ʷ}}\<%
\\
\>[2]\AgdaInductiveConstructor{comm₂}\AgdaSpace{}%
\AgdaSymbol{:}\AgdaSpace{}%
\AgdaSymbol{(}\AgdaBound{h}\AgdaSpace{}%
\AgdaSymbol{:}\AgdaSpace{}%
\AgdaDatatype{Gate}\AgdaSpace{}%
\AgdaNumber{2}\AgdaSymbol{)}\AgdaSpace{}%
\AgdaSymbol{(}\AgdaBound{g}\AgdaSpace{}%
\AgdaSymbol{:}\AgdaSpace{}%
\AgdaDatatype{Gen}\AgdaSpace{}%
\AgdaGeneralizable{n}\AgdaSymbol{)}\AgdaSpace{}%
\AgdaSymbol{→}\AgdaSpace{}%
\AgdaSymbol{(}\AgdaInductiveConstructor{₂₊}\AgdaSpace{}%
\AgdaGeneralizable{n}\AgdaSymbol{)}\AgdaSpace{}%
\AgdaOperator{\AgdaDatatype{VRel,}}\AgdaSpace{}%
\AgdaOperator{\AgdaInductiveConstructor{[}}\AgdaSpace{}%
\AgdaBound{g}\AgdaSpace{}%
\AgdaOperator{\AgdaInductiveConstructor{↥}}\AgdaSpace{}%
\AgdaOperator{\AgdaInductiveConstructor{↥}}\AgdaSpace{}%
\AgdaOperator{\AgdaInductiveConstructor{]ʷ}}\AgdaSpace{}%
\AgdaOperator{\AgdaInductiveConstructor{•}}\AgdaSpace{}%
\AgdaOperator{\AgdaInductiveConstructor{[}}\AgdaSpace{}%
\AgdaInductiveConstructor{gate₂}\AgdaSpace{}%
\AgdaBound{h}\AgdaSpace{}%
\AgdaOperator{\AgdaInductiveConstructor{]ʷ}}\AgdaSpace{}%
\AgdaOperator{\AgdaDatatype{===}}\AgdaSpace{}%
\AgdaOperator{\AgdaInductiveConstructor{[}}\AgdaSpace{}%
\AgdaInductiveConstructor{gate₂}\AgdaSpace{}%
\AgdaBound{h}\AgdaSpace{}%
\AgdaOperator{\AgdaInductiveConstructor{]ʷ}}\AgdaSpace{}%
\AgdaOperator{\AgdaInductiveConstructor{•}}\AgdaSpace{}%
\AgdaOperator{\AgdaInductiveConstructor{[}}\AgdaSpace{}%
\AgdaBound{g}\AgdaSpace{}%
\AgdaOperator{\AgdaInductiveConstructor{↥}}\AgdaSpace{}%
\AgdaOperator{\AgdaInductiveConstructor{↥}}\AgdaSpace{}%
\AgdaOperator{\AgdaInductiveConstructor{]ʷ}}\<%
\end{code}

From here on,
$\approx$ denotes the monoid congruence generated by
$n\;\aid{VRel,{\_}==={\_}}$ (\S\ref{sec:presentations}): the least relation
containing the axioms and closed under reflexivity, symmetry, transitivity,
congruence of \aidop{•}, associativity, and the unit laws; deciding it is
the word problem.  Because $\sigma$ is its own inverse and shifts of
invertible generators are invertible, the presented monoid is a group; we
witness this by a \aid{Grouplike} proof (every generator has a left
inverse), by induction on generators.

\subsection{Cosets and the Staircase Normal Form}\label{sec:sn-cosets}

The engine behind the normal form is the Reidemeister--Schreier method
along the chain
$S_1 < S_2 < \dots < S_n < S_{n+1}$, where $S_n$ sits inside $S_{n+1}$ as
the permutations fixing the bottom wire (syntactically, the circuits
of the form $c\;{\uparrow}$).  Read as permutations, two circuits lie in the
same right coset of $S_n$ exactly when they send the bottom input wire
to the same position; there are $n{+}1$ choices, so $S_n$ has index $n{+}1$,
and a natural representative of each coset is the \emph{ascending
staircase} of swaps that carries that wire up.  In the library a coset
is not a set of circuits but a label naming one, an element of the finite type
$\aid{C}\;n$; here the label is the length of the staircase, as a unary
numeral bounded by the width, and the \emph{section} $[\,\_\,]^{\mathsf
c}$ maps it to the staircase circuit.  The representatives on three,
two, and one wires are drawn on the right:

\begin{center}
\begin{minipage}[c]{0.30\linewidth}
\begingroup\setlength{\mathindent}{0pt}
\begin{code}%
\>[0]\AgdaKeyword{data}\AgdaSpace{}%
\AgdaDatatype{C}\AgdaSpace{}%
\AgdaSymbol{:}\AgdaSpace{}%
\AgdaDatatype{ℕ}\AgdaSpace{}%
\AgdaSymbol{→}\AgdaSpace{}%
\AgdaPrimitive{Set}\AgdaSpace{}%
\AgdaKeyword{where}\<%
\\
\>[0][@{}l@{\AgdaIndent{0}}]%
\>[2]\AgdaInductiveConstructor{ε}%
\>[6]\AgdaSymbol{:}\AgdaSpace{}%
\AgdaDatatype{C}\AgdaSpace{}%
\AgdaGeneralizable{n}\<%
\\
\>[2]\AgdaOperator{\AgdaInductiveConstructor{σ•\AgdaUnderscore{}}}\AgdaSpace{}%
\AgdaSymbol{:}\AgdaSpace{}%
\AgdaDatatype{C}\AgdaSpace{}%
\AgdaGeneralizable{n}\AgdaSpace{}%
\AgdaSymbol{→}\AgdaSpace{}%
\AgdaDatatype{C}\AgdaSpace{}%
\AgdaSymbol{(}\AgdaInductiveConstructor{₁₊}\AgdaSpace{}%
\AgdaGeneralizable{n}\AgdaSymbol{)}\<%
\\
\>[0]\AgdaOperator{\AgdaFunction{[\AgdaUnderscore{}]ᶜ}}\AgdaSpace{}%
\AgdaSymbol{:}\AgdaSpace{}%
\AgdaDatatype{C}\AgdaSpace{}%
\AgdaGeneralizable{n}\AgdaSpace{}%
\AgdaSymbol{→}\AgdaSpace{}%
\AgdaFunction{Circuit}\AgdaSpace{}%
\AgdaSymbol{(}\AgdaInductiveConstructor{₁₊}\AgdaSpace{}%
\AgdaGeneralizable{n}\AgdaSymbol{)}\<%
\\
\>[0]\AgdaOperator{\AgdaFunction{[}}\AgdaSpace{}%
\AgdaInductiveConstructor{ε}\AgdaSpace{}%
\AgdaOperator{\AgdaFunction{]ᶜ}}%
\>[11]\AgdaSymbol{=}\AgdaSpace{}%
\AgdaInductiveConstructor{ε}\<%
\\
\>[0]\AgdaOperator{\AgdaFunction{[}}\AgdaSpace{}%
\AgdaOperator{\AgdaInductiveConstructor{σ•}}\AgdaSpace{}%
\AgdaBound{c}\AgdaSpace{}%
\AgdaOperator{\AgdaFunction{]ᶜ}}%
\>[11]\AgdaSymbol{=}\AgdaSpace{}%
\AgdaFunction{σ}\AgdaSpace{}%
\AgdaOperator{\AgdaInductiveConstructor{•}}\AgdaSpace{}%
\AgdaSymbol{(}\AgdaOperator{\AgdaFunction{[}}\AgdaSpace{}%
\AgdaBound{c}\AgdaSpace{}%
\AgdaOperator{\AgdaFunction{]ᶜ}}\AgdaSpace{}%
\AgdaOperator{\AgdaFunction{↑}}\AgdaSymbol{)}\<%
\end{code}
\endgroup
\end{minipage}
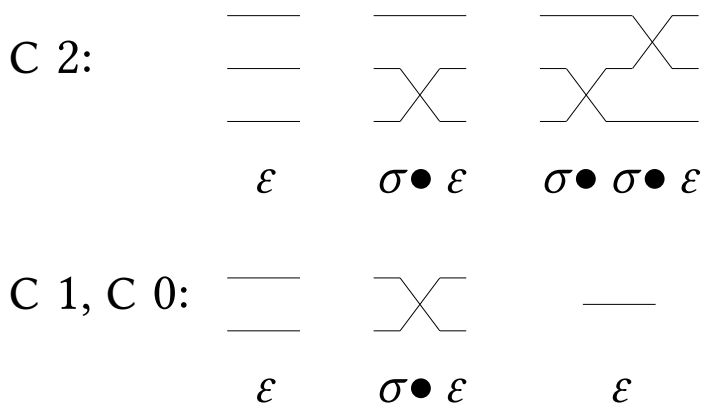
\hspace{2.5em}%
\begin{minipage}[c]{0.42\linewidth}\centering
\begin{tabular}{@{}l@{\;\;}ccc@{}}
  $\aid{C}\;2$: & \ufig[0.35]{c2-0} & \ufig[0.35]{c2-1} & \ufig[0.35]{c2-2}\\
  & $\varepsilon$ & $\sigma{\bullet}\:\varepsilon$ &
    $\sigma{\bullet}\:\sigma{\bullet}\:\varepsilon$\\[4pt]
  $\aid{C}\;1$, $\aid{C}\;0$: & \ufig[0.35]{c1-0} & \ufig[0.35]{c1-1} & \ufig[0.35]{c0-0}\\
  & $\varepsilon$ & $\sigma{\bullet}\:\varepsilon$ & $\varepsilon$
\end{tabular}
\end{minipage}
\end{center}

\noindent So $\aid{C}\;n$ has exactly the $n{+}1$ elements drawn above
for $n = 2, 1, 0$, matching the index of $S_n$ in $S_{n+1}$.  Why this yields
a normal form is the classical argument of \S\ref{sec:preliminaries}:
every $f \in S_{n+1}$ lies in one coset, say with representative $c_n$,
so $f = f' \cdot c_n$ with $f' \in S_n$, and recursing on $f'$ writes $f$
uniquely as $c_1 c_2 \cdots c_n$ with $c_k \in \aid{C}\;k$, a numeral
in the \emph{factorial number system}, one digit per level, since
$|S_{n+1}| = (n{+}1) \cdot n \cdots 2$.

\begin{figure}[h]
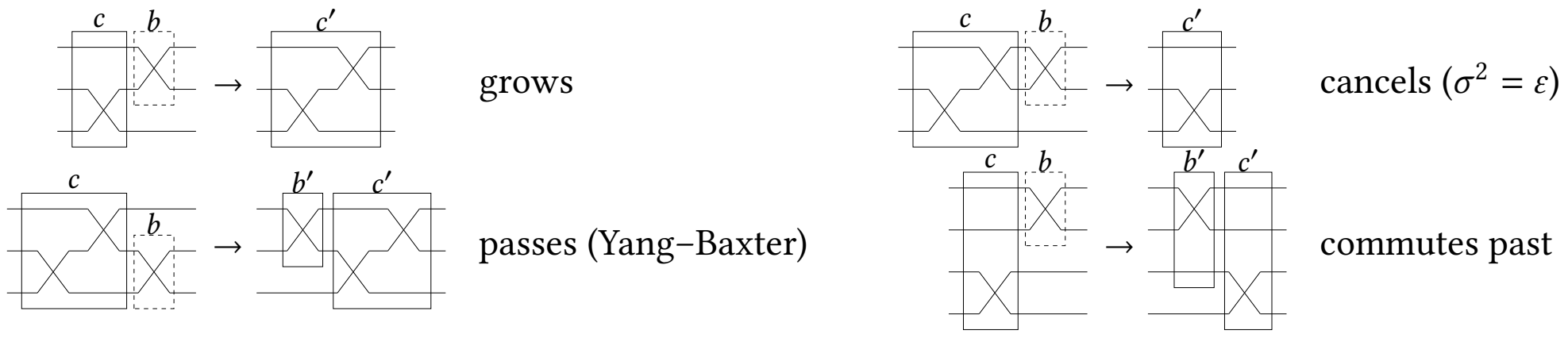

\centering
\begin{tabular}{@{}r@{\,}c@{\,}l@{\;\;}l@{\qquad}r@{\,}c@{\,}l@{\;\;}l@{}}
  \ufig[0.35]{push4-l} & \scalebox{0.7}{$\rightarrow$} & \ufig[0.35]{push4-r} &
    \small grows &
  \ufig[0.35]{push3-l} & \scalebox{0.7}{$\rightarrow$} & \ufig[0.35]{push3-r} &
    \small cancels ($\sigma^2 = \varepsilon$)\\[4pt]
  \ufig[0.35]{push2-l} & \scalebox{0.7}{$\rightarrow$} & \ufig[0.35]{push2-r} &
    \small passes (Yang--Baxter) &
  \ufig[0.35]{push1-l} & \scalebox{0.7}{$\rightarrow$} & \ufig[0.35]{push1-r} &
    \small commutes past
\end{tabular}
\caption{Pushing a swap through a staircase: the four cases of the coset
table \aid{ract} as circuit rewrites.}
\Description{Four circuit rewrites: a swap pushed into a staircase grows
it, cancels against it, passes through it by Yang--Baxter, or commutes
past it.}
\label{fig:pushing}
\end{figure}

Computing this factorization is the job of the coset table.
Given a coset, represented by its staircase, and one generator, the
table says where their product lands (the updated coset) and what
is left over on the wires above (the residual circuit), so that
staircase followed by generator equals residual (shifted up) followed by
updated staircase.
Pushing $\sigma$ into a staircase can lengthen it, cancel with it ($\sigma^2 =
\varepsilon$), or pass it through and shift it up (Yang--Baxter); a
shifted generator commutes past the empty staircase, and meets a longer
one by recursing into its tail.  The residual is empty or a single swap here, but in richer
gate sets it can be a longer circuit, which is why the table returns a
circuit on one wire fewer rather than a generator
(Figure~\ref{fig:pushing} draws the four cases):

\begin{center}
\begin{minipage}[c]{0.65\linewidth}
\begingroup\setlength{\mathindent}{0pt}
\begin{code}%
\>[0]\AgdaFunction{ract}\AgdaSpace{}%
\AgdaSymbol{:}\AgdaSpace{}%
\AgdaDatatype{C}\AgdaSpace{}%
\AgdaSymbol{(}\AgdaInductiveConstructor{₁₊}\AgdaSpace{}%
\AgdaGeneralizable{n}\AgdaSymbol{)}\AgdaSpace{}%
\AgdaSymbol{→}\AgdaSpace{}%
\AgdaDatatype{Gen}\AgdaSpace{}%
\AgdaSymbol{(}\AgdaInductiveConstructor{₂₊}\AgdaSpace{}%
\AgdaGeneralizable{n}\AgdaSymbol{)}\AgdaSpace{}%
\AgdaSymbol{→}\AgdaSpace{}%
\AgdaFunction{Circuit}\AgdaSpace{}%
\AgdaSymbol{(}\AgdaInductiveConstructor{₁₊}\AgdaSpace{}%
\AgdaGeneralizable{n}\AgdaSymbol{)}\AgdaSpace{}%
\AgdaOperator{\AgdaFunction{×}}\AgdaSpace{}%
\AgdaDatatype{C}\AgdaSpace{}%
\AgdaSymbol{(}\AgdaInductiveConstructor{₁₊}\AgdaSpace{}%
\AgdaGeneralizable{n}\AgdaSymbol{)}\<%
\\
\>[0]\AgdaFunction{ract}\AgdaSpace{}%
\AgdaInductiveConstructor{ε}\AgdaSpace{}%
\AgdaInductiveConstructor{σ-gen}\AgdaSpace{}%
\AgdaSymbol{=}\AgdaSpace{}%
\AgdaInductiveConstructor{ε}\AgdaSpace{}%
\AgdaOperator{\AgdaInductiveConstructor{,}}\AgdaSpace{}%
\AgdaOperator{\AgdaInductiveConstructor{σ•}}\AgdaSpace{}%
\AgdaInductiveConstructor{ε}%
\>[36]\AgdaComment{--\ grows}\<%
\\
\>[0]\AgdaFunction{ract}\AgdaSpace{}%
\AgdaSymbol{(}\AgdaOperator{\AgdaInductiveConstructor{σ•}}\AgdaSpace{}%
\AgdaInductiveConstructor{ε}\AgdaSymbol{)}\AgdaSpace{}%
\AgdaInductiveConstructor{σ-gen}\AgdaSpace{}%
\AgdaSymbol{=}\AgdaSpace{}%
\AgdaInductiveConstructor{ε}\AgdaSpace{}%
\AgdaOperator{\AgdaInductiveConstructor{,}}\AgdaSpace{}%
\AgdaInductiveConstructor{ε}%
\>[36]\AgdaComment{--\ cancels\ (σ²\ =\ ε)}\<%
\\
\>[0]\AgdaFunction{ract}\AgdaSpace{}%
\AgdaSymbol{(}\AgdaOperator{\AgdaInductiveConstructor{σ•}}\AgdaSpace{}%
\AgdaOperator{\AgdaInductiveConstructor{σ•}}\AgdaSpace{}%
\AgdaBound{c}\AgdaSymbol{)}\AgdaSpace{}%
\AgdaInductiveConstructor{σ-gen}\AgdaSpace{}%
\AgdaSymbol{=}\AgdaSpace{}%
\AgdaFunction{σ}\AgdaSpace{}%
\AgdaOperator{\AgdaInductiveConstructor{,}}\AgdaSpace{}%
\AgdaOperator{\AgdaInductiveConstructor{σ•}}\AgdaSpace{}%
\AgdaOperator{\AgdaInductiveConstructor{σ•}}\AgdaSpace{}%
\AgdaBound{c}%
\>[36]\AgdaComment{--\ passes\ (Yang–Baxter)}\<%
\\
\>[0]\AgdaFunction{ract}\AgdaSpace{}%
\AgdaInductiveConstructor{ε}\AgdaSpace{}%
\AgdaSymbol{(}\AgdaBound{g}\AgdaSpace{}%
\AgdaOperator{\AgdaInductiveConstructor{↥}}\AgdaSymbol{)}\AgdaSpace{}%
\AgdaSymbol{=}\AgdaSpace{}%
\AgdaOperator{\AgdaInductiveConstructor{[}}\AgdaSpace{}%
\AgdaBound{g}\AgdaSpace{}%
\AgdaOperator{\AgdaInductiveConstructor{]ʷ}}\AgdaSpace{}%
\AgdaOperator{\AgdaInductiveConstructor{,}}\AgdaSpace{}%
\AgdaInductiveConstructor{ε}%
\>[36]\AgdaComment{--\ commutes\ past}\<%
\\
\>[0]\AgdaFunction{ract}\AgdaSpace{}%
\AgdaSymbol{(}\AgdaOperator{\AgdaInductiveConstructor{σ•}}\AgdaSpace{}%
\AgdaBound{c}\AgdaSymbol{)}\AgdaSpace{}%
\AgdaSymbol{(}\AgdaBound{g}\AgdaSpace{}%
\AgdaOperator{\AgdaInductiveConstructor{↥}}\AgdaSymbol{)}\AgdaSpace{}%
\AgdaSymbol{=}\AgdaSpace{}%
\AgdaField{proj₁}\AgdaSpace{}%
\AgdaSymbol{(}\AgdaFunction{ract}\AgdaSpace{}%
\AgdaBound{c}\AgdaSpace{}%
\AgdaBound{g}\AgdaSymbol{)}\AgdaSpace{}%
\AgdaOperator{\AgdaFunction{↑}}\AgdaSpace{}%
\AgdaOperator{\AgdaInductiveConstructor{,}}\AgdaSpace{}%
\AgdaOperator{\AgdaInductiveConstructor{σ•}}\AgdaSpace{}%
\AgdaSymbol{(}\AgdaField{proj₂}\AgdaSpace{}%
\AgdaSymbol{(}\AgdaFunction{ract}\AgdaSpace{}%
\AgdaBound{c}\AgdaSpace{}%
\AgdaBound{g}\AgdaSymbol{))}%
\>[64]\AgdaComment{--\ recurses}\<%
\end{code}
\endgroup
\end{minipage}\hspace{0.5em}%
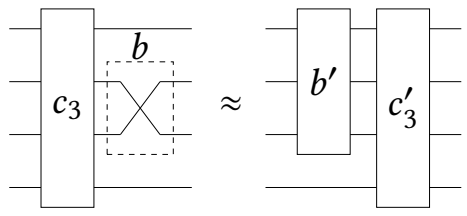
\begin{minipage}[c]{0.30\linewidth}\raggedright
  \ufig[0.35]{ract-l}\;\scalebox{0.7}{$\approx$}\;\ufig[0.35]{ract-r}
\end{minipage}
\end{center}

\noindent
\aid{σ-gen} abbreviates $\aid{gate₂}\;\aid{σ-gate}$.  The last clause is the recursion: a
shifted generator meets a staircase by meeting its tail one wire down.
Each clause is certified by the \emph{soundness law}, pictured on the
right and proved once per table by a short equational derivation from
the axioms:

\begin{code}%
\>[0]\AgdaFunction{ract-sound}\AgdaSpace{}%
\AgdaSymbol{:}\AgdaSpace{}%
\AgdaSymbol{∀}\AgdaSpace{}%
\AgdaBound{c}\AgdaSpace{}%
\AgdaBound{b}\AgdaSpace{}%
\AgdaSymbol{→}\AgdaSpace{}%
\AgdaKeyword{let}\AgdaSpace{}%
\AgdaSymbol{(}\AgdaBound{b'}\AgdaSpace{}%
\AgdaOperator{\AgdaInductiveConstructor{,}}\AgdaSpace{}%
\AgdaBound{c'}\AgdaSymbol{)}\AgdaSpace{}%
\AgdaSymbol{=}\AgdaSpace{}%
\AgdaFunction{ract}\AgdaSpace{}%
\AgdaBound{c}\AgdaSpace{}%
\AgdaBound{b}\AgdaSpace{}%
\AgdaKeyword{in}\AgdaSpace{}%
\AgdaOperator{\AgdaFunction{[}}\AgdaSpace{}%
\AgdaBound{c}\AgdaSpace{}%
\AgdaOperator{\AgdaFunction{]ᶜ}}\AgdaSpace{}%
\AgdaOperator{\AgdaInductiveConstructor{•}}\AgdaSpace{}%
\AgdaOperator{\AgdaInductiveConstructor{[}}\AgdaSpace{}%
\AgdaBound{b}\AgdaSpace{}%
\AgdaOperator{\AgdaInductiveConstructor{]ʷ}}\AgdaSpace{}%
\AgdaOperator{\AgdaPostulate{≈}}\AgdaSpace{}%
\AgdaBound{b'}\AgdaSpace{}%
\AgdaOperator{\AgdaFunction{↑}}\AgdaSpace{}%
\AgdaOperator{\AgdaInductiveConstructor{•}}\AgdaSpace{}%
\AgdaOperator{\AgdaFunction{[}}\AgdaSpace{}%
\AgdaBound{c'}\AgdaSpace{}%
\AgdaOperator{\AgdaFunction{]ᶜ}}\<%
\end{code}
\begin{code}[hide]%
\>[0]\AgdaFunction{ract-sound}\AgdaSpace{}%
\AgdaSymbol{=}\AgdaSpace{}%
\AgdaPostulate{proof-elided}\<%
\end{code}

The stateful word traversal \aid{{\_}ᵗ} (\S\ref{sec:words}) extends
\aid{ract} to whole circuits, threading the coset from left to right and
concatenating the residuals; run from the identity coset, it maps an
$(n{+}2)$-wire circuit to an $(n{+}1)$-wire circuit and a coset, one
level of normalization, the normal-form transport of
\S\ref{sec:engine}.  Iterating this step gives the coset tower, whose
carrier grows by one digit per level above the one-wire base ($\top$ is
the unit type, with sole element $\mathsf{tt}$):

\begin{center}
\begin{minipage}[c]{0.66\linewidth}
\begingroup\setlength{\mathindent}{0pt}
\begin{code}%
\>[0]\AgdaFunction{NF}\AgdaSpace{}%
\AgdaSymbol{:}\AgdaSpace{}%
\AgdaDatatype{ℕ}\AgdaSpace{}%
\AgdaSymbol{→}\AgdaSpace{}%
\AgdaPrimitive{Set}\<%
\\
\>[0]\AgdaFunction{NF}\AgdaSpace{}%
\AgdaNumber{0}%
\>[11]\AgdaSymbol{=}\AgdaSpace{}%
\AgdaRecord{⊤}\<%
\\
\>[0]\AgdaFunction{NF}\AgdaSpace{}%
\AgdaNumber{1}%
\>[11]\AgdaSymbol{=}\AgdaSpace{}%
\AgdaRecord{⊤}\<%
\\
\>[0]\AgdaFunction{NF}\AgdaSpace{}%
\AgdaSymbol{(}\AgdaInductiveConstructor{₂₊}\AgdaSpace{}%
\AgdaBound{k}\AgdaSymbol{)}%
\>[11]\AgdaSymbol{=}\AgdaSpace{}%
\AgdaFunction{NF}\AgdaSpace{}%
\AgdaSymbol{(}\AgdaInductiveConstructor{₁₊}\AgdaSpace{}%
\AgdaBound{k}\AgdaSymbol{)}\AgdaSpace{}%
\AgdaOperator{\AgdaFunction{×}}\AgdaSpace{}%
\AgdaDatatype{C}\AgdaSpace{}%
\AgdaSymbol{(}\AgdaInductiveConstructor{₁₊}\AgdaSpace{}%
\AgdaBound{k}\AgdaSymbol{)}%
\>[36]\AgdaComment{--\ ⊤\ ×\ C\ 1\ ×\ ⋯\ ×\ C\ (₁₊\ k)}\<%
\end{code}
\endgroup
\end{minipage}\hspace{1.5em}%
\begin{minipage}[c]{0.16\linewidth}\centering
  \ufig[0.35]{fig3_staircase}
\end{minipage}
\end{center}

\noindent so $\aid{NF}\;n$ has exactly $n!$ elements.  The normal-form
function $\aid{nf} : \aid{Circuit}\;n \to \aid{NF}\;n$ runs the tables;
its \emph{section} $\aid{inv-nf} : \aid{NF}\;n \to \aid{Circuit}\;n$
rebuilds a circuit from its digits by concatenating the staircases,
suitably shifted, as drawn on the right for $(\mathsf{tt}, c_1, c_2, c_3)
\in \aid{NF}\;4$: one staircase per level, each shifted up to the
wires of its level, a ``staircase of staircases''.

\subsection{Two Semantics, Uniqueness, and Completeness}

So far everything was syntax.  The development interprets circuits twice:
by a \emph{loose} semantics $\aid{⟦\_⟧} : \aid{Circuit}\;n \to
(\mathsf{Fin}\;n \to \mathsf{Fin}\;n)$ into endofunctions of
$\mathsf{Fin}\;n$ ($\sigma$ swaps the bottom two elements), a monoid that
is not a group; and by a \emph{tight} semantics into
$\aid{Permutation′-group}\;n$, the standard library's permutations of
$\mathsf{Fin}\;n$ bundled as a \aid{Group} by our \texttt{ForStdlib}
layer (\S\ref{sec:forstdlib}).  \emph{Soundness}, that $w \approx v$
implies $\aid{⟦}w\aid{⟧} \doteq \aid{⟦}v\aid{⟧}$, is a small case
analysis over the axioms.  The main step is the \emph{unique
normal form} property (\S\ref{sec:nfproperty}), that normal forms with
equal denotations are equal: $\aid{⟦}\;\aid{inv-nf}\;u\,\aid{⟧} \doteq
\aid{⟦}\;\aid{inv-nf}\;v\,\aid{⟧}$ implies $u \equiv v$, proved by
induction on the tower: the lower digits fix the bottom point ($0$), so the top
digit alone determines where the permutation sends it; equal denotations
therefore force equal top digits, and cancelling them reduces the claim
to the level below.  Soundness and uniqueness are the premises of the generic
lemma \aid{by-normalization}, which concludes \emph{completeness}:
$\aid{⟦}w\aid{⟧} \doteq \aid{⟦}v\aid{⟧}$ implies $w \approx v$.
Unfolding the lemma on this example shows where the axioms enter.  Each
step of \aid{nf} is one of the rewrites of Figure~\ref{fig:pushing}, and
each of those is an instance of \aid{order}, of \aid{yang-baxter}, or of
the structural rules (the far commutation \aid{comm₂}, lifted to the
wires above by \aid{cong↑}); so $w \approx \aid{inv-nf}\;(\aid{nf}\;w)$
(the law of the \aid{NormalForm} witness) is a derivation from
those rules.  Now suppose $\aid{⟦}w\aid{⟧} \doteq
\aid{⟦}v\aid{⟧}$.  By soundness the normal forms of $w$ and $v$ have the
same denotation, so by uniqueness they are equal, and hence $w \approx
\aid{inv-nf}\;(\aid{nf}\;w) = \aid{inv-nf}\;(\aid{nf}\;v) \approx v$, a
derivation that uses only the two normal-form derivations, and hence only
the axioms and structural rules above.  The index module states these for every $n$ (\aid{symmetric-soundness},
\aid{symmetric-completeness}, \aid{symmetric-unique-nf}).  Completeness
for the \emph{loose} semantics says that, even read as mere endofunctions,
circuits equal as functions are provably equal.  For the tight semantics,
soundness, completeness, and
surjectivity (every permutation is a product of adjacent transpositions)
assemble into the \emph{presentation record} of \S\ref{sec:presentations}.
The record packages two things: a \aid{Grouplike} witness, which makes
the presented monoid a group, and a proof that the semantics
$\aid{⟦\_⟧}$ is a group isomorphism from that group onto the target.  We quote
this one statement in Agda because every presentation theorem in the
paper has its shape:

\begin{code}%
\>[0]\AgdaFunction{symmetric-presentation}\AgdaSpace{}%
\AgdaSymbol{:}\AgdaSpace{}%
\AgdaSymbol{∀}\AgdaSpace{}%
\AgdaBound{n}\AgdaSpace{}%
\AgdaSymbol{→}\AgdaSpace{}%
\AgdaSymbol{(}\AgdaBound{n}\AgdaSpace{}%
\AgdaOperator{\AgdaDatatype{VRel,\AgdaUnderscore{}===\AgdaUnderscore{}}}\AgdaSymbol{)}\AgdaSpace{}%
\AgdaOperator{\AgdaRecord{IsPresentationOf}}\AgdaSpace{}%
\AgdaSymbol{(}\AgdaFunction{Permutation′-group}\AgdaSpace{}%
\AgdaBound{n}\AgdaSymbol{)}\<%
\end{code}
\begin{code}[hide]%
\>[0]\AgdaFunction{symmetric-presentation}\AgdaSpace{}%
\AgdaBound{n}\AgdaSpace{}%
\AgdaSymbol{=}\AgdaSpace{}%
\AgdaFunction{SymThm.Tight.presentation}\AgdaSpace{}%
\AgdaBound{n}\<%
\end{code}

\noindent read: the monoid of swap circuits modulo \{\aid{order},
\aid{yang-baxter}\} and the structural rules is the group of
permutations of $\mathsf{Fin}\;n$ (\aid{IsPresentationOf} is the mixfix
record name \aid{{\_}IsPresentationOf{\_}} used infix).  Later tower-based
case studies repeat this pipeline, with richer gates, longer towers,
and amalgamations where the chain branches.
\section{Preliminaries: Normal Forms from Subgroup Chains}\label{sec:preliminaries}

The library's coset-tower normal forms follow one classical recipe from
computational group theory.  This section recalls it in mathematical
vocabulary and fixes the intuitions (transversals as digits, towers as
positional number systems) that the Agda modules of
\S\ref{sec:design} implement.  Standard references are Sims's monograph on
computation with finitely presented groups~\cite{sims1994computation} and
the Handbook of Computational Group Theory~\cite{holt2005handbook}; the
subgroup-presentation method goes back to Reidemeister and Schreier, in the
form given by Magnus, Karrass, and Solitar~\cite{magnus2004combinatorial}.

\subsection{Transversals and Digits}

Let $H \le G$ be a subgroup.  A \emph{(right) transversal} is a set of
representatives for the right cosets $Hg$, one per coset; write $\bar g$
for the representative of $Hg$.  Every $g \in G$ then factors
\emph{uniquely} as
\[
  g \;=\; h \cdot \bar g, \qquad h \in H,\ \bar g \in \text{transversal},
\]
and the map $g \mapsto (h, \bar g)$ is a bijection
$G \cong H \times (H\backslash G)$ of sets (not of groups).  If one now
fixes a \emph{chain} of subgroups
\[
  1 = G_0 \;\le\; G_1 \;\le\; \dots \;\le\; G_n = G ,
\]
with a transversal $T_k$ for each step $G_{k-1} \le G_k$, iterating the
factorization expresses every $g \in G$ uniquely as
\[
  g \;=\; t_1 \, t_2 \cdots t_n , \qquad t_k \in T_k .
\]
The tuple $(t_1,\dots,t_n)$ is a \emph{normal form} for $g$: existence is
the iterated factorization, uniqueness is uniqueness of coset
representatives at each level.  When the indices $[G_k : G_{k-1}] = r_k$
are finite, $|G| = r_1 r_2 \cdots r_n$, and the normal form is a
\emph{mixed-radix positional numeral}: the $k$-th digit ranges over $r_k$
values, and enumerating tuples lexicographically enumerates the group.  For
the chain $S_1 < S_2 < \dots < S_n$ of symmetric groups the radices are
$2, 3, \dots, n$ and the numeral system is the factorial number system,
which is the carrier $\aid{NF}\;n = \top \times C\,1 \times \dots
\times C\,(n{-}1)$ built by the coset tower in \S\ref{sec:permutations}.
The Schreier--Sims method for permutation
groups~\cite{holt2005handbook} reads stabilizer-chain normal forms in the
same way, with the $G_k$ point stabilizers; our chains are chosen
syntactically, but the counting argument is identical.

\subsection{Words, Presentations, and the Word Problem}\label{sec:wordproblem}

A \emph{presentation} $\langle X \mid R \rangle$ consists of a set $X$ of
generators and a relation $R$ on the free monoid $X^{*}$, that is, a set of
pairs of words over $X$, the axioms (we work with monoid
presentations, so words are finite sequences with concatenation, and there
are no formal inverses).  The presented monoid is the quotient of the free
monoid $X^{*}$ by the least congruence $\approx$ containing $R$; the
\emph{word problem} asks to decide $w \approx v$.  A presentation
\emph{presents} a monoid or group $G$ (relative to an interpretation of
the generators) when the induced homomorphism $X^{*}/{\approx} \to G$ is
an isomorphism, where the interpretation extends by freeness to
$X^{*} \to G$ and descends to the quotient when every axiom holds in $G$
(\emph{soundness}); its injectivity is \emph{completeness} of
$R$ for $G$, its surjectivity says the images of the generators
generate $G$.  A group presentation in the
classical sense is recovered by adding, for each generator, a generator
playing the role of its inverse together with the cancellation relations;
in the library this is the \aid{Grouplike} condition of
\S\ref{sec:presentations}, which asks every generator to have a left
inverse up to $\approx$ and derives the group structure from it.

Given $H \le G$ with $G = \langle Y \mid S \rangle$, the
\emph{Reidemeister--Schreier} method produces a presentation of $H$: its
generators are the \emph{Schreier generators} $\bar t y\,\overline{\bar t y}^{\,-1}$
($t$ a coset label, $y \in Y$), and its relations are rewrites of the
relators of $G$ through the coset table.  For finite index and finite $Y$,
the data one must supply is finite and concrete: the coset table
$\tau : \text{cosets} \times Y \to \text{cosets}$ recording
$H\bar t y = H\overline{\tau(t,y)}$,
together with, for each step, the element of $H$ by which $\bar t y$ differs from
its representative, the residual $h$ with $\bar t\, y = h
\cdot \overline{\tau(t,y)}$.  These are the two components that
\aid{ract} returns in \S\ref{sec:sn-cosets}, and the equation is its
soundness law.  Classical treatments run this as an algorithm
(Todd--Coxeter coset enumeration~\cite{sims1994computation,
holt2005handbook}); we instead take a claimed coset table
as input and verify it, by exhibiting five equational
hypotheses relating the table, the section, and the relations
(\S\ref{sec:engine}).  This suits a proof assistant: tables are found
offline, by computation or on paper, and Agda only checks them.
Bian and Selinger introduced this monoid-level, verified use of
Reidemeister--Schreier in Agda for their gate-group
presentations~\cite{bian2023cliffordt2,bian2023cliffordcs3,bian2023thesis};
we turn it into a reusable module with the hypotheses packaged as
a record.

\subsection{Specialization to Circuits}\label{sec:circuit-chain}

For circuits, the wire structure supplies a candidate chain.  Writing $\mathsf{Cir}\,k$ for the group (or monoid) presented
by $k$-wire circuits modulo a relation set, the shift operation
$c \mapsto c{\uparrow}$ is a homomorphism.  Verified coset tables establish
its injectivity, giving
$\mathsf{Cir}\,k \le \mathsf{Cir}\,(k{+}1)$
as ``the circuits that do not touch the bottom wire''.  The digit that a
normal form for $\mathsf{Cir}\,(k{+}1)$ adds to one for $\mathsf{Cir}\,k$
then ranges over a transversal of
this subgroup (for permutations, the staircases of
\S\ref{sec:sn-cosets}; for Clifford-style groups, the staircase-shaped
representatives of Selinger~\cite{selinger2015clifford} and their qutrit
analogues~\cite{li2025qutrit}), and a normal form for the whole group is
a tower of such levels.  The library's \aid{CosetTower}
(\S\ref{sec:engine}) implements this specialization.  Where the subgroup chain branches instead of nesting (two
subgroups sharing a common base, as for the Clifford+$T$ groups),
the appropriate gluing is the \emph{amalgamated free product}, whose
classical alternating normal form~\cite{magnus2004combinatorial} we
also verify (\S\ref{sec:constructions}).

\begin{code}[hide]%
\>[0]\AgdaKeyword{module}\AgdaSpace{}%
\AgdaModule{paper.sections.design}\AgdaSpace{}%
\AgdaKeyword{where}\<%
\\
\\[\AgdaEmptyExtraSkip]%
\>[0]\AgdaKeyword{open}\AgdaSpace{}%
\AgdaKeyword{import}\AgdaSpace{}%
\AgdaModule{Data.Nat}\AgdaSpace{}%
\AgdaKeyword{using}\AgdaSpace{}%
\AgdaSymbol{(}\AgdaDatatype{ℕ}\AgdaSpace{}%
\AgdaSymbol{;}\AgdaSpace{}%
\AgdaOperator{\AgdaPrimitive{\AgdaUnderscore{}+\AgdaUnderscore{}}}\AgdaSymbol{)}\<%
\\
\>[0]\AgdaKeyword{open}\AgdaSpace{}%
\AgdaKeyword{import}\AgdaSpace{}%
\AgdaModule{Data.Product}\AgdaSpace{}%
\AgdaKeyword{using}\AgdaSpace{}%
\AgdaSymbol{(}\AgdaOperator{\AgdaFunction{\AgdaUnderscore{}×\AgdaUnderscore{}}}\AgdaSpace{}%
\AgdaSymbol{;}\AgdaSpace{}%
\AgdaOperator{\AgdaInductiveConstructor{\AgdaUnderscore{},\AgdaUnderscore{}}}\AgdaSymbol{)}\<%
\\
\>[0]\AgdaKeyword{open}\AgdaSpace{}%
\AgdaKeyword{import}\AgdaSpace{}%
\AgdaModule{Data.Sum}\AgdaSpace{}%
\AgdaKeyword{using}\AgdaSpace{}%
\AgdaSymbol{(}\AgdaOperator{\AgdaDatatype{\AgdaUnderscore{}⊎\AgdaUnderscore{}}}\AgdaSymbol{)}\<%
\\
\>[0]\AgdaKeyword{open}\AgdaSpace{}%
\AgdaKeyword{import}\AgdaSpace{}%
\AgdaModule{Data.List}\AgdaSpace{}%
\AgdaKeyword{using}\AgdaSpace{}%
\AgdaSymbol{(}\AgdaDatatype{List}\AgdaSymbol{)}\<%
\\
\>[0]\AgdaKeyword{open}\AgdaSpace{}%
\AgdaKeyword{import}\AgdaSpace{}%
\AgdaModule{Data.Unit}\AgdaSpace{}%
\AgdaKeyword{using}\AgdaSpace{}%
\AgdaSymbol{(}\AgdaRecord{⊤}\AgdaSymbol{)}\<%
\\
\>[0]\AgdaKeyword{open}\AgdaSpace{}%
\AgdaKeyword{import}\AgdaSpace{}%
\AgdaModule{Level}\AgdaSpace{}%
\AgdaKeyword{using}\AgdaSpace{}%
\AgdaSymbol{(}\AgdaPostulate{Level}\AgdaSpace{}%
\AgdaSymbol{;}\AgdaSpace{}%
\AgdaOperator{\AgdaPrimitive{\AgdaUnderscore{}⊔\AgdaUnderscore{}}}\AgdaSpace{}%
\AgdaSymbol{;}\AgdaSpace{}%
\AgdaFunction{0ℓ}\AgdaSymbol{)}\<%
\\
\>[0]\AgdaKeyword{open}\AgdaSpace{}%
\AgdaKeyword{import}\AgdaSpace{}%
\AgdaModule{Relation.Binary}\AgdaSpace{}%
\AgdaKeyword{using}\AgdaSpace{}%
\AgdaSymbol{(}\AgdaFunction{Rel}\AgdaSymbol{)}\<%
\\
\>[0]\AgdaKeyword{open}\AgdaSpace{}%
\AgdaKeyword{import}\AgdaSpace{}%
\AgdaModule{Relation.Binary.Definitions}\AgdaSpace{}%
\AgdaKeyword{using}\AgdaSpace{}%
\AgdaSymbol{(}\AgdaFunction{Decidable}\AgdaSymbol{)}\<%
\\
\>[0]\AgdaKeyword{open}\AgdaSpace{}%
\AgdaKeyword{import}\AgdaSpace{}%
\AgdaModule{Function.Definitions}\AgdaSpace{}%
\AgdaKeyword{using}\AgdaSpace{}%
\AgdaSymbol{(}\AgdaFunction{Congruent}\AgdaSpace{}%
\AgdaSymbol{;}\AgdaSpace{}%
\AgdaFunction{Injective}\AgdaSymbol{)}\<%
\\
\>[0]\AgdaKeyword{open}\AgdaSpace{}%
\AgdaKeyword{import}\AgdaSpace{}%
\AgdaModule{Algebra.Bundles}\AgdaSpace{}%
\AgdaKeyword{using}\AgdaSpace{}%
\AgdaSymbol{(}\AgdaRecord{Group}\AgdaSymbol{)}\<%
\\
\>[0]\AgdaKeyword{open}\AgdaSpace{}%
\AgdaKeyword{import}\AgdaSpace{}%
\AgdaModule{Algebra.Morphism.Structures}\AgdaSpace{}%
\AgdaKeyword{using}\AgdaSpace{}%
\AgdaSymbol{(}\AgdaKeyword{module}\AgdaSpace{}%
\AgdaModule{GroupMorphisms}\AgdaSymbol{)}\<%
\\
\\[\AgdaEmptyExtraSkip]%
\>[0]\AgdaKeyword{open}\AgdaSpace{}%
\AgdaKeyword{import}\AgdaSpace{}%
\AgdaModule{Notations}\AgdaSpace{}%
\AgdaKeyword{using}\AgdaSpace{}%
\AgdaSymbol{(}\AgdaInductiveConstructor{₁₊}\AgdaSpace{}%
\AgdaSymbol{;}\AgdaSpace{}%
\AgdaInductiveConstructor{₂₊}\AgdaSymbol{)}\<%
\\
\>[0]\AgdaKeyword{open}\AgdaSpace{}%
\AgdaKeyword{import}\AgdaSpace{}%
\AgdaModule{Word.Base}\AgdaSpace{}%
\AgdaKeyword{using}\AgdaSpace{}%
\AgdaSymbol{(}\AgdaFunction{wconcatmap}\AgdaSymbol{)}\<%
\\
\>[0]\AgdaKeyword{import}\AgdaSpace{}%
\AgdaModule{Word.Base}\AgdaSpace{}%
\AgdaSymbol{as}\AgdaSpace{}%
\AgdaModule{WB}\<%
\\
\>[0]\AgdaKeyword{open}\AgdaSpace{}%
\AgdaKeyword{import}\AgdaSpace{}%
\AgdaModule{Presentation.GroupLike}\AgdaSpace{}%
\AgdaKeyword{using}\AgdaSpace{}%
\AgdaSymbol{(}\AgdaFunction{Grouplike}\AgdaSpace{}%
\AgdaSymbol{;}\AgdaSpace{}%
\AgdaKeyword{module}\AgdaSpace{}%
\AgdaModule{Group-Lemmas}\AgdaSymbol{)}\<%
\\
\>[0]\AgdaKeyword{open}\AgdaSpace{}%
\AgdaKeyword{import}\AgdaSpace{}%
\AgdaModule{Presentation.Construct.Base}\AgdaSpace{}%
\AgdaKeyword{using}\AgdaSpace{}%
\AgdaSymbol{(}\AgdaOperator{\AgdaFunction{[\AgdaUnderscore{}]ₗ}}\AgdaSpace{}%
\AgdaSymbol{;}\AgdaSpace{}%
\AgdaOperator{\AgdaFunction{[\AgdaUnderscore{}]ᵣ}}\AgdaSymbol{)}\<%
\\
\\[\AgdaEmptyExtraSkip]%
\>[0]\AgdaKeyword{postulate}\<%
\\
\>[0][@{}l@{\AgdaIndent{0}}]%
\>[2]\AgdaPostulate{proof-elided}\AgdaSpace{}%
\AgdaSymbol{:}\AgdaSpace{}%
\AgdaSymbol{∀}\AgdaSpace{}%
\AgdaSymbol{\{}\AgdaBound{ℓ}\AgdaSymbol{\}}\AgdaSpace{}%
\AgdaSymbol{\{}\AgdaBound{A}\AgdaSpace{}%
\AgdaSymbol{:}\AgdaSpace{}%
\AgdaPrimitive{Set}\AgdaSpace{}%
\AgdaBound{ℓ}\AgdaSymbol{\}}\AgdaSpace{}%
\AgdaSymbol{→}\AgdaSpace{}%
\AgdaBound{A}\<%
\\
\>[2]\AgdaPostulate{X}\AgdaSpace{}%
\AgdaPostulate{Y}\AgdaSpace{}%
\AgdaPostulate{A}\AgdaSpace{}%
\AgdaPostulate{B}\AgdaSpace{}%
\AgdaPostulate{C}\AgdaSpace{}%
\AgdaPostulate{D}%
\>[15]\AgdaSymbol{:}\AgdaSpace{}%
\AgdaPrimitive{Set}\<%
\\
\>[2]\AgdaPostulate{a}\AgdaSpace{}%
\AgdaPostulate{ℓ}\AgdaSpace{}%
\AgdaPostulate{c}\AgdaSpace{}%
\AgdaPostulate{d}%
\>[15]\AgdaSymbol{:}\AgdaSpace{}%
\AgdaPostulate{Level}\<%
\\
\>[2]\AgdaPostulate{n}\AgdaSpace{}%
\AgdaPostulate{k}%
\>[15]\AgdaSymbol{:}\AgdaSpace{}%
\AgdaDatatype{ℕ}\<%
\end{code}
\section{Library Design}\label{sec:design}

This section presents the library bottom-up.  Each subsection names its
home module; code is quoted from the library, lightly abridged (implicit
arguments and absurd clauses omitted).  Three constraints guide the design:
use the standard library's abstractions (setoids, function bundles, algebra
bundles), keep everything safe and axiom-free, and arrange definitions so
that concrete instances are checked by Agda's evaluator.
Concretely, words stay unquotiented so that normal forms and sections
compute (\S\ref{sec:words}); finite coset tables reduce their five
verification obligations to case splits (\S\ref{sec:engine}); and
successor-only wire indices avoid arithmetic in coverage checking
(\S\ref{sec:circuits}).

\subsection{Words: The Free Monoid}\label{sec:words}

A word over a generator set $X$ (\texttt{Word.Base}) is an unquotiented
syntax tree rather than a list:

\begin{code}%
\>[0]\AgdaKeyword{data}\AgdaSpace{}%
\AgdaDatatype{Word}\AgdaSpace{}%
\AgdaSymbol{(}\AgdaBound{X}\AgdaSpace{}%
\AgdaSymbol{:}\AgdaSpace{}%
\AgdaPrimitive{Set}\AgdaSymbol{)}\AgdaSpace{}%
\AgdaSymbol{:}\AgdaSpace{}%
\AgdaPrimitive{Set}\AgdaSpace{}%
\AgdaKeyword{where}\<%
\\
\>[0][@{}l@{\AgdaIndent{0}}]%
\>[2]\AgdaOperator{\AgdaInductiveConstructor{[\AgdaUnderscore{}]ʷ}}\AgdaSpace{}%
\AgdaSymbol{:}\AgdaSpace{}%
\AgdaBound{X}\AgdaSpace{}%
\AgdaSymbol{→}\AgdaSpace{}%
\AgdaDatatype{Word}\AgdaSpace{}%
\AgdaBound{X}\<%
\\
\>[2]\AgdaInductiveConstructor{ε}%
\>[7]\AgdaSymbol{:}\AgdaSpace{}%
\AgdaDatatype{Word}\AgdaSpace{}%
\AgdaBound{X}\<%
\\
\>[2]\AgdaOperator{\AgdaInductiveConstructor{\AgdaUnderscore{}•\AgdaUnderscore{}}}%
\>[7]\AgdaSymbol{:}\AgdaSpace{}%
\AgdaDatatype{Word}\AgdaSpace{}%
\AgdaBound{X}\AgdaSpace{}%
\AgdaSymbol{→}\AgdaSpace{}%
\AgdaDatatype{Word}\AgdaSpace{}%
\AgdaBound{X}\AgdaSpace{}%
\AgdaSymbol{→}\AgdaSpace{}%
\AgdaDatatype{Word}\AgdaSpace{}%
\AgdaBound{X}\<%
\end{code}
\begin{code}[hide]%
\>[0]\AgdaKeyword{infixr}\AgdaSpace{}%
\AgdaNumber{7}\AgdaSpace{}%
\AgdaOperator{\AgdaInductiveConstructor{\AgdaUnderscore{}•\AgdaUnderscore{}}}\<%
\\
\>[0]\AgdaKeyword{infixl}\AgdaSpace{}%
\AgdaNumber{9}\AgdaSpace{}%
\AgdaOperator{\AgdaFunction{\AgdaUnderscore{}ʷ}}\AgdaSpace{}%
\AgdaOperator{\AgdaFunction{\AgdaUnderscore{}ᵗ}}\<%
\\
\>[0]\AgdaKeyword{infix}\AgdaSpace{}%
\AgdaNumber{4}\AgdaSpace{}%
\AgdaOperator{\AgdaDatatype{\AgdaUnderscore{}≈\AgdaUnderscore{}}}\AgdaSpace{}%
\AgdaOperator{\AgdaPostulate{\AgdaUnderscore{}===\AgdaUnderscore{}}}\AgdaSpace{}%
\AgdaOperator{\AgdaPostulate{\AgdaUnderscore{}≈ₙ\AgdaUnderscore{}}}\AgdaSpace{}%
\AgdaOperator{\AgdaPostulate{\AgdaUnderscore{}≈₂\AgdaUnderscore{}}}\AgdaSpace{}%
\AgdaOperator{\AgdaPostulate{\AgdaUnderscore{}\textasciitilde{}\AgdaUnderscore{}}}\AgdaSpace{}%
\AgdaOperator{\AgdaPostulate{\AgdaUnderscore{}===₁\AgdaUnderscore{}}}\AgdaSpace{}%
\AgdaOperator{\AgdaPostulate{\AgdaUnderscore{}===₂\AgdaUnderscore{}}}\<%
\\
\>[0]\AgdaKeyword{infix}\AgdaSpace{}%
\AgdaNumber{5}\AgdaSpace{}%
\AgdaOperator{\AgdaDatatype{\AgdaUnderscore{}⋄\AgdaUnderscore{}⋄\AgdaUnderscore{}}}\<%
\\
\\[\AgdaEmptyExtraSkip]%
\>[0]\AgdaFunction{WRel}\AgdaSpace{}%
\AgdaSymbol{:}\AgdaSpace{}%
\AgdaPrimitive{Set}\AgdaSpace{}%
\AgdaSymbol{→}\AgdaSpace{}%
\AgdaPrimitive{Set₁}\<%
\\
\>[0]\AgdaFunction{WRel}\AgdaSpace{}%
\AgdaBound{X}\AgdaSpace{}%
\AgdaSymbol{=}\AgdaSpace{}%
\AgdaFunction{Rel}\AgdaSpace{}%
\AgdaSymbol{(}\AgdaDatatype{Word}\AgdaSpace{}%
\AgdaBound{X}\AgdaSymbol{)}\AgdaSpace{}%
\AgdaFunction{0ℓ}\<%
\\
\\[\AgdaEmptyExtraSkip]%
\>[0]\AgdaKeyword{postulate}\<%
\\
\>[0][@{}l@{\AgdaIndent{0}}]%
\>[2]\AgdaPostulate{w}\AgdaSpace{}%
\AgdaPostulate{v}\AgdaSpace{}%
\AgdaPostulate{u}\AgdaSpace{}%
\AgdaPostulate{w'}\AgdaSpace{}%
\AgdaPostulate{v'}\AgdaSpace{}%
\AgdaSymbol{:}\AgdaSpace{}%
\AgdaDatatype{Word}\AgdaSpace{}%
\AgdaPostulate{X}\<%
\\
\>[2]\AgdaOperator{\AgdaPostulate{\AgdaUnderscore{}===\AgdaUnderscore{}}}%
\>[14]\AgdaSymbol{:}\AgdaSpace{}%
\AgdaFunction{WRel}\AgdaSpace{}%
\AgdaPostulate{X}\<%
\end{code}

Keeping composition as a binary node means a word remembers its
bracketing, so terms mirror pen-and-paper calculations and rules may be
applied to any parenthesized subterm; associativity is then one of the
congruence generators rather than a property of the carrier, and a
solver discharges re-bracketing obligations
(\S\ref{sec:solvers}).  The functorial and monadic structure is standard:
\aid{wmap}, \aid{wconcat}, and the extension of a generator-level
substitution to words, written postfix,

\begin{code}%
\>[0]\AgdaOperator{\AgdaFunction{\AgdaUnderscore{}ʷ}}\AgdaSpace{}%
\AgdaSymbol{:}\AgdaSpace{}%
\AgdaSymbol{(}\AgdaPostulate{X}\AgdaSpace{}%
\AgdaSymbol{→}\AgdaSpace{}%
\AgdaDatatype{Word}\AgdaSpace{}%
\AgdaPostulate{Y}\AgdaSymbol{)}\AgdaSpace{}%
\AgdaSymbol{→}\AgdaSpace{}%
\AgdaSymbol{(}\AgdaDatatype{Word}\AgdaSpace{}%
\AgdaPostulate{X}\AgdaSpace{}%
\AgdaSymbol{→}\AgdaSpace{}%
\AgdaDatatype{Word}\AgdaSpace{}%
\AgdaPostulate{Y}\AgdaSymbol{)}\<%
\\
\>[0]\AgdaOperator{\AgdaFunction{\AgdaUnderscore{}ʷ}}\AgdaSpace{}%
\AgdaSymbol{=}\AgdaSpace{}%
\AgdaFunction{wconcatmap}\<%
\end{code}

\noindent so \aid{(f ʷ)} is ``the unique monoid homomorphism extending $f$''.  Two
stateful traversals drive everything in \S\ref{sec:engine}: a \emph{coset
action} on letters extends to words by threading the coset left to right,

\begin{code}%
\>[0]\AgdaOperator{\AgdaFunction{\AgdaUnderscore{}ᵗ}}\AgdaSpace{}%
\AgdaSymbol{:}\AgdaSpace{}%
\AgdaSymbol{(}\AgdaPostulate{C}\AgdaSpace{}%
\AgdaSymbol{→}\AgdaSpace{}%
\AgdaPostulate{Y}\AgdaSpace{}%
\AgdaSymbol{→}\AgdaSpace{}%
\AgdaDatatype{Word}\AgdaSpace{}%
\AgdaPostulate{X}\AgdaSpace{}%
\AgdaOperator{\AgdaFunction{×}}\AgdaSpace{}%
\AgdaPostulate{C}\AgdaSymbol{)}\AgdaSpace{}%
\AgdaSymbol{→}\AgdaSpace{}%
\AgdaSymbol{(}\AgdaPostulate{C}\AgdaSpace{}%
\AgdaSymbol{→}\AgdaSpace{}%
\AgdaDatatype{Word}\AgdaSpace{}%
\AgdaPostulate{Y}\AgdaSpace{}%
\AgdaSymbol{→}\AgdaSpace{}%
\AgdaDatatype{Word}\AgdaSpace{}%
\AgdaPostulate{X}\AgdaSpace{}%
\AgdaOperator{\AgdaFunction{×}}\AgdaSpace{}%
\AgdaPostulate{C}\AgdaSymbol{)}\<%
\end{code}
\begin{code}[hide]%
\>[0]\AgdaOperator{\AgdaFunction{\AgdaUnderscore{}ᵗ}}\AgdaSpace{}%
\AgdaSymbol{=}\AgdaSpace{}%
\AgdaOperator{\AgdaFunction{WB.\AgdaUnderscore{}ᵗ}}\<%
\end{code}

\noindent together with its right-to-left mirror, and conjugation
helpers lift an action of one alphabet on another through words, which
is what the semidirect-product construction of \S\ref{sec:constructions}
uses to state its word-valued conjugation rule.  A relation set is simply
$\aid{WRel}\;X = \aid{Rel}\;(\aid{Word}\;X)\;0\ell$.

\subsection{Presented Monoids as Inductive Congruences}\label{sec:presentations}

Presentations (\texttt{Presentation.Base}, parameterized by
$\Gamma : \aid{WRel}\;X$) follow a two-level convention:
\aid{{\_}==={\_}} denotes the raw axioms ($\Gamma$ itself), and
\aid{{\_}≈{\_}} the monoid congruence they generate,
an inductive family with eight constructors:

\begin{code}%
\>[0]\AgdaKeyword{data}\AgdaSpace{}%
\AgdaOperator{\AgdaDatatype{\AgdaUnderscore{}≈\AgdaUnderscore{}}}\AgdaSpace{}%
\AgdaSymbol{:}\AgdaSpace{}%
\AgdaFunction{WRel}\AgdaSpace{}%
\AgdaPostulate{X}\AgdaSpace{}%
\AgdaKeyword{where}\<%
\\
\>[0][@{}l@{\AgdaIndent{0}}]%
\>[2]\AgdaInductiveConstructor{refl}%
\>[8]\AgdaSymbol{:}\AgdaSpace{}%
\AgdaPostulate{w}\AgdaSpace{}%
\AgdaOperator{\AgdaDatatype{≈}}\AgdaSpace{}%
\AgdaPostulate{w}\<%
\\
\>[2]\AgdaInductiveConstructor{sym}%
\>[8]\AgdaSymbol{:}\AgdaSpace{}%
\AgdaPostulate{w}\AgdaSpace{}%
\AgdaOperator{\AgdaDatatype{≈}}\AgdaSpace{}%
\AgdaPostulate{v}\AgdaSpace{}%
\AgdaSymbol{→}\AgdaSpace{}%
\AgdaPostulate{v}\AgdaSpace{}%
\AgdaOperator{\AgdaDatatype{≈}}\AgdaSpace{}%
\AgdaPostulate{w}\<%
\\
\>[2]\AgdaInductiveConstructor{trans}\AgdaSpace{}%
\AgdaSymbol{:}\AgdaSpace{}%
\AgdaPostulate{w}\AgdaSpace{}%
\AgdaOperator{\AgdaDatatype{≈}}\AgdaSpace{}%
\AgdaPostulate{v}\AgdaSpace{}%
\AgdaSymbol{→}\AgdaSpace{}%
\AgdaPostulate{v}\AgdaSpace{}%
\AgdaOperator{\AgdaDatatype{≈}}\AgdaSpace{}%
\AgdaPostulate{u}\AgdaSpace{}%
\AgdaSymbol{→}\AgdaSpace{}%
\AgdaPostulate{w}\AgdaSpace{}%
\AgdaOperator{\AgdaDatatype{≈}}\AgdaSpace{}%
\AgdaPostulate{u}\<%
\\
\>[2]\AgdaInductiveConstructor{cong}%
\>[8]\AgdaSymbol{:}\AgdaSpace{}%
\AgdaPostulate{w}\AgdaSpace{}%
\AgdaOperator{\AgdaDatatype{≈}}\AgdaSpace{}%
\AgdaPostulate{w'}\AgdaSpace{}%
\AgdaSymbol{→}\AgdaSpace{}%
\AgdaPostulate{v}\AgdaSpace{}%
\AgdaOperator{\AgdaDatatype{≈}}\AgdaSpace{}%
\AgdaPostulate{v'}\AgdaSpace{}%
\AgdaSymbol{→}\AgdaSpace{}%
\AgdaPostulate{w}\AgdaSpace{}%
\AgdaOperator{\AgdaInductiveConstructor{•}}\AgdaSpace{}%
\AgdaPostulate{v}\AgdaSpace{}%
\AgdaOperator{\AgdaDatatype{≈}}\AgdaSpace{}%
\AgdaPostulate{w'}\AgdaSpace{}%
\AgdaOperator{\AgdaInductiveConstructor{•}}\AgdaSpace{}%
\AgdaPostulate{v'}\<%
\\
\>[2]\AgdaInductiveConstructor{assoc}%
\>[13]\AgdaSymbol{:}\AgdaSpace{}%
\AgdaSymbol{(}\AgdaPostulate{w}\AgdaSpace{}%
\AgdaOperator{\AgdaInductiveConstructor{•}}\AgdaSpace{}%
\AgdaPostulate{v}\AgdaSymbol{)}\AgdaSpace{}%
\AgdaOperator{\AgdaInductiveConstructor{•}}\AgdaSpace{}%
\AgdaPostulate{u}\AgdaSpace{}%
\AgdaOperator{\AgdaDatatype{≈}}\AgdaSpace{}%
\AgdaPostulate{w}\AgdaSpace{}%
\AgdaOperator{\AgdaInductiveConstructor{•}}\AgdaSpace{}%
\AgdaSymbol{(}\AgdaPostulate{v}\AgdaSpace{}%
\AgdaOperator{\AgdaInductiveConstructor{•}}\AgdaSpace{}%
\AgdaPostulate{u}\AgdaSymbol{)}\<%
\\
\>[2]\AgdaInductiveConstructor{left-unit}%
\>[13]\AgdaSymbol{:}\AgdaSpace{}%
\AgdaInductiveConstructor{ε}\AgdaSpace{}%
\AgdaOperator{\AgdaInductiveConstructor{•}}\AgdaSpace{}%
\AgdaPostulate{w}\AgdaSpace{}%
\AgdaOperator{\AgdaDatatype{≈}}\AgdaSpace{}%
\AgdaPostulate{w}\<%
\\
\>[2]\AgdaInductiveConstructor{right-unit}\AgdaSpace{}%
\AgdaSymbol{:}\AgdaSpace{}%
\AgdaPostulate{w}\AgdaSpace{}%
\AgdaOperator{\AgdaInductiveConstructor{•}}\AgdaSpace{}%
\AgdaInductiveConstructor{ε}\AgdaSpace{}%
\AgdaOperator{\AgdaDatatype{≈}}\AgdaSpace{}%
\AgdaPostulate{w}\<%
\\
\>[2]\AgdaInductiveConstructor{axiom}%
\>[13]\AgdaSymbol{:}\AgdaSpace{}%
\AgdaPostulate{w}\AgdaSpace{}%
\AgdaOperator{\AgdaPostulate{===}}\AgdaSpace{}%
\AgdaPostulate{v}\AgdaSpace{}%
\AgdaSymbol{→}\AgdaSpace{}%
\AgdaPostulate{w}\AgdaSpace{}%
\AgdaOperator{\AgdaDatatype{≈}}\AgdaSpace{}%
\AgdaPostulate{v}\<%
\end{code}

Deciding $\approx$ is the word problem of $\langle X \mid \Gamma \rangle$.
A derivation is an ordinary inductive value; the congruence-lifting lemmas of
\S\ref{sec:constructions} are inductions on it.  Compared to
quotient-based designs (e.g.\ mathlib's \aid{PresentedGroup} as a quotient
of a free group by a normal
closure~\cite{mathlib2020,mathlib-doc-presented}, or higher inductive
types in cubical Agda~\cite{vezzosi2019cubical,cubicallibrary}), the
setoid design requires a congruence proof for each function out of the type,
but the normal-form maps of \S\ref{sec:nfproperty} remain ordinary
computable functions on words, with no quotient to eliminate through.
The case studies depend on this, since they prove equations by evaluation,
and it keeps the library compatible with the standard library
(cf.~\cite{barthe2003setoids} for the tradition).
\texttt{Presentation.Properties} packages $\approx$ as an equivalence, a
setoid (the \emph{word setoid}, the domain of every normal form in
\S\ref{sec:nfproperty}), and the monoid $\aid{•-ε-monoid}$ (the
presented monoid), and \texttt{Presentation.GroupLike} upgrades it to a
group when a \aid{Grouplike} witness supplies a left inverse for every
generator: inverses of words are then derived, cancellation and
uniqueness of inverses are lemmas, and $\aid{•-ε-group}$ is the presented
group.

Presentations of concrete algebras are records
(\texttt{Presentation.Definitions}).  The central one:

\begin{code}%
\>[0]\AgdaKeyword{record}\AgdaSpace{}%
\AgdaOperator{\AgdaRecord{\AgdaUnderscore{}IsPresentationOf\AgdaUnderscore{}}}\AgdaSpace{}%
\AgdaSymbol{(}\AgdaOperator{\AgdaBound{\AgdaUnderscore{}===\AgdaUnderscore{}}}\AgdaSpace{}%
\AgdaSymbol{:}\AgdaSpace{}%
\AgdaFunction{WRel}\AgdaSpace{}%
\AgdaPostulate{X}\AgdaSymbol{)}\AgdaSpace{}%
\AgdaSymbol{(}\AgdaBound{G}\AgdaSpace{}%
\AgdaSymbol{:}\AgdaSpace{}%
\AgdaRecord{Group}\AgdaSpace{}%
\AgdaPostulate{a}\AgdaSpace{}%
\AgdaPostulate{ℓ}\AgdaSymbol{)}\AgdaSpace{}%
\AgdaSymbol{:}\AgdaSpace{}%
\AgdaPrimitive{Set}\AgdaSpace{}%
\AgdaSymbol{(}\AgdaPostulate{a}\AgdaSpace{}%
\AgdaOperator{\AgdaPrimitive{⊔}}\AgdaSpace{}%
\AgdaPostulate{ℓ}\AgdaSymbol{)}\AgdaSpace{}%
\AgdaKeyword{where}\<%
\\
\>[0][@{}l@{\AgdaIndent{0}}]%
\>[2]\AgdaKeyword{field}\AgdaSpace{}%
\AgdaField{gl}\AgdaSpace{}%
\AgdaSymbol{:}\AgdaSpace{}%
\AgdaFunction{Grouplike}\AgdaSpace{}%
\AgdaOperator{\AgdaBound{\AgdaUnderscore{}===\AgdaUnderscore{}}}\<%
\\
\>[2]\AgdaKeyword{module}\AgdaSpace{}%
\AgdaModule{GL}\AgdaSpace{}%
\AgdaSymbol{=}\AgdaSpace{}%
\AgdaModule{Group-Lemmas}\AgdaSpace{}%
\AgdaOperator{\AgdaBound{\AgdaUnderscore{}===\AgdaUnderscore{}}}\AgdaSpace{}%
\AgdaField{gl}\<%
\\
\>[2]\AgdaKeyword{open}\AgdaSpace{}%
\AgdaModule{GroupMorphisms}\AgdaSpace{}%
\AgdaSymbol{(}\AgdaFunction{Group.rawGroup}\AgdaSpace{}%
\AgdaPostulate{GL.•-ε-group}\AgdaSymbol{)}\AgdaSpace{}%
\AgdaSymbol{(}\AgdaFunction{Group.rawGroup}\AgdaSpace{}%
\AgdaBound{G}\AgdaSymbol{)}\<%
\\
\>[2]\AgdaKeyword{field}%
\>[295I]\AgdaOperator{\AgdaField{⟦\AgdaUnderscore{}⟧}}\AgdaSpace{}%
\AgdaSymbol{:}\AgdaSpace{}%
\AgdaDatatype{Word}\AgdaSpace{}%
\AgdaPostulate{X}\AgdaSpace{}%
\AgdaSymbol{→}\AgdaSpace{}%
\AgdaField{Group.Carrier}\AgdaSpace{}%
\AgdaBound{G}\<%
\\
\>[.][@{}l@{}]\<[295I]%
\>[8]\AgdaField{iso}\AgdaSpace{}%
\AgdaSymbol{:}\AgdaSpace{}%
\AgdaPostulate{IsGroupIsomorphism}\AgdaSpace{}%
\AgdaOperator{\AgdaField{⟦\AgdaUnderscore{}⟧}}\<%
\end{code}
\begin{code}[hide]%
\>[0]\AgdaKeyword{postulate}\<%
\\
\>[0][@{}l@{\AgdaIndent{0}}]%
\>[2]\AgdaPostulate{|NF|}%
\>[13]\AgdaSymbol{:}\AgdaSpace{}%
\AgdaPrimitive{Set}\<%
\\
\>[2]\AgdaPostulate{nf}%
\>[13]\AgdaSymbol{:}\AgdaSpace{}%
\AgdaDatatype{Word}\AgdaSpace{}%
\AgdaPostulate{X}\AgdaSpace{}%
\AgdaSymbol{→}\AgdaSpace{}%
\AgdaPostulate{|NF|}\<%
\\
\>[2]\AgdaOperator{\AgdaPostulate{\AgdaUnderscore{}≈ₙ\AgdaUnderscore{}}}%
\>[13]\AgdaSymbol{:}\AgdaSpace{}%
\AgdaPostulate{|NF|}\AgdaSpace{}%
\AgdaSymbol{→}\AgdaSpace{}%
\AgdaPostulate{|NF|}\AgdaSpace{}%
\AgdaSymbol{→}\AgdaSpace{}%
\AgdaPrimitive{Set}\<%
\\
\>[2]\AgdaPostulate{NormalForm}\AgdaSpace{}%
\AgdaSymbol{:}\AgdaSpace{}%
\AgdaPrimitive{Set}\<%
\\
\>[2]\AgdaPostulate{inv-nf}%
\>[13]\AgdaSymbol{:}\AgdaSpace{}%
\AgdaPostulate{|NF|}\AgdaSpace{}%
\AgdaSymbol{→}\AgdaSpace{}%
\AgdaDatatype{Word}\AgdaSpace{}%
\AgdaPostulate{X}\<%
\\
\>[2]\AgdaPostulate{Sem}%
\>[13]\AgdaSymbol{:}\AgdaSpace{}%
\AgdaPrimitive{Set}\<%
\\
\>[2]\AgdaOperator{\AgdaPostulate{⟦\AgdaUnderscore{}⟧}}%
\>[13]\AgdaSymbol{:}\AgdaSpace{}%
\AgdaDatatype{Word}\AgdaSpace{}%
\AgdaPostulate{X}\AgdaSpace{}%
\AgdaSymbol{→}\AgdaSpace{}%
\AgdaPostulate{Sem}\<%
\\
\>[2]\AgdaOperator{\AgdaPostulate{\AgdaUnderscore{}≈₂\AgdaUnderscore{}}}%
\>[13]\AgdaSymbol{:}\AgdaSpace{}%
\AgdaPostulate{Sem}\AgdaSpace{}%
\AgdaSymbol{→}\AgdaSpace{}%
\AgdaPostulate{Sem}\AgdaSpace{}%
\AgdaSymbol{→}\AgdaSpace{}%
\AgdaPrimitive{Set}\<%
\end{code}

\noindent i.e.\ a grouplike witness plus a group isomorphism from the presented
group onto $G$ whose underlying function is the interpretation.  Weaker
records (\aid{{\_}IsSubPresentationOf{\_}} with a monomorphism;
monoid-level variants) come with upgrade lemmas: a surjective
sub-presentation is a presentation, and a grouplike monoid presentation of
$G$'s underlying monoid is a group presentation; the latter uses the
\texttt{ForStdlib} fact that monoid homomorphisms between groups preserve
inverses (\S\ref{sec:forstdlib}).  These upgrades structure every proof in
the catalogue: one proves soundness, injectivity (by normalization), and
surjectivity separately, and the records assemble the isomorphism.

\subsection{The Zoo of Normal-Form Witnesses}\label{sec:nfproperty}

Normal forms (\texttt{Normalization.NormalForm.Setoid}), parameterized by
$\Gamma$ and a codomain setoid $\mathit{NF}$, are defined as what it means
for a function out of words to be a normal form.  The definitions are
abbreviations for the standard library's function bundles out of the word
setoid:
\[
\begin{array}{lcl}
\aid{NormalFormInjective} & = & \aid{Injection}\ \text{(word setoid)}\ \mathit{NF} \\
\aid{NormalForm}          & = & \aid{RightInverse}\ \text{(word setoid)}\ \mathit{NF} \\
\aid{BijectiveNormalForm} & = & \aid{Bijection}\ \text{(word setoid)}\ \mathit{NF}
\end{array}
\]
An \aid{Injection} bundles the map $\aid{nf}$, its congruence, and
injectivity: $w \approx v \iff \aid{nf}\,w \approx_{\mathit{NF}}
\aid{nf}\,v$.  Two consequences follow.  The lemma
\begin{code}%
\>[0]\AgdaFunction{by-equal-nf}\AgdaSpace{}%
\AgdaSymbol{:}\AgdaSpace{}%
\AgdaPostulate{nf}\AgdaSpace{}%
\AgdaPostulate{w}\AgdaSpace{}%
\AgdaOperator{\AgdaPostulate{≈ₙ}}\AgdaSpace{}%
\AgdaPostulate{nf}\AgdaSpace{}%
\AgdaPostulate{v}\AgdaSpace{}%
\AgdaSymbol{→}\AgdaSpace{}%
\AgdaPostulate{w}\AgdaSpace{}%
\AgdaOperator{\AgdaDatatype{≈}}\AgdaSpace{}%
\AgdaPostulate{v}\<%
\end{code}
\begin{code}[hide]%
\>[0]\AgdaFunction{by-equal-nf}\AgdaSpace{}%
\AgdaSymbol{=}\AgdaSpace{}%
\AgdaPostulate{proof-elided}\<%
\end{code}
proves word equations by computing both normal forms
(\S\ref{sec:solvers}), and decidability transports along the map,
\begin{code}%
\>[0]\AgdaFunction{≈-dec}\AgdaSpace{}%
\AgdaSymbol{:}\AgdaSpace{}%
\AgdaFunction{Decidable}\AgdaSpace{}%
\AgdaOperator{\AgdaPostulate{\AgdaUnderscore{}≈ₙ\AgdaUnderscore{}}}\AgdaSpace{}%
\AgdaSymbol{→}\AgdaSpace{}%
\AgdaFunction{Decidable}\AgdaSpace{}%
\AgdaOperator{\AgdaDatatype{\AgdaUnderscore{}≈\AgdaUnderscore{}}}\<%
\end{code}
\begin{code}[hide]%
\>[0]\AgdaFunction{≈-dec}\AgdaSpace{}%
\AgdaSymbol{=}\AgdaSpace{}%
\AgdaPostulate{proof-elided}\<%
\end{code}
so a normal form with decidable carrier decides the word problem.
A \aid{RightInverse} additionally carries a section
$\aid{inv-nf} : |\mathit{NF}| \to \aid{Word}\;X$ with
$\aid{inv-nf} \circ \aid{nf} \approx \mathrm{id}$; from it we
derive $\aid{normalize} = \aid{inv-nf} \circ \aid{nf}$, its idempotence,
and injectivity of \aid{nf}, so every \aid{NormalForm} is a
\aid{NormalFormInjective}.  The bijective witness is a \aid{Bijection}
rather than the interderivable \aid{Inverse}: its section is read off the
surjectivity witness, as in the extension construction
(\S\ref{sec:constructions}).

The converse does not hold constructively.  Classically, any
injective $\aid{nf}$ whose image can be identified has a section, obtained
by choosing preimages; constructively, the section must be supplied.  An
\aid{Injection} gives no algorithm producing, for a normal form $u$ in
its image, a word realizing it.  That algorithm is the extra data of
\aid{RightInverse}, and the completeness lemma below uses it
($\aid{inv-nf}$ recovers each word from its normal form).  The two
witnesses therefore differ in strength, and we keep both: the weaker
suffices for word-problem reasoning (\aid{by-equal-nf}, decidability),
while completeness and presentations require the stronger.  Exactness,
$\aid{nf} \circ \aid{inv-nf} \approx \mathrm{id}$, is a separate hypothesis
for \aid{by-completeness}; it excludes unused carrier elements.  A fourth
witness, \aid{WeakNormalForm}, also drops the
congruence of the map; an injective map suffices to separate words,
which is enough for some transports along unions of
relation sets (\S\ref{sec:constructions}).

Completeness itself is one generic lemma.  Fix a semantics
$\aid{⟦\_⟧} : \aid{Word}\;X \to \mathit{Sem}$ into any setoid.  A
\aid{NormalForm} is \emph{unique for} $\aid{⟦\_⟧}$ when representatives
with equal denotations are equal normal forms:

\begin{code}%
\>[0]\AgdaKeyword{record}\AgdaSpace{}%
\AgdaRecord{UniqueNormalForm}\AgdaSpace{}%
\AgdaSymbol{(}\AgdaBound{normalForm}\AgdaSpace{}%
\AgdaSymbol{:}\AgdaSpace{}%
\AgdaPostulate{NormalForm}\AgdaSymbol{)}\AgdaSpace{}%
\AgdaSymbol{:}\AgdaSpace{}%
\AgdaPrimitive{Set}\AgdaSpace{}%
\AgdaSymbol{(}\AgdaPostulate{c}\AgdaSpace{}%
\AgdaOperator{\AgdaPrimitive{⊔}}\AgdaSpace{}%
\AgdaPostulate{d}\AgdaSymbol{)}\AgdaSpace{}%
\AgdaKeyword{where}\<%
\\
\>[0][@{}l@{\AgdaIndent{0}}]%
\>[2]\AgdaKeyword{field}\AgdaSpace{}%
\AgdaField{unique}\AgdaSpace{}%
\AgdaSymbol{:}\AgdaSpace{}%
\AgdaSymbol{∀}\AgdaSpace{}%
\AgdaSymbol{\{}\AgdaBound{u}\AgdaSpace{}%
\AgdaBound{v}\AgdaSpace{}%
\AgdaSymbol{:}\AgdaSpace{}%
\AgdaPostulate{|NF|}\AgdaSymbol{\}}\AgdaSpace{}%
\AgdaSymbol{→}\AgdaSpace{}%
\AgdaOperator{\AgdaPostulate{⟦}}\AgdaSpace{}%
\AgdaPostulate{inv-nf}\AgdaSpace{}%
\AgdaBound{u}\AgdaSpace{}%
\AgdaOperator{\AgdaPostulate{⟧}}\AgdaSpace{}%
\AgdaOperator{\AgdaPostulate{≈₂}}\AgdaSpace{}%
\AgdaOperator{\AgdaPostulate{⟦}}\AgdaSpace{}%
\AgdaPostulate{inv-nf}\AgdaSpace{}%
\AgdaBound{v}\AgdaSpace{}%
\AgdaOperator{\AgdaPostulate{⟧}}\AgdaSpace{}%
\AgdaSymbol{→}\AgdaSpace{}%
\AgdaBound{u}\AgdaSpace{}%
\AgdaOperator{\AgdaPostulate{≈ₙ}}\AgdaSpace{}%
\AgdaBound{v}\<%
\\
\>[0]\AgdaFunction{by-normalization}\AgdaSpace{}%
\AgdaSymbol{:}\AgdaSpace{}%
\AgdaSymbol{\{}\AgdaBound{normalForm}\AgdaSpace{}%
\AgdaSymbol{:}\AgdaSpace{}%
\AgdaPostulate{NormalForm}\AgdaSymbol{\}}\AgdaSpace{}%
\AgdaSymbol{→}\AgdaSpace{}%
\AgdaRecord{UniqueNormalForm}\AgdaSpace{}%
\AgdaBound{normalForm}\AgdaSpace{}%
\AgdaSymbol{→}\<%
\\
\>[0][@{}l@{\AgdaIndent{0}}]%
\>[2]\AgdaFunction{Congruent}\AgdaSpace{}%
\AgdaOperator{\AgdaDatatype{\AgdaUnderscore{}≈\AgdaUnderscore{}}}\AgdaSpace{}%
\AgdaOperator{\AgdaPostulate{\AgdaUnderscore{}≈₂\AgdaUnderscore{}}}\AgdaSpace{}%
\AgdaOperator{\AgdaPostulate{⟦\AgdaUnderscore{}⟧}}\AgdaSpace{}%
\AgdaSymbol{→}\AgdaSpace{}%
\AgdaFunction{Injective}\AgdaSpace{}%
\AgdaOperator{\AgdaDatatype{\AgdaUnderscore{}≈\AgdaUnderscore{}}}\AgdaSpace{}%
\AgdaOperator{\AgdaPostulate{\AgdaUnderscore{}≈₂\AgdaUnderscore{}}}\AgdaSpace{}%
\AgdaOperator{\AgdaPostulate{⟦\AgdaUnderscore{}⟧}}\<%
\end{code}
\begin{code}[hide]%
\>[0]\AgdaFunction{by-normalization}\AgdaSpace{}%
\AgdaSymbol{=}\AgdaSpace{}%
\AgdaPostulate{proof-elided}\<%
\end{code}

\emph{Soundness plus a unique normal form yields completeness}: given
$\aid{⟦}w\aid{⟧} \approx_2 \aid{⟦}v\aid{⟧}$, soundness transports along
$w \approx \aid{normalize}\,w$, uniqueness identifies the normal forms,
and injectivity of \aid{nf} gives $w \approx v$.  The proof is five lines.
Its use is that completeness reduces to \aid{unique}, a
statement about normal forms (concrete first-order data)
rather than about arbitrary words.  A companion lemma
(\aid{by-normalization-wsurj}) similarly reduces surjectivity of the
semantics to surjectivity on normal forms.  Sub-presentations follow
(\texttt{Normalization.StarPresentation}): given soundness on
axioms, a \aid{NormalForm}, and a \aid{UniqueNormalForm}, the module
assembles the monoid (or, with \aid{Grouplike}, group) monomorphism,
whose injectivity field is \aid{by-normalization}.

\subsection{Inductively Defined Circuits}\label{sec:circuits}

The circuit layer (\texttt{Circuit.Base}) is parameterized by a gate
family $\aid{Gate} : \mathbb{N} \to \mathsf{Set}$ and defines the
wire-indexed generators \aid{Gen} and circuits $\aid{Circuit}\;n =
\aid{Word}\;(\aid{Gen}\;n)$ as in \S\ref{sec:permutations}.
A generator is a 1- or 2-ary gate at the bottom of the wires, shifted up
by some number of \aid{{\_}↥}s; $c{\uparrow} = \aid{wmap}\;\aid{{\_}↥}$
shifts a whole circuit.  Two design choices are worth explaining.
First, the indices are built from successors only: \aid{gate₂} lands in
$\aid{Gen}\,(2{+}n)$, not in $\aid{Gen}\,(k)$ with a side condition
$k \ge 2$, and the $k$-fold shift has type
$\aid{Gen}\;n \to \aid{Gen}\;(k + n)$ with $k$ on the \emph{left} of
$+$: addition on naturals recurses on its first argument, so $k + n$
computes as soon as $k$ is a successor pattern while $n + k$ is stuck.
This lets the coverage checker
solve its unification problems by injectivity of \aid{₁₊} alone,
avoiding stuck Diophantine equations of the form
$\mathit{arity} + k \doteq \mathit{target}$; downstream, it is
why the coset tables of the case studies can be defined by direct pattern
matching.  Second, the structural rules common
to all circuit theories are factored out once, by
\aid{Lift-Relation}, whose permutation instance appeared in
\S\ref{sec:permutations}.  The accompanying lemma \aid{lemma-cong↑}
lifts the whole congruence one wire up, by
induction over the eight constructors of $\approx$, so that equational
proofs transfer along the wire homomorphisms
$\mathsf{Cir}\,k \to \mathsf{Cir}\,(k{+}1)$ of
\S\ref{sec:circuit-chain}; injectivity follows from the coset-table hypotheses.  A new circuit theory is defined by instantiating \aid{Gate}
and giving its gate-specific relation set; the structural rules and their
lemmas are inherited.

\subsection{The Reidemeister--Schreier Engine}\label{sec:engine}

The module \texttt{Normalization.Reidemeister-Schreier} provides
injectivity theorems for extensions of generator maps, in two strengths.
The \emph{simplified} form needs no cosets: let $\Gamma$ be a relation
set over $X$ and $\Delta$ one over $Y$, with congruences $\approx_\Gamma$
and $\approx_\Delta$; if $f : X \to \aid{Word}\;Y$ has a
generator-level retraction $g : Y \to \aid{Word}\;X$ (meaning that
$(g\,\aid{ʷ})$ respects $\Delta$'s axioms and
$[x]\aid{ʷ} \approx_\Gamma (g\,\aid{ʷ})(f\,x)$), then $(f\,\aid{ʷ})$ is
injective and $(g\,\aid{ʷ})$ surjective.  This is the tool for
isomorphisms of presentations, and it is used in the final step of each
amalgamation case study via the \aid{StarIsomorphism} builder of
\texttt{Presentation.\allowbreak Morphism}, which turns the four
generator-level conditions into an \aid{IsMonoidIsomorphism} of the
presented monoids.

The \emph{full} form works from a coset table.  Its packaging
(\texttt{Normalization.CosetNF}) is the library's central module
(\aid{SingleLevel} shown; a setoid-coset variant exists):

\begin{code}%
\>[0]\AgdaKeyword{module}\AgdaSpace{}%
\AgdaModule{SingleLevel}\AgdaSpace{}%
\AgdaSymbol{(}\AgdaBound{Γ}\AgdaSpace{}%
\AgdaSymbol{:}\AgdaSpace{}%
\AgdaFunction{WRel}\AgdaSpace{}%
\AgdaPostulate{X}\AgdaSymbol{)}\AgdaSpace{}%
\AgdaSymbol{(}\AgdaBound{Δ}\AgdaSpace{}%
\AgdaSymbol{:}\AgdaSpace{}%
\AgdaFunction{WRel}\AgdaSpace{}%
\AgdaPostulate{Y}\AgdaSymbol{)}\AgdaSpace{}%
\AgdaSymbol{(}\AgdaBound{C}\AgdaSpace{}%
\AgdaSymbol{:}\AgdaSpace{}%
\AgdaPrimitive{Set}\AgdaSymbol{)}\AgdaSpace{}%
\AgdaSymbol{(}\AgdaBound{I}\AgdaSpace{}%
\AgdaSymbol{:}\AgdaSpace{}%
\AgdaBound{C}\AgdaSymbol{)}\<%
\\
\>[0][@{}l@{\AgdaIndent{0}}]%
\>[2]\AgdaSymbol{(}\AgdaBound{f}\AgdaSpace{}%
\AgdaSymbol{:}\AgdaSpace{}%
\AgdaPostulate{X}\AgdaSpace{}%
\AgdaSymbol{→}\AgdaSpace{}%
\AgdaDatatype{Word}\AgdaSpace{}%
\AgdaPostulate{Y}\AgdaSymbol{)}\AgdaSpace{}%
\AgdaSymbol{(}\AgdaBound{h}\AgdaSpace{}%
\AgdaSymbol{:}\AgdaSpace{}%
\AgdaBound{C}\AgdaSpace{}%
\AgdaSymbol{→}\AgdaSpace{}%
\AgdaPostulate{Y}\AgdaSpace{}%
\AgdaSymbol{→}\AgdaSpace{}%
\AgdaDatatype{Word}\AgdaSpace{}%
\AgdaPostulate{X}\AgdaSpace{}%
\AgdaOperator{\AgdaFunction{×}}\AgdaSpace{}%
\AgdaBound{C}\AgdaSymbol{)}\AgdaSpace{}%
\AgdaSymbol{(}\AgdaOperator{\AgdaBound{[\AgdaUnderscore{}]}}\AgdaSpace{}%
\AgdaSymbol{:}\AgdaSpace{}%
\AgdaBound{C}\AgdaSpace{}%
\AgdaSymbol{→}\AgdaSpace{}%
\AgdaDatatype{Word}\AgdaSpace{}%
\AgdaPostulate{Y}\AgdaSymbol{)}\<%
\end{code}
\begin{code}[hide]%
\>[2]\AgdaKeyword{where}\<%
\\
\\[\AgdaEmptyExtraSkip]%
\>[0]\AgdaKeyword{postulate}\<%
\\
\>[0][@{}l@{\AgdaIndent{0}}]%
\>[2]\AgdaPostulate{I}%
\>[9]\AgdaSymbol{:}\AgdaSpace{}%
\AgdaPostulate{C}\<%
\\
\>[2]\AgdaPostulate{f}%
\>[9]\AgdaSymbol{:}\AgdaSpace{}%
\AgdaPostulate{X}\AgdaSpace{}%
\AgdaSymbol{→}\AgdaSpace{}%
\AgdaDatatype{Word}\AgdaSpace{}%
\AgdaPostulate{Y}\<%
\\
\>[2]\AgdaPostulate{h}%
\>[9]\AgdaSymbol{:}\AgdaSpace{}%
\AgdaPostulate{C}\AgdaSpace{}%
\AgdaSymbol{→}\AgdaSpace{}%
\AgdaPostulate{Y}\AgdaSpace{}%
\AgdaSymbol{→}\AgdaSpace{}%
\AgdaDatatype{Word}\AgdaSpace{}%
\AgdaPostulate{X}\AgdaSpace{}%
\AgdaOperator{\AgdaFunction{×}}\AgdaSpace{}%
\AgdaPostulate{C}\<%
\\
\>[2]\AgdaOperator{\AgdaPostulate{[\AgdaUnderscore{}]}}%
\>[9]\AgdaSymbol{:}\AgdaSpace{}%
\AgdaPostulate{C}\AgdaSpace{}%
\AgdaSymbol{→}\AgdaSpace{}%
\AgdaDatatype{Word}\AgdaSpace{}%
\AgdaPostulate{Y}\<%
\\
\>[2]\AgdaOperator{\AgdaPostulate{\AgdaUnderscore{}\textasciitilde{}\AgdaUnderscore{}}}%
\>[9]\AgdaSymbol{:}\AgdaSpace{}%
\AgdaDatatype{Word}\AgdaSpace{}%
\AgdaPostulate{X}\AgdaSpace{}%
\AgdaOperator{\AgdaFunction{×}}\AgdaSpace{}%
\AgdaPostulate{C}\AgdaSpace{}%
\AgdaSymbol{→}\AgdaSpace{}%
\AgdaDatatype{Word}\AgdaSpace{}%
\AgdaPostulate{X}\AgdaSpace{}%
\AgdaOperator{\AgdaFunction{×}}\AgdaSpace{}%
\AgdaPostulate{C}\AgdaSpace{}%
\AgdaSymbol{→}\AgdaSpace{}%
\AgdaPrimitive{Set}\<%
\\
\>[2]\AgdaOperator{\AgdaPostulate{\AgdaUnderscore{}===₁\AgdaUnderscore{}}}\AgdaSpace{}%
\AgdaSymbol{:}\AgdaSpace{}%
\AgdaFunction{WRel}\AgdaSpace{}%
\AgdaPostulate{X}\<%
\\
\>[2]\AgdaOperator{\AgdaPostulate{\AgdaUnderscore{}===₂\AgdaUnderscore{}}}\AgdaSpace{}%
\AgdaSymbol{:}\AgdaSpace{}%
\AgdaFunction{WRel}\AgdaSpace{}%
\AgdaPostulate{Y}\<%
\end{code}

Think of $\Gamma$ as presenting a subgroup $H$ of the group $G$ presented
by $\Delta$, over the letters $X$ and $Y$ respectively; $C$ is the set of
right cosets, $I$ the identity coset, $f$ the embedding of generators, and
$[\,\_\,]$ the Schreier section.  The table $h$ pushes one letter of $G$
past a coset, emitting a word of $H$ and the successor coset, so
$\aid{nf} = (h\,\aid{ᵗ})\,I : \aid{Word}\;Y \to \aid{Word}\;X \times C$
runs the whole word from the identity coset.  A submodule
\aid{Transfer} takes the five hypotheses under which the data form a
coset presentation (in the code, $\aid{===₁}$, $\approx_1$ and
$\aid{===₂}$, $\approx_2$ are the axioms and congruences of $\Gamma$ and
$\Delta$, and $\sim$ compares word-and-coset pairs, the words up
to $\approx_1$):

\begin{code}%
\>[0]\AgdaKeyword{module}\AgdaSpace{}%
\AgdaModule{Transfer}\<%
\\
\>[0][@{}l@{\AgdaIndent{0}}]%
\>[2]\AgdaSymbol{(}\AgdaBound{h=⁻¹f-gen}\AgdaSpace{}%
\AgdaSymbol{:}\AgdaSpace{}%
\AgdaSymbol{∀}\AgdaSpace{}%
\AgdaSymbol{(}\AgdaBound{x}\AgdaSpace{}%
\AgdaSymbol{:}\AgdaSpace{}%
\AgdaPostulate{X}\AgdaSymbol{)}\AgdaSpace{}%
\AgdaSymbol{→}\AgdaSpace{}%
\AgdaSymbol{(}\AgdaOperator{\AgdaInductiveConstructor{[}}\AgdaSpace{}%
\AgdaBound{x}\AgdaSpace{}%
\AgdaOperator{\AgdaInductiveConstructor{]ʷ}}\AgdaSpace{}%
\AgdaOperator{\AgdaInductiveConstructor{,}}\AgdaSpace{}%
\AgdaPostulate{I}\AgdaSymbol{)}\AgdaSpace{}%
\AgdaOperator{\AgdaPostulate{\textasciitilde{}}}\AgdaSpace{}%
\AgdaSymbol{((}\AgdaPostulate{h}\AgdaSpace{}%
\AgdaOperator{\AgdaFunction{ᵗ}}\AgdaSymbol{)}\AgdaSpace{}%
\AgdaPostulate{I}\AgdaSpace{}%
\AgdaSymbol{(}\AgdaPostulate{f}\AgdaSpace{}%
\AgdaBound{x}\AgdaSymbol{)))}%
\>[83]\AgdaComment{--\ 1}\<%
\\
\>[2]\AgdaSymbol{(}\AgdaBound{h-wd-ax}\AgdaSpace{}%
\AgdaSymbol{:}\AgdaSpace{}%
\AgdaSymbol{∀}\AgdaSpace{}%
\AgdaSymbol{(}\AgdaBound{c}\AgdaSpace{}%
\AgdaSymbol{:}\AgdaSpace{}%
\AgdaPostulate{C}\AgdaSymbol{)\{}\AgdaBound{u}\AgdaSpace{}%
\AgdaBound{t}\AgdaSpace{}%
\AgdaSymbol{:}\AgdaSpace{}%
\AgdaDatatype{Word}\AgdaSpace{}%
\AgdaPostulate{Y}\AgdaSymbol{\}}\AgdaSpace{}%
\AgdaSymbol{→}\AgdaSpace{}%
\AgdaBound{u}\AgdaSpace{}%
\AgdaOperator{\AgdaPostulate{===₂}}\AgdaSpace{}%
\AgdaBound{t}\AgdaSpace{}%
\AgdaSymbol{→}\AgdaSpace{}%
\AgdaSymbol{((}\AgdaPostulate{h}\AgdaSpace{}%
\AgdaOperator{\AgdaFunction{ᵗ}}\AgdaSymbol{)}\AgdaSpace{}%
\AgdaBound{c}\AgdaSpace{}%
\AgdaBound{u}\AgdaSymbol{)}\AgdaSpace{}%
\AgdaOperator{\AgdaPostulate{\textasciitilde{}}}\AgdaSpace{}%
\AgdaSymbol{((}\AgdaPostulate{h}\AgdaSpace{}%
\AgdaOperator{\AgdaFunction{ᵗ}}\AgdaSymbol{)}\AgdaSpace{}%
\AgdaBound{c}\AgdaSpace{}%
\AgdaBound{t}\AgdaSymbol{))}%
\>[83]\AgdaComment{--\ 2}\<%
\\
\>[2]\AgdaSymbol{(}\AgdaBound{f-wd-ax}\AgdaSpace{}%
\AgdaSymbol{:}\AgdaSpace{}%
\AgdaSymbol{∀}\AgdaSpace{}%
\AgdaSymbol{\{}\AgdaBound{w}\AgdaSpace{}%
\AgdaBound{v}\AgdaSymbol{\}}\AgdaSpace{}%
\AgdaSymbol{→}\AgdaSpace{}%
\AgdaBound{w}\AgdaSpace{}%
\AgdaOperator{\AgdaPostulate{===₁}}\AgdaSpace{}%
\AgdaBound{v}\AgdaSpace{}%
\AgdaSymbol{→}\AgdaSpace{}%
\AgdaSymbol{(}\AgdaPostulate{f}\AgdaSpace{}%
\AgdaOperator{\AgdaFunction{ʷ}}\AgdaSymbol{)}\AgdaSpace{}%
\AgdaBound{w}\AgdaSpace{}%
\AgdaOperator{\AgdaPostulate{≈₂}}\AgdaSpace{}%
\AgdaSymbol{(}\AgdaPostulate{f}\AgdaSpace{}%
\AgdaOperator{\AgdaFunction{ʷ}}\AgdaSymbol{)}\AgdaSpace{}%
\AgdaBound{v}\AgdaSymbol{)}%
\>[83]\AgdaComment{--\ 3}\<%
\\
\>[2]\AgdaSymbol{(}\AgdaBound{[I]≈ε}\AgdaSpace{}%
\AgdaSymbol{:}\AgdaSpace{}%
\AgdaOperator{\AgdaPostulate{[}}\AgdaSpace{}%
\AgdaPostulate{I}\AgdaSpace{}%
\AgdaOperator{\AgdaPostulate{]}}\AgdaSpace{}%
\AgdaOperator{\AgdaPostulate{≈₂}}\AgdaSpace{}%
\AgdaInductiveConstructor{ε}\AgdaSymbol{)}%
\>[83]\AgdaComment{--\ 4}\<%
\\
\>[2]\AgdaSymbol{(}\AgdaBound{h=ract}\AgdaSpace{}%
\AgdaSymbol{:}\AgdaSpace{}%
\AgdaSymbol{∀}\AgdaSpace{}%
\AgdaBound{c}\AgdaSpace{}%
\AgdaBound{b}\AgdaSpace{}%
\AgdaSymbol{→}\AgdaSpace{}%
\AgdaKeyword{let}\AgdaSpace{}%
\AgdaSymbol{(}\AgdaBound{b'}\AgdaSpace{}%
\AgdaOperator{\AgdaInductiveConstructor{,}}\AgdaSpace{}%
\AgdaBound{c'}\AgdaSymbol{)}\AgdaSpace{}%
\AgdaSymbol{=}\AgdaSpace{}%
\AgdaPostulate{h}\AgdaSpace{}%
\AgdaBound{c}\AgdaSpace{}%
\AgdaBound{b}\AgdaSpace{}%
\AgdaKeyword{in}\AgdaSpace{}%
\AgdaOperator{\AgdaPostulate{[}}\AgdaSpace{}%
\AgdaBound{c}\AgdaSpace{}%
\AgdaOperator{\AgdaPostulate{]}}\AgdaSpace{}%
\AgdaOperator{\AgdaInductiveConstructor{•}}\AgdaSpace{}%
\AgdaOperator{\AgdaInductiveConstructor{[}}\AgdaSpace{}%
\AgdaBound{b}\AgdaSpace{}%
\AgdaOperator{\AgdaInductiveConstructor{]ʷ}}\AgdaSpace{}%
\AgdaOperator{\AgdaPostulate{≈₂}}\AgdaSpace{}%
\AgdaSymbol{(}\AgdaPostulate{f}\AgdaSpace{}%
\AgdaOperator{\AgdaFunction{ʷ}}\AgdaSymbol{)}\AgdaSpace{}%
\AgdaBound{b'}\AgdaSpace{}%
\AgdaOperator{\AgdaInductiveConstructor{•}}\AgdaSpace{}%
\AgdaOperator{\AgdaPostulate{[}}\AgdaSpace{}%
\AgdaBound{c'}\AgdaSpace{}%
\AgdaOperator{\AgdaPostulate{]}}\AgdaSymbol{)}%
\>[83]\AgdaComment{--\ 5}\<%
\end{code}
\begin{code}[hide]%
\>[2]\AgdaKeyword{where}\<%
\end{code}

\noindent (1) the table inverts the embedding at the identity coset; (2) the
table respects $\Delta$'s axioms, coset by coset; (3) the embedding
respects $\Gamma$'s axioms; (4) the identity coset is represented by
$\varepsilon$; (5) sliding one letter past a coset's representative
matches the table (the law \aid{ract-sound} of \S\ref{sec:permutations}).
From these, \aid{Transfer} derives well-definedness of
\aid{nf} on $\approx_2$, the factorization
$[c] \aid{•} b \approx_2 (f\,\aid{ʷ})\,b' \aid{•} [c']$ for whole words,
injectivity, and the \emph{normal-form transport}
\[
  \aid{nfp'} : \mathrm{NormalForm}\;\Gamma\;\mathit{NF}
    \to \mathrm{NormalForm}\;\Delta\;(\mathit{NF} \times C),
\]
a normal form for the subgroup yields one for the group, with one more
coset digit.  A record \aid{CosetTower.Extension} bundles the data and
hypotheses as a single value, and \aid{CosetTower} iterates levels up an
$\mathbb{N}$-indexed family of presentations, with carrier
$\aid{tower-carrier}\;\mathit{NF}_0\;(k{+}1) =
 \aid{tower-carrier}\;\mathit{NF}_0\;k \times C_k$: the mixed-radix normal
form of \S\ref{sec:preliminaries}, one verified digit per level.
The stronger \aid{PackedCosetTable}, used for amalgamation, additionally
requires preservation of nonidentity cosets and exact traversal of
embedded generators at the identity.

Two properties keep the verification effort manageable.  First, each
hypothesis quantifies only over cosets and generators (finite
at each level in our case studies) and axioms (a finite inductive family),
so the obligations are finite case splits.  Second, the cases of an obligation like
(2) close by computation: with $\Gamma$'s normal form in scope,
each becomes \aid{by-equal-nf Eq.refl}: both sides normalize to the
same value, and Agda's evaluator checks it.  In the qutrit Clifford+$T$
tower (\S\ref{sec:qutrit}), the nine-coset level imposes seventy-two
such obligations; every
one is discharged by \aid{refl}.  The other hypotheses, the soundness
law (5) above all, are equational derivations done by hand, one per
case, as \aid{ract-sound} was in \S\ref{sec:permutations}.  In this
sense the framework mechanizes part of the case analysis described in
\S\ref{sec:intro}.

\subsection{Products of Presentations, and Their Presentation Theorems}\label{sec:constructions}

Relation sets over a disjoint union of alphabets
(\texttt{Presentation.Construct.Base}) are built from a single primitive,
a three-way join whose \emph{mixed} component is the design parameter:

\begin{code}%
\>[0]\AgdaKeyword{data}\AgdaSpace{}%
\AgdaOperator{\AgdaDatatype{\AgdaUnderscore{}⋄\AgdaUnderscore{}⋄\AgdaUnderscore{}}}\AgdaSpace{}%
\AgdaSymbol{(}\AgdaBound{Γ₁}\AgdaSpace{}%
\AgdaSymbol{:}\AgdaSpace{}%
\AgdaFunction{WRel}\AgdaSpace{}%
\AgdaPostulate{A}\AgdaSymbol{)}\AgdaSpace{}%
\AgdaSymbol{(}\AgdaBound{Γ₂}\AgdaSpace{}%
\AgdaSymbol{:}\AgdaSpace{}%
\AgdaFunction{WRel}\AgdaSpace{}%
\AgdaPostulate{B}\AgdaSymbol{)}\AgdaSpace{}%
\AgdaSymbol{(}\AgdaBound{Γ₃}\AgdaSpace{}%
\AgdaSymbol{:}\AgdaSpace{}%
\AgdaFunction{WRel}\AgdaSpace{}%
\AgdaSymbol{(}\AgdaPostulate{A}\AgdaSpace{}%
\AgdaOperator{\AgdaDatatype{⊎}}\AgdaSpace{}%
\AgdaPostulate{B}\AgdaSymbol{))}\AgdaSpace{}%
\AgdaSymbol{:}\AgdaSpace{}%
\AgdaFunction{WRel}\AgdaSpace{}%
\AgdaSymbol{(}\AgdaPostulate{A}\AgdaSpace{}%
\AgdaOperator{\AgdaDatatype{⊎}}\AgdaSpace{}%
\AgdaPostulate{B}\AgdaSymbol{)}\AgdaSpace{}%
\AgdaKeyword{where}\<%
\\
\>[0][@{}l@{\AgdaIndent{0}}]%
\>[2]\AgdaInductiveConstructor{left}%
\>[8]\AgdaSymbol{:}\AgdaSpace{}%
\AgdaBound{Γ₁}\AgdaSpace{}%
\AgdaPostulate{u}\AgdaSpace{}%
\AgdaPostulate{v}\AgdaSpace{}%
\AgdaSymbol{→}\AgdaSpace{}%
\AgdaSymbol{(}\AgdaBound{Γ₁}\AgdaSpace{}%
\AgdaOperator{\AgdaDatatype{⋄}}\AgdaSpace{}%
\AgdaBound{Γ₂}\AgdaSpace{}%
\AgdaOperator{\AgdaDatatype{⋄}}\AgdaSpace{}%
\AgdaBound{Γ₃}\AgdaSymbol{)}\AgdaSpace{}%
\AgdaOperator{\AgdaFunction{[}}\AgdaSpace{}%
\AgdaPostulate{u}\AgdaSpace{}%
\AgdaOperator{\AgdaFunction{]ₗ}}\AgdaSpace{}%
\AgdaOperator{\AgdaFunction{[}}\AgdaSpace{}%
\AgdaPostulate{v}\AgdaSpace{}%
\AgdaOperator{\AgdaFunction{]ₗ}}\<%
\\
\>[2]\AgdaInductiveConstructor{right}\AgdaSpace{}%
\AgdaSymbol{:}\AgdaSpace{}%
\AgdaBound{Γ₂}\AgdaSpace{}%
\AgdaPostulate{u}\AgdaSpace{}%
\AgdaPostulate{v}\AgdaSpace{}%
\AgdaSymbol{→}\AgdaSpace{}%
\AgdaSymbol{(}\AgdaBound{Γ₁}\AgdaSpace{}%
\AgdaOperator{\AgdaDatatype{⋄}}\AgdaSpace{}%
\AgdaBound{Γ₂}\AgdaSpace{}%
\AgdaOperator{\AgdaDatatype{⋄}}\AgdaSpace{}%
\AgdaBound{Γ₃}\AgdaSymbol{)}\AgdaSpace{}%
\AgdaOperator{\AgdaFunction{[}}\AgdaSpace{}%
\AgdaPostulate{u}\AgdaSpace{}%
\AgdaOperator{\AgdaFunction{]ᵣ}}\AgdaSpace{}%
\AgdaOperator{\AgdaFunction{[}}\AgdaSpace{}%
\AgdaPostulate{v}\AgdaSpace{}%
\AgdaOperator{\AgdaFunction{]ᵣ}}\<%
\\
\>[2]\AgdaInductiveConstructor{mid}%
\>[8]\AgdaSymbol{:}\AgdaSpace{}%
\AgdaBound{Γ₃}\AgdaSpace{}%
\AgdaPostulate{u}\AgdaSpace{}%
\AgdaPostulate{v}\AgdaSpace{}%
\AgdaSymbol{→}\AgdaSpace{}%
\AgdaSymbol{(}\AgdaBound{Γ₁}\AgdaSpace{}%
\AgdaOperator{\AgdaDatatype{⋄}}\AgdaSpace{}%
\AgdaBound{Γ₂}\AgdaSpace{}%
\AgdaOperator{\AgdaDatatype{⋄}}\AgdaSpace{}%
\AgdaBound{Γ₃}\AgdaSymbol{)}\AgdaSpace{}%
\AgdaPostulate{u}\AgdaSpace{}%
\AgdaPostulate{v}\<%
\end{code}

Choosing the mixed relation recovers the classical constructions:
free product ($\aid{EmptyRel}$), direct product ($\aid{CommRel}$:
generators of the two sides commute), semidirect product
($\aid{ConjRel}$/$\aid{ConjRelʷ}$: moving $h$ past $n$ conjugates $n$,
by a generator or by a word), amalgamated free product
($\aid{AmalgRel}\,f_1\,f_2$: the two images of a shared alphabet are
identified), and definitional extension ($\aid{SugarRel}$: a new
generator desugars to a word), plus $n$-fold iterations
$\Gamma\,\aid{⊕\char94}\,n$.  Congruence-lifting lemmas (\aid{lefts},
\aid{rights}) embed the factors' congruences into the join, and
normal-form transports move witnesses along unions and monomorphisms.

For each construction, \texttt{Presentation.Construct.Properties} proves
two kinds of theorems.  \emph{Normal-form lifting}: from NF witnesses for
$\Gamma$ and $\Delta$, an NF witness for the product relation set, with
product carrier ($\mathit{NF}_1 \times \mathit{NF}_2$ for direct and
semidirect products; for amalgamations the base normal form times the
classical alternating carrier
\begin{code}[hide]%
\>[0]\AgdaFunction{amalNFC}\AgdaSpace{}%
\AgdaSymbol{:}\AgdaSpace{}%
\AgdaPrimitive{Set}\AgdaSpace{}%
\AgdaSymbol{→}\AgdaSpace{}%
\AgdaPrimitive{Set}\AgdaSpace{}%
\AgdaSymbol{→}\AgdaSpace{}%
\AgdaPrimitive{Set}\<%
\end{code}
\begin{code}%
\>[0]\AgdaFunction{amalNFC}\AgdaSpace{}%
\AgdaBound{C}\AgdaSpace{}%
\AgdaBound{D}\AgdaSpace{}%
\AgdaSymbol{=}\AgdaSpace{}%
\AgdaSymbol{(}\AgdaBound{D}\AgdaSpace{}%
\AgdaOperator{\AgdaDatatype{⊎}}\AgdaSpace{}%
\AgdaRecord{⊤}\AgdaSymbol{)}\AgdaSpace{}%
\AgdaOperator{\AgdaFunction{×}}\AgdaSpace{}%
\AgdaDatatype{List}\AgdaSpace{}%
\AgdaSymbol{(}\AgdaBound{C}\AgdaSpace{}%
\AgdaOperator{\AgdaFunction{×}}\AgdaSpace{}%
\AgdaBound{D}\AgdaSymbol{)}\AgdaSpace{}%
\AgdaOperator{\AgdaFunction{×}}\AgdaSpace{}%
\AgdaSymbol{(}\AgdaBound{C}\AgdaSpace{}%
\AgdaOperator{\AgdaDatatype{⊎}}\AgdaSpace{}%
\AgdaRecord{⊤}\AgdaSymbol{)}\<%
\end{code}
that is, a word of alternating nontrivial coset representatives between two
tails~\cite{magnus2004combinatorial}, with $C$ and $D$ the nonidentity
coset types).  \emph{Presentation}: if the factors present concrete groups,
the constructed relation set presents the constructed group.  Quoting the direct, semidirect, $n$-fold, and
amalgamated statements (premise telescopes elided; each \aid{dpres} lives
in its own construction's module, $P_1 \mathrel{\aid{*}} P_2
\mathrel{\aid{⋆}} f_1 \mathrel{\aid{⋆}} f_2$ is the amalgamated relation
set over the embeddings $f_1, f_2$ of the shared alphabet, and
$\otimes\aid{-group}\;n$ the $n$-fold product group):
\[
\begin{array}{l}
\aid{dpres} : (\Gamma \mathrel{\aidop{⋄}} \Delta \mathrel{\aidop{⋄}} \aid{CommRel})
   \ \aid{IsPresentationOf}\ (G_1 \times G_2) \\[3pt]
\aid{dpres} : (\Gamma \mathrel{\aidop{⋄}} \Delta \mathrel{\aidop{⋄}} \aid{ConjRelʷ}\;\mathit{conj})
   \ \aid{IsPresentationOf}\ (G_1 \rtimes G_2) \\[3pt]
\aid{presentation} : \forall n \to (\Gamma \mathrel{\aid{⊕\char94}} n)
   \ \aid{IsPresentationOf}\ (\otimes\aid{-group}\;n) \\[3pt]
\aid{dpres} : (P_1 \mathrel{\aid{*}} P_2 \mathrel{\aid{⋆}} f_1 \mathrel{\aid{⋆}} f_2)
   \ \aid{IsPresentationOf}\ (G_1 \ast_{G_0} G_2)
\end{array}
\]
where the semantic groups on the right are built by the
\texttt{ForStdlib} constructions below (for the amalgamated product, the
premises are two \aid{PackedCosetTable}s over a common base presentation,
bundled as the record \aid{AmalDataNF}, plus presentations of the three
groups; UNF additionally requires exactness of the base and coset sections.
The amalgamating homomorphisms $\varphi, \psi$ are then
derived from the syntactic embeddings through the presentation
isomorphisms).  A further module proves the recipe for arbitrary group
\emph{extensions}: given a short exact sequence
$1 \to N \to G \to Q \to 1$ realizing conjugation and relator-correction
data, the union of $N$'s relations, word-valued conjugation rules, and
twisted lifts of $Q$'s relators presents $G$
(cf.\ Proposition~2.55 of~\cite{holt2005handbook}); semidirect products are
the special case with trivial corrections.  The formalization assumes
bijective factor normal forms over propositional equality and
$\aid{inv-nf}\,(\aid{nf}\,\varepsilon) \equiv \varepsilon$ for the quotient.
These construction theorems
keep the catalogue of \S\ref{sec:examples} short: the Pauli
presentation, for instance, is assembled from cyclic factors, one
binary product, and one $n$-fold product, and uses no coset table
of its own.

\subsection{The Semantic Side: Contributions to the Standard Library}\label{sec:forstdlib}

The right-hand sides of the presentation theorems require the constructed
groups as standard-library-style bundles.  We provide them
in a \texttt{ForStdlib} hierarchy whose module paths mirror the standard
library's.
The module \path{Algebra.Construct.SemiDirectProduct} builds $N \rtimes H$
for raw monoids through groups, driven by a bundled \aid{Action} record
whose six fields (the action, its congruence, and its preservation of
units and products on both sides) are the laws needed to make the
twisted multiplication associative and unital.
\path{Algebra.Construct.Amalgamation} builds the amalgamated free product
$M_1 \ast_{M_0} M_2$ of monoids, as a setoid rather than a set:
elements are words over $|M_1| \uplus |M_2|$ and equality is the least
congruence multiplying out same-factor letters, deleting units, and gluing
$\varphi\,c$ to $\psi\,c$ for $c \in M_0$; when $\varphi, \psi$ are
homomorphisms this is the pushout in the category of monoids.
\path{Algebra.Construct.Extension}
and \path{.SplitExtension} define short exact sequences of groups, without
and with a homomorphic section, semidirect products being the canonical
split model.  \path{Algebra.Morphism.Consequences} proves that a monoid
homomorphism between groups is a group homomorphism (previously derivable
only via a deprecated stdlib module), and
\path{Algebra.Morphism.Structures} adds \aid{IsMonoidEpimorphism} (stdlib
has monomorphisms and isomorphisms, but no epimorphisms).  Finally,
\path{Data.Fin.Permutation.Properties} bundles stdlib's
\aid{Permutation′}~$n$ with composition, identity, and \aid{flip} as the
\aid{Group} used by the tight semantics of \S\ref{sec:permutations}.

\subsection{Proof Engineering: Solvers and Evaluation}\label{sec:solvers}

Three small tools (\texttt{Presentation.Tactic}, re-exported by
\texttt{Presentation.Properties}) are used throughout the equational
proofs.
\aid{by-assoc} proves $w \approx v$ whenever the two words have the same
letter sequence: it normalizes both sides to lists and compares them
with \aid{refl}, so all re-bracketing and unit shuffling in a
calculation is one appeal to the solver; the pattern-guided variant
\aid{by-passoc} re-brackets toward a shape specified by a pattern word
with holes, useful when the next rule application needs a
particular association.  \aid{by-equal-nf} (\S\ref{sec:nfproperty})
closes any goal whose two sides have the same computed normal form; since
coset tables and normal-form maps are structurally recursive programs,
this turns the well-definedness half of table verification into
evaluation.  Finally, a shared \texttt{Notations}
module of numeral patterns
(\aid{₀} … \aid{₁₅}, successor patterns \aid{₁₊} … \aid{₄₊} overloaded
over $\mathbb{N}$ and $\mathsf{Fin}$, and \aid{auto} for \aid{refl})
keeps indices readable, and equational proofs use the standard
library's setoid-reasoning combinators, so that a proof such as
\aid{lemma-XS} in the Clifford+$T$ study follows the pen-and-paper
derivation line by line.
\begin{code}[hide]%
\>[0]\AgdaKeyword{module}\AgdaSpace{}%
\AgdaModule{paper.sections.examples}\AgdaSpace{}%
\AgdaKeyword{where}\<%
\\
\\[\AgdaEmptyExtraSkip]%
\>[0]\AgdaKeyword{open}\AgdaSpace{}%
\AgdaKeyword{import}\AgdaSpace{}%
\AgdaModule{Algebra.Bundles}\AgdaSpace{}%
\AgdaKeyword{using}\AgdaSpace{}%
\AgdaSymbol{(}\AgdaRecord{Group}\AgdaSpace{}%
\AgdaSymbol{;}\AgdaSpace{}%
\AgdaRecord{Monoid}\AgdaSymbol{)}\<%
\\
\>[0]\AgdaKeyword{open}\AgdaSpace{}%
\AgdaKeyword{import}\AgdaSpace{}%
\AgdaModule{Algebra.Morphism.Structures}\AgdaSpace{}%
\AgdaKeyword{using}\AgdaSpace{}%
\AgdaSymbol{(}\AgdaKeyword{module}\AgdaSpace{}%
\AgdaModule{MonoidMorphisms}\AgdaSymbol{)}\<%
\\
\>[0]\AgdaKeyword{open}\AgdaSpace{}%
\AgdaKeyword{import}\AgdaSpace{}%
\AgdaModule{Data.Nat}\AgdaSpace{}%
\AgdaKeyword{using}\AgdaSpace{}%
\AgdaSymbol{(}\AgdaDatatype{ℕ}\AgdaSpace{}%
\AgdaSymbol{;}\AgdaSpace{}%
\AgdaInductiveConstructor{suc}\AgdaSpace{}%
\AgdaSymbol{;}\AgdaSpace{}%
\AgdaInductiveConstructor{2+}\AgdaSymbol{)}\<%
\\
\>[0]\AgdaKeyword{open}\AgdaSpace{}%
\AgdaKeyword{import}\AgdaSpace{}%
\AgdaModule{Data.Nat.Primality}\AgdaSpace{}%
\AgdaKeyword{using}\AgdaSpace{}%
\AgdaSymbol{(}\AgdaRecord{Prime}\AgdaSymbol{)}\<%
\\
\>[0]\AgdaKeyword{open}\AgdaSpace{}%
\AgdaKeyword{import}\AgdaSpace{}%
\AgdaModule{Level}\AgdaSpace{}%
\AgdaKeyword{using}\AgdaSpace{}%
\AgdaSymbol{(}\AgdaFunction{0ℓ}\AgdaSymbol{)}\<%
\\
\\[\AgdaEmptyExtraSkip]%
\>[0]\AgdaKeyword{open}\AgdaSpace{}%
\AgdaKeyword{import}\AgdaSpace{}%
\AgdaModule{Notations}\AgdaSpace{}%
\AgdaKeyword{using}\AgdaSpace{}%
\AgdaSymbol{(}\AgdaInductiveConstructor{₁₊}\AgdaSymbol{)}\<%
\\
\>[0]\AgdaKeyword{open}\AgdaSpace{}%
\AgdaKeyword{import}\AgdaSpace{}%
\AgdaModule{Word.Base}\AgdaSpace{}%
\AgdaKeyword{using}\AgdaSpace{}%
\AgdaSymbol{(}\AgdaOperator{\AgdaFunction{\AgdaUnderscore{}ʷ}}\AgdaSymbol{)}\<%
\\
\>[0]\AgdaKeyword{open}\AgdaSpace{}%
\AgdaKeyword{import}\AgdaSpace{}%
\AgdaModule{Presentation.Definitions}\AgdaSpace{}%
\AgdaKeyword{using}\AgdaSpace{}%
\AgdaSymbol{(}\AgdaOperator{\AgdaRecord{\AgdaUnderscore{}IsPresentationOf\AgdaUnderscore{}}}\AgdaSymbol{)}\<%
\\
\>[0]\AgdaKeyword{open}\AgdaSpace{}%
\AgdaKeyword{import}\AgdaSpace{}%
\AgdaModule{Presentation.Construct.Base}\AgdaSpace{}%
\AgdaKeyword{using}\AgdaSpace{}%
\AgdaSymbol{(}\AgdaOperator{\AgdaDatatype{\AgdaUnderscore{}⋄\AgdaUnderscore{}⋄\AgdaUnderscore{}}}\AgdaSpace{}%
\AgdaSymbol{;}\AgdaSpace{}%
\AgdaDatatype{CommRel}\AgdaSpace{}%
\AgdaSymbol{;}\AgdaSpace{}%
\AgdaDatatype{ConjRelʷ}\AgdaSpace{}%
\AgdaSymbol{;}\AgdaSpace{}%
\AgdaOperator{\AgdaFunction{\AgdaUnderscore{}⊕\textasciicircum{}\AgdaUnderscore{}}}\AgdaSymbol{)}\<%
\\
\>[0]\AgdaKeyword{import}\AgdaSpace{}%
\AgdaModule{Presentation.Properties}\AgdaSpace{}%
\AgdaSymbol{as}\AgdaSpace{}%
\AgdaModule{PP}\<%
\\
\>[0]\AgdaKeyword{import}\AgdaSpace{}%
\AgdaModule{Presentation.Construct.Properties.DirectProduct}\AgdaSpace{}%
\AgdaSymbol{as}\AgdaSpace{}%
\AgdaModule{DP}\<%
\\
\>[0]\AgdaKeyword{import}\AgdaSpace{}%
\AgdaModule{Presentation.Construct.Properties.NDirectProduct}\AgdaSpace{}%
\AgdaSymbol{as}\AgdaSpace{}%
\AgdaModule{NDP}\<%
\\
\>[0]\AgdaKeyword{open}\AgdaSpace{}%
\AgdaKeyword{import}\AgdaSpace{}%
\AgdaModule{Examples.Groups.Cyclic.Normalization}\AgdaSpace{}%
\AgdaKeyword{using}\AgdaSpace{}%
\AgdaSymbol{(}\AgdaOperator{\AgdaDatatype{\AgdaUnderscore{}Cn,\AgdaUnderscore{}===\AgdaUnderscore{}}}\AgdaSymbol{)}\<%
\\
\>[0]\AgdaKeyword{open}\AgdaSpace{}%
\AgdaKeyword{import}\AgdaSpace{}%
\AgdaModule{Examples.Groups.Cyclic.Semantics}\AgdaSpace{}%
\AgdaKeyword{using}\AgdaSpace{}%
\AgdaSymbol{(}\AgdaFunction{Cn-group}\AgdaSymbol{)}\<%
\\
\>[0]\AgdaKeyword{import}\AgdaSpace{}%
\AgdaModule{Examples.Groups.Cyclic.Presentation}\AgdaSpace{}%
\AgdaSymbol{as}\AgdaSpace{}%
\AgdaModule{CyP}\<%
\\
\>[0]\AgdaKeyword{open}\AgdaSpace{}%
\AgdaKeyword{import}\AgdaSpace{}%
\AgdaModule{Examples.Groups.Symmetric.Syntactics}\AgdaSpace{}%
\AgdaKeyword{using}\AgdaSpace{}%
\AgdaSymbol{(}\AgdaOperator{\AgdaDatatype{\AgdaUnderscore{}VRel,\AgdaUnderscore{}===\AgdaUnderscore{}}}\AgdaSymbol{)}\<%
\\
\>[0]\AgdaKeyword{import}\AgdaSpace{}%
\AgdaModule{Examples.Groups.Pauli.Presentation}\AgdaSpace{}%
\AgdaSymbol{as}\AgdaSpace{}%
\AgdaModule{Pauli}\<%
\\
\>[0]\AgdaKeyword{import}\AgdaSpace{}%
\AgdaModule{Examples.Construct.SemiDirectProduct.SnD}\AgdaSpace{}%
\AgdaSymbol{as}\AgdaSpace{}%
\AgdaModule{SnD}\<%
\\
\>[0]\AgdaKeyword{import}\AgdaSpace{}%
\AgdaModule{Examples.Amalgamations.CliffordT1}\AgdaSpace{}%
\AgdaSymbol{as}\AgdaSpace{}%
\AgdaModule{CliffordT1}\<%
\\
\\[\AgdaEmptyExtraSkip]%
\>[0]\AgdaKeyword{open}\AgdaSpace{}%
\AgdaModule{SnD}\AgdaSpace{}%
\AgdaKeyword{using}\AgdaSpace{}%
\AgdaSymbol{(}\AgdaFunction{conj}\AgdaSpace{}%
\AgdaSymbol{;}\AgdaSpace{}%
\AgdaKeyword{module}\AgdaSpace{}%
\AgdaModule{Wreath}\AgdaSymbol{)}\<%
\\
\>[0]\AgdaKeyword{open}\AgdaSpace{}%
\AgdaModule{Pauli}\AgdaSpace{}%
\AgdaKeyword{using}\AgdaSpace{}%
\AgdaSymbol{(}\AgdaFunction{Pauli-group}\AgdaSymbol{)}\AgdaSpace{}%
\AgdaKeyword{renaming}\AgdaSpace{}%
\AgdaSymbol{(}\AgdaFunction{Γ-H}\AgdaSpace{}%
\AgdaSymbol{to}\AgdaSpace{}%
\AgdaFunction{Γ-Pauli₁}\AgdaSymbol{)}\<%
\\
\>[0]\AgdaKeyword{open}\AgdaSpace{}%
\AgdaModule{Monoid}\AgdaSpace{}%
\AgdaKeyword{using}\AgdaSpace{}%
\AgdaSymbol{(}\AgdaFunction{rawMonoid}\AgdaSymbol{)}\<%
\\
\>[0]\AgdaKeyword{open}\AgdaSpace{}%
\AgdaModule{MonoidMorphisms}\AgdaSpace{}%
\AgdaKeyword{using}\AgdaSpace{}%
\AgdaSymbol{(}\AgdaRecord{IsMonoidIsomorphism}\AgdaSymbol{)}\<%
\\
\>[0]\AgdaKeyword{open}\AgdaSpace{}%
\AgdaModule{PP}\AgdaSpace{}%
\AgdaKeyword{using}\AgdaSpace{}%
\AgdaSymbol{(}\AgdaFunction{•-ε-monoid}\AgdaSymbol{)}\<%
\\
\>[0]\AgdaKeyword{open}\AgdaSpace{}%
\AgdaModule{CliffordT1.CliffordT1}\AgdaSpace{}%
\AgdaKeyword{using}\AgdaSpace{}%
\AgdaSymbol{(}\AgdaOperator{\AgdaDatatype{\AgdaUnderscore{}===\AgdaUnderscore{}}}\AgdaSpace{}%
\AgdaSymbol{;}\AgdaSpace{}%
\AgdaFunction{amalPres}\AgdaSpace{}%
\AgdaSymbol{;}\AgdaSpace{}%
\AgdaFunction{f}\AgdaSymbol{)}\<%
\end{code}
\section{Examples: A Catalogue of Verified Presentations}\label{sec:examples}

Table~\ref{tab:presentations} lists
every concrete \aid{{\_}IsPresentationOf{\_}} theorem in the catalogue,
and its monoid version for the order-$0$ cyclic case;
Table~\ref{tab:nfs} lists the normal-form and uniqueness witnesses behind
them.  The headline results are restated with full types in the index module
\texttt{MainTheorems.agda}, which also re-exports the conditional
construction theorems of \S\ref{sec:constructions} with their premise
telescopes.  We then discuss each family, ending with the three
amalgamation case studies, whose final theorems are monoid
isomorphisms between gate-set presentations and amalgamated
products of presentations.

\begin{table}
\caption{Concrete presentation theorems (all verified; names as in
\texttt{MainTheorems.agda}, except where noted).  $p$ ranges over primes,
$n, m$ over naturals, and $N$ over positive naturals.}
\label{tab:presentations}
\centering
\small
\begin{tabular}{@{}lll@{}}
\toprule
Relation set & Presents & Index name \\
\midrule
\aid{EmptyRel} over $\bot$ & trivial group & \aid{trivial-presentation} \\
\aid{TrivialRel} over any $A$ & trivial group & \aid{trivial-presentation′} \\
$T^{\,n+1} = \varepsilon$ & $\mathbb{Z}/(n{+}1)\mathbb{Z}$ & \aid{cyclic-presentation} \\
$T^{\,0} = \varepsilon$ & the \emph{monoid} $(\mathbb{N}, +, 0)$ & \aid{cyclic-monoid-presentation} \\
\aid{order}, \aid{yang-baxter} + structural & permutations of $\mathsf{Fin}\;n$ & \aid{symmetric-presentation} \\
$C_p \aid{⋄} C_p \aid{⋄} \aid{CommRel}$ & $\mathbb{Z}_p \times \mathbb{Z}_p$ & (\aid{H-pres}, Pauli development) \\
$(C_p \aid{⋄} C_p \aid{⋄} \aid{CommRel})\,\aid{⊕\char94}\,n$ & Pauli quotient $(\mathbb{Z}_p{\times}\mathbb{Z}_p)^n$ & \aid{pauli-presentation} \\
$(C_N\,\aid{⊕\char94}\,n) \aid{⋄} S_n\text{-circuits} \aid{⋄} \aid{ConjRelʷ}$ & $\mathbb{Z}_N \wr S_n$ (generalized symmetric groups) & \aid{wreath-presentation} \\
\bottomrule
\end{tabular}
\end{table}

\begin{table}
\caption{Normal-form and uniqueness witnesses (selection).  NFI =
\aid{NormalFormInjective}, NF = \aid{NormalForm} (with section), UNF =
\aid{UniqueNormalForm}.  For products, UNF is lifted from the factor
witnesses; amalgamation also requires exact base and coset sections
(\S\ref{sec:constructions}).}
\label{tab:nfs}
\centering
\footnotesize
\setlength{\tabcolsep}{4pt}
\begin{tabular}{@{}llll@{}}
\toprule
Presentation & Carrier & Witnesses & Note \\
\midrule
cyclic, order $n{+}1$ & $\mathsf{Fin}\,(n{+}1)$ & NFI, NF, UNF & presents $\mathbb{Z}/(n{+}1)\mathbb{Z}$ \\
cyclic, order $0$ & $\mathbb{N}$ & NFI, NF, UNF & presents the monoid $(\mathbb{N}, +, 0)$ (\S\ref{sec:cyclic}) \\
$S_n$ circuits & $\top \times C\,1 \times \dots \times C\,(n{-}1)$ & NFI, NF, UNF$\times 2$ & factorial system, $|{\cdot}| = n!$ \\
$\Gamma \aid{⋄} \Delta \aid{⋄} \aid{CommRel}$ & $\mathit{NF}_1 \times \mathit{NF}_2$ & NFI, NF, UNF & lifted from factors \\
$\Gamma \aid{⋄} \Delta \aid{⋄} \aid{ConjRelʷ}\,\mathit{conj}$ & $\mathit{NF}_1 \times \mathit{NF}_2$ & NFI, NF, UNF & twist on the left factor \\
$P_1 \mathrel{\aid{*}} P_2 \mathrel{\aid{⋆}} f_1 \mathrel{\aid{⋆}} f_2$ & $\mathit{NF}_0 \times \aid{amalNFC}\;C\;D$ & NFI, NF, UNF & base NF, then alternating cosets (\S\ref{sec:constructions}) \\
Clifford+$T$ base $\langle \omega \rangle {\times} \langle S \rangle$ & $\mathsf{Fin}\,8 \times \mathsf{Fin}\,4$ & NFI, NF, UNF & radices $8, 4$; unique for $\mathbb{Z}_8 {\times} \mathbb{Z}_4$ \\
\bottomrule
\end{tabular}
\end{table}

\subsection{Cyclic Groups, and the Trivial Group Twice}\label{sec:cyclic}

The cyclic development presents $\langle\, T \mid T^{\,n+1} =
\varepsilon\,\rangle$ with normal-form carrier $\mathsf{Fin}\,(n{+}1)$,
for every positive order at once (\aid{cyclic-presentation}): the normal
form of a word is its letter count modulo $n{+}1$, the section sends the
residue $k$ to $T^{k}$, and uniqueness for
$\mathbb{Z}/(n{+}1)\mathbb{Z}$ is immediate.  Order $0$ gets a theorem
of its own, \aid{cyclic-monoid-presentation}: with carrier $\mathbb{N}$
(a word count with no wraparound), $T^{\,0} = \varepsilon$
\aid{IsMonoidPresentationOf} $(\mathbb{N}, +, 0)$, the free monoid
rather than the group $\mathbb{Z}$, since, as \S\ref{sec:wordproblem}
explained, the generator has no inverse there, so the \aid{Grouplike}
condition fails.  The
trivial group is presented twice (empty alphabet with no relations, and
arbitrary alphabet with the everything-is-$\varepsilon$ relation), and
the two presentations are proved isomorphic.

\subsection{Wreath Products: Signed Permutations and Beyond}\label{sec:wreath}

The wreath-product presentation uses most of the library's
constructions.  Fix a base order $N = m{+}1$ and $n$ wires.  The $n$-fold
direct power $(C_N)^{\oplus n}$ presents $(\mathbb{Z}_N)^n$; the circuit
presentation of $S_n$ acts on it by permuting coordinates, an action
specified syntactically as a word-valued conjugation
$\mathit{conj} : \aid{Gen}\;n \to (\top \uplus^n) \to
\aid{Word}\,(\top \uplus^n)$ sending a generator and a coordinate to the
permuted coordinate.  Two finite compatibility lemmas (the action
respects the Coxeter relations pointwise, and respects the product
relations) feed the semidirect-product presentation theorem, giving
\begin{code}%
\>[0]\AgdaFunction{wreath-presentation}\AgdaSpace{}%
\AgdaSymbol{:}\AgdaSpace{}%
\AgdaSymbol{∀}\AgdaSpace{}%
\AgdaBound{n}\AgdaSpace{}%
\AgdaBound{m}\AgdaSpace{}%
\AgdaSymbol{→}%
\>[118I]\AgdaSymbol{(((}\AgdaInductiveConstructor{suc}\AgdaSpace{}%
\AgdaBound{m}\AgdaSpace{}%
\AgdaOperator{\AgdaDatatype{Cn,\AgdaUnderscore{}===\AgdaUnderscore{}}}\AgdaSymbol{)}\AgdaSpace{}%
\AgdaOperator{\AgdaFunction{⊕\textasciicircum{}}}\AgdaSpace{}%
\AgdaBound{n}\AgdaSymbol{)}\AgdaSpace{}%
\AgdaOperator{\AgdaDatatype{⋄}}\AgdaSpace{}%
\AgdaSymbol{(}\AgdaBound{n}\AgdaSpace{}%
\AgdaOperator{\AgdaDatatype{VRel,\AgdaUnderscore{}===\AgdaUnderscore{}}}\AgdaSymbol{)}\AgdaSpace{}%
\AgdaOperator{\AgdaDatatype{⋄}}\AgdaSpace{}%
\AgdaDatatype{ConjRelʷ}\AgdaSpace{}%
\AgdaFunction{conj}\AgdaSymbol{)}\<%
\\
\>[118I][@{}l@{\AgdaIndent{0}}]%
\>[32]\AgdaOperator{\AgdaRecord{IsPresentationOf}}\AgdaSpace{}%
\AgdaSymbol{(}\AgdaFunction{Wreath.wreath-group}\AgdaSpace{}%
\AgdaBound{n}\AgdaSpace{}%
\AgdaBound{m}\AgdaSymbol{)}\<%
\end{code}
\begin{code}[hide]%
\>[0]\AgdaFunction{wreath-presentation}\AgdaSpace{}%
\AgdaBound{n}\AgdaSpace{}%
\AgdaBound{m}\AgdaSpace{}%
\AgdaSymbol{=}\AgdaSpace{}%
\AgdaFunction{Wreath.presentation}\AgdaSpace{}%
\AgdaBound{n}\AgdaSpace{}%
\AgdaBound{m}\<%
\end{code}
where $\aid{wreath-group}\;n\;m$ is the semantic
$(\mathbb{Z}_N)^n \rtimes S_n = \mathbb{Z}_N \wr S_n$ built from the
\texttt{ForStdlib} semidirect product and the bundled permutation group.
The lifted normal form is a pair of $n$ residues modulo $N$ and the
factorial tuple of a permutation, so its section is a word of $T$-powers,
one per coordinate, followed by the staircases of
\S\ref{sec:permutations}.  At $N = 2$ this is the group of
\emph{signed permutations} (the hyperoctahedral group $B_n$, the
symmetry group of the hypercube); at
general $N$ it is the generalized symmetric group $G(N,1,n)$ of monomial
matrices with $N$-th-root-of-unity entries.  These groups occur throughout
the qudit setting (Pauli-frame changes, permutation-and-phase
subgroups of Clifford groups).  The theorem is uniform in $N$ and $n$:
one proof gives a completeness result for every pair.

\subsection{Pauli Groups Modulo Phases in All Prime Dimensions}\label{sec:pauli}

The Pauli development is entirely compositional and uses no coset table:

\begin{code}%
\>[0]\AgdaFunction{pauli-presentation}\AgdaSpace{}%
\AgdaSymbol{:}\AgdaSpace{}%
\AgdaSymbol{∀}\AgdaSpace{}%
\AgdaBound{m}\AgdaSpace{}%
\AgdaSymbol{(}\AgdaBound{prime}\AgdaSpace{}%
\AgdaSymbol{:}\AgdaSpace{}%
\AgdaRecord{Prime}\AgdaSpace{}%
\AgdaSymbol{(}\AgdaInductiveConstructor{2+}\AgdaSpace{}%
\AgdaBound{m}\AgdaSymbol{))}\AgdaSpace{}%
\AgdaBound{n}\AgdaSpace{}%
\AgdaSymbol{→}\<%
\\
\>[0][@{}l@{\AgdaIndent{0}}]%
\>[2]\AgdaSymbol{(}\AgdaFunction{Γ-Pauli₁}\AgdaSpace{}%
\AgdaBound{m}\AgdaSpace{}%
\AgdaBound{prime}\AgdaSpace{}%
\AgdaOperator{\AgdaFunction{⊕\textasciicircum{}}}\AgdaSpace{}%
\AgdaBound{n}\AgdaSymbol{)}\AgdaSpace{}%
\AgdaOperator{\AgdaRecord{IsPresentationOf}}\AgdaSpace{}%
\AgdaSymbol{(}\AgdaFunction{Pauli-group}\AgdaSpace{}%
\AgdaBound{m}\AgdaSpace{}%
\AgdaBound{prime}\AgdaSpace{}%
\AgdaBound{n}\AgdaSymbol{)}\<%
\end{code}
\begin{code}[hide]%
\>[0]\AgdaFunction{pauli-presentation}\AgdaSpace{}%
\AgdaBound{m}\AgdaSpace{}%
\AgdaBound{prime}\AgdaSpace{}%
\AgdaSymbol{=}\AgdaSpace{}%
\AgdaFunction{Pauli.Pauli-presentation}\AgdaSpace{}%
\AgdaBound{m}\AgdaSpace{}%
\AgdaBound{prime}\<%
\end{code}

For a prime $p = 2 + m$, with \aid{prime} its primality proof (a
\emph{qupit} is a qudit whose dimension is a prime), the single-qupit
relation set $\aid{Γ-Pauli₁}$ (\aid{Γ-H} in the index module) is the
direct-product
join of two order-$p$ cyclic presentations ($X$- and $Z$-type generators,
which commute in the quotient of the Pauli group by its phase subgroup),
and the $n$-qupit relation set is its $n$-fold power.  The semantic group
$(\mathbb{Z}_p \times \mathbb{Z}_p)^n$ is assembled by the standard
library's binary direct product and our $n$-fold iteration, and the
presentation is the composition of three construction theorems (the
cyclic presentation, the binary product, and the $n$-fold product).  The
normal form is a tuple of $n$ pairs of exponents in
$\mathsf{Fin}\,p \times \mathsf{Fin}\,p$, with section
$X^{a_1} Z^{b_1} \cdots X^{a_n} Z^{b_n}$.  The
full Pauli group with phases, and its role beneath the Clifford
hierarchy, is left to future work (\S\ref{sec:conclusion}).

\subsection{Single-Qubit Clifford+\texorpdfstring{$T$}{T}, as an Amalgamated Free Product}\label{sec:cliffordt1}

The first amalgamation case study presents the single-qubit Clifford+$T$
\emph{monoid}, a standard example in the completeness literature:
Clifford+$T$ is a widely used universal gate set, its single-qubit
fragment has the Matsumoto--Amano normal
form~\cite{matsumoto2008representation,giles2013remarks}, which underlies
exact synthesis and $T$-count optimality, and the later normal forms
for single-qutrit and single-qudit
Clifford+$T$~\cite{glaudell2019qutrit,prakash2018qutrit,jain2020qudit}
follow the same syllable pattern.  The relation set (over generators
$T, H, S, \omega$, with $X := H S S H$ as an abbreviation) is
\[
  \omega^8 = \varepsilon,\quad
  T^2 = S,\quad
  S^4 = \varepsilon,\quad
  H^2 = \varepsilon,\quad
  (SH)^3 = \omega,\quad
  (TX)^2 = \omega,\quad
  \omega\ \text{central},
\]
and the verified structure exhibits the presented monoid, written
$\mathrm{Clifford}{+}T$ below, as an
amalgamated free product over the subgroup $M$ generated by $X$, $S$, and
$\omega$:
\[
  \mathrm{Clifford}{+}T \;\cong\;
  \langle T, M \rangle \,\ast_{M}\, \langle H, M \rangle,
  \qquad M = \langle X, S, \omega \rangle .
\]
The development builds a tower whose base is the direct product
$\langle \omega \rangle \times \langle S \rangle$ of cyclic groups of
orders $8$ and $4$ (normal-form radices $8 \cdot 4$), extends it by one
verified coset table to the subgroup $M$ (two cosets:
$\varepsilon$ and $X$), and then extends twice from $M$:
by a two-coset table (cosets $\varepsilon, T$) to
$\mathit{TXS\omega} = \langle T, X, S, \omega \rangle$, and by a
three-coset table (cosets $\varepsilon, H, HS$) to the Clifford group
$\langle H, X, S, \omega \rangle$.  The two \aid{PackedCosetTable}s over
the shared $M$ form an \aid{AmalDataNF}, and the
amalgamation module produces the normal form for the amalgamated
product.  Concretely, with a one-element $C$ ($T$-coset) and two-element $D$
($H, HS$), a normal form is an element of $\mathit{NF}_0 \times
\aid{amalNFC}\;C\;D$, and its section spells it out, in the library's
left-to-right word order, as the regular expression
\[
  \mathcal{M}\;(\varepsilon \mid H \mid HS)\;(TH \mid THS)^{*}\;(\varepsilon \mid T),
\]
where $\mathcal{M}$ stands for the base normal form of the $M$-part; an
element of $\aid{amalNFC}\;C\;D$ is an optional leading
$H$-syllable, a list of alternations, and an optional trailing $T$,
which, read right to left, is the Matsumoto--Amano normal
form~\cite{matsumoto2008representation,giles2013remarks},
with its single Clifford factor $\mathcal{C}$ split into a coset
representative and an element of $M$:
the alternating normal form of an amalgamated free product, as observed
for this group by Bian and Selinger~\cite{bian2023cliffordcs3}.  The
final theorem is the
syntactic isomorphism assembled by \aid{StarIsomorphism} from
generator-level translations both ways (with \aid{{\_}==={\_}} the
gate-set relations, \aid{amalPres} the amalgamated presentation, and
\aid{f} the translation):

\begin{code}%
\>[0]\AgdaFunction{clifford+T-qubit-isomorphism}\AgdaSpace{}%
\AgdaSymbol{:}\AgdaSpace{}%
\AgdaRecord{IsMonoidIsomorphism}\<%
\\
\>[0][@{}l@{\AgdaIndent{0}}]%
\>[2]\AgdaSymbol{(}\AgdaFunction{rawMonoid}\AgdaSpace{}%
\AgdaSymbol{(}\AgdaFunction{•-ε-monoid}\AgdaSpace{}%
\AgdaOperator{\AgdaDatatype{\AgdaUnderscore{}===\AgdaUnderscore{}}}\AgdaSymbol{))}\AgdaSpace{}%
\AgdaSymbol{(}\AgdaFunction{rawMonoid}\AgdaSpace{}%
\AgdaSymbol{(}\AgdaFunction{•-ε-monoid}\AgdaSpace{}%
\AgdaFunction{amalPres}\AgdaSymbol{))}\AgdaSpace{}%
\AgdaSymbol{(}\AgdaFunction{f}\AgdaSpace{}%
\AgdaOperator{\AgdaFunction{ʷ}}\AgdaSymbol{)}\<%
\end{code}
\begin{code}[hide]%
\>[0]\AgdaFunction{clifford+T-qubit-isomorphism}\AgdaSpace{}%
\AgdaSymbol{=}\AgdaSpace{}%
\AgdaFunction{CliffordT1.CliffordT1.CliffordT1-isomorphism}\<%
\end{code}

Every well-definedness obligation in the tower (on all axioms at all
cosets, 59 cases) is discharged by
\aid{(by-equal-nf Eq.refl) , Eq.refl}: the typechecker runs the table
on both sides of the axiom from the given coset, the resulting cosets
coincide, and the residual words have the same normal form one level
down.

\subsection{Single-Qutrit Clifford+\texorpdfstring{$T$}{T}}\label{sec:qutrit}

The qutrit analogue replaces $\omega$ by a ninth root $\zeta$ and runs a
longer tower: from the cyclic base $\langle \zeta \rangle$,
$\zeta^9 = \varepsilon$, through
$\langle S, \zeta \rangle$ ($S^3 = \zeta^6$; three cosets), then a
nine-coset table to $\langle S, X, \zeta \rangle$
($X^3 = \varepsilon$, $(SX)^3 = \varepsilon$, and a twisted commutation
$(XS)(SX) = \zeta^6 (SX)(XS)$), then a two-coset table adjoining the
squared Hadamard $H\!H$, and finally an amalgamation of a $T$-side and an
$H$-side extension over the common subgroup.  The presented gate-set
relations (generators $T, H, S, \zeta$; abbreviations
$X := \zeta^3 (HSH)(HSSH)$ and $Z := \zeta^3 S^2 H^2 S H^2$):
\[
  \zeta^9 = \varepsilon,\;\;
  S^3 = \zeta^6,\;\;
  H^4 = \varepsilon,\;\;
  T^3 = Z,\;\;
  (SH)^3 = \varepsilon,\;\;
  (THH)^2 = \varepsilon,
\]
\[
  HHSHHS = SHHSHH,\;\;
  TS = ST,\;\;
  TX = \zeta^3 S^2 X\, T,\;\;
  \zeta\ \text{central},
\]
a presentation derived and verified in this development.  The verified
structure is again an amalgamated free product of the presented monoid,
\[
  \mathrm{Clifford}{+}T_3 \;\cong\;
  \langle T, M \rangle \,\ast_{M}\, \langle H, M \rangle,
  \qquad M = \langle S, X, \zeta, H^2 \rangle.
\]
The alternating normal form reads
\[
  \mathcal{M}\;(\varepsilon \mid H \mid HS \mid HS^2)\;
  \bigl((T \mid TH^2)(H \mid HS \mid HS^2)\bigr)^{*}\;
  (\varepsilon \mid T \mid TH^2),
\]
and right to left it is the $T$-optimal canonical form of Glaudell,
Ross, and Taylor~\cite{glaudell2019qutrit}
(Prakash, Jain, Kapur, and Seth independently found a similar
form~\cite{prakash2018qutrit}, extended to qudits
in~\cite{jain2020qudit}), with $\mathcal{C}$ split as in the qubit
case.  The final theorem, \aid{clifford+T-qutrit-isomorphism} in the
index, has the same shape as the qubit one.  The nine-coset level alone
has seventy-two axiom-preservation obligations (nine cosets against eight
axiom cases), each closed by \aid{refl} after evaluation.  We know of no prior
machine-checked normal form for a qutrit gate-set presentation.

\subsection{\texorpdfstring{$U_3(\mathbb{Z}[\tfrac12,i])$}{U3(Z[1/2,i])}}\label{sec:u3}

The third case study formalizes the case $n = 3$ of the presentation, due
to Bian and Selinger, of the groups
$U_n(\mathbb{Z}[\tfrac12,i])$~\cite{bian2021un}, the unitary matrices over
the ring $\mathbb{Z}[\tfrac12,i]$.
Generators are $i_0$ (a local phase), $K_{01}$ (a scaled $2{\times}2$
Hadamard-type block), and the two transpositions $X_{01}, X_{12}$;
derived generators $i_1, i_2, K_{02}, K_{12}$ are conjugates.  The eleven
relations (labelled $\mathrm{S1}$--$\mathrm{S14}$ in the development, a
numbering with gaps) include the orders $i_0^4 = X_{01}^2 = X_{12}^2 =
\varepsilon$, braid and commutation laws for the transpositions and
phases, and the interactions of $K_{01}$ with the
phases, e.g.
$K_{01} i_1^2 = X_{01} K_{01}$ and
$K_{01}^2\, i_0\, i_1 = \varepsilon$.  The verified structure, with
$U_3(\mathbb{Z}[\tfrac12,i])$ standing for the presented group
(\S\ref{sec:threats}), is
\[
  U_3(\mathbb{Z}[\tfrac12,i]) \;\cong\;
  (\mathbb{Z}_4 \wr S_3) \,\ast_{M}\, \langle M, K_{01} \rangle,
  \qquad M = (\mathbb{Z}_4 \wr S_2) \times \mathbb{Z}_4 .
\]
The right factor adjoins
$K_{01}$ to $M$: a six-coset Reidemeister--Schreier level presents
the $\langle K_{01}, i_0, i_1 \rangle$ subgroup over the product
of two order-$4$ cyclic groups, a direct product then adjoins
$\langle i_2 \rangle$, and $X_{01}$ enters as a word in $K_{01}$, $i_0$,
$i_1$.  The alternating normal form reads
\[
  \mathcal{M}\;(\varepsilon \mid K_{01} \mid K_{01} i_0)\;
  \bigl((X_{12} \mid X_{12}X_{01})(K_{01} \mid K_{01} i_0)\bigr)^{*}\;
  (\varepsilon \mid X_{12} \mid X_{12}X_{01}),
\]
with $\mathcal{M}$ the normal form of $M$.
The final theorem,
\aid{U₃-presentation-isomorphism}, is again a monoid isomorphism between
the flat eleven-relation presentation and the amalgamated-product
presentation, with the alternating normal form on the amalgam side.  At
$2\,868$ and $1\,394$ source lines, the qutrit and $U_3$ studies are the
largest developments in the library; both consist mostly of
first-order tables and obligations discharged by \aid{refl}.
\section{Related Work and Comparison}\label{sec:related}

We compare along the three axes the library combines, namely (i) presentations
and normal forms of groups and monoids, (ii) completeness of equational
theories, and (iii) circuits, across Agda, Lean, Rocq, and Isabelle.
Table~\ref{tab:compare} summarizes; the text gives the details and the
precise differences.  Claims of absence are stated to the best of our
knowledge.

\begin{table}
\caption{Mechanized work adjacent to this library (citations in the
text).  ``Pres.''\ = concrete presentations by generators and relations;
``NF'' = verified normal forms for presented structures; ``Compl.''\ =
completeness of a relation set for an independent semantics.}
\label{tab:compare}
\centering
\footnotesize
\setlength{\tabcolsep}{4pt}
\begin{tabular}{@{}llcccc@{}}
\toprule
Development & Prover & Circuits & Pres. & NF & Compl. \\
\midrule
This library & Agda & \ding{51} & \ding{51} & \ding{51} & \ding{51} \\
Bian--Selinger developments & Agda & \ding{51} & \ding{51} & \ding{51} & syntactic \\
\textsc{sqir}/\textsc{voqc} & Rocq & \ding{51} & -- & -- & sound only \\
\textsc{Qwire} & Rocq & \ding{51} & -- & -- & -- \\
VyZX & Rocq & \ding{51} & -- & -- & sound only \\
CoqQ & Rocq & \ding{51} & -- & -- & -- \\
Qbricks & Why3 & \ding{51} & -- & -- & -- \\
QHLProver, Isabelle-Marries-Dirac & Isabelle & \ding{51} & -- & -- & -- \\
$\Pi$ as free rig groupoid & Agda (HoTT) & \ding{51} & categorical & \ding{51} & classical \\
Quantum $\Pi$ with square roots & Agda (partial) & \ding{51} & categorical & -- & 2q Clifford \\
mathlib group theory & Lean & -- & \ding{51} & \ding{51} & -- \\
MathComp / Odd Order & Rocq & -- & partial & -- & -- \\
Free groups (AFP, CICM~2023) & Isabelle & -- & -- & \ding{51} & -- \\
IsaFoR completion & Isabelle & -- & -- & \ding{51} & -- \\
cubical / agda-unimath & Agda & -- & partial & \ding{51} & -- \\
\bottomrule
\end{tabular}
\end{table}

\subsection{The Bian--Selinger Verifications: Syntax to Syntax}\label{sec:rel-bs}

The direct ancestors of this library are the Agda developments that Bian
and Selinger published alongside their presentations of the $2$-qubit
Clifford+$T$ group~\cite{bian2023cliffordt2} and the $3$-qubit
Clifford+CS group~\cite{bian2023cliffordcs3}, systematized in Bian's
thesis~\cite{bian2023thesis}.  These developments introduced
the methods we build on: monoid-level presentations (no formal inverses),
the Reidemeister--Schreier theorem applied (to Greylyn's presentation
of $U_4(\mathbb{Z}[1/\sqrt2,i])$~\cite{greylyn2014} in the $2$-qubit
case) to derive subgroup presentations,
amalgamated products of monoids, and machine-checked simplification of
``thousands of relations'' down to a small set (seventeen, for
Clifford+CS).  However, they
work entirely on the syntactic side: each theorem is an isomorphism
between two finitely
presented monoids, and no independent semantic object (a matrix group,
a permutation group, a product of groups) occurs in any statement.
Soundness and completeness in the model-theoretic sense are therefore not
expressible there, only relative completeness of one relation set against
another.  The present library keeps the syntactic technology (our
amalgamation case studies end in such isomorphisms) and adds the
semantic half: the \aid{{\_}IsPresentationOf{\_}} records of
Table~\ref{tab:presentations} target groups built independently of the
syntax (stdlib permutations, \texttt{ForStdlib} products and
extensions), and the generic \aid{by-normalization} lemma
connects the two halves.  A second difference is reusability: coset tables,
towers, and the product constructions are library modules
with explicit hypothesis records, where the earlier developments were
structured as per-paper proofs.

\subsection{Quantum Circuits in Proof Assistants: Soundness, Not Completeness}\label{sec:rel-qc}

Several systems verify quantum programs and compilers.
\textsc{Qwire}~\cite{paykin2017qwire,rand2018qwirepractice} embeds a circuit language in Rocq
with a denotational semantics; \textsc{sqir}/\textsc{voqc}
\cite{hietala2021voqc,hietala2021proving} defines a quantum IR in Rocq and
verifies an optimizing compiler: every rewrite the optimizer performs
is proved sound against the matrix semantics.
VyZX~\cite{lehmann2025vyzx} inductively represents ZX diagrams in Rocq and
proves the ZX rewrite rules sound.  CoqQ~\cite{zhou2023coqq} builds
Dirac-notation program logics over MathComp;
Qbricks~\cite{chareton2021qbricks} verifies circuit-building programs
with automated first-order reasoning over path-sums; on the Isabelle
side, QHLProver formalizes quantum Hoare logic~\cite{liu2019quantum} and
Isabelle-Marries-Dirac verifies textbook algorithms~\cite{bordg2020dirac}.
None of these addresses the completeness of the calculus itself, that is,
whether a finite rule set derives all true equations; they establish the
correctness of particular transformations.  The two lines of work are
complementary: one could connect our presented
gate groups to the matrix semantics of \textsc{sqir} or VyZX and obtain
end-to-end completeness against matrices rather than against abstract
product groups.

\subsection{Completeness Theorems on Paper: Gate Sets and ZX}\label{sec:rel-paper}

The mathematical literature proves such theorems on
paper~\cite{selinger2015clifford,clement2023complete,clement2024minimal,
li2025qutrit,clement2026realcliffordch}.  Further examples include the
CNOT-dihedral fragment~\cite{amy2018cnotdihedral} and its
prime-dimensional generalization~\cite{blake2026phaseaffine}.  Likewise for the graphical
side, where the completeness results for the ZX-calculus, from the qubit
stabilizer fragment through Clifford+$T$ to the odd-prime-dimensional
qupit calculi~\cite{backens2014zx,jeandel2018zx,booth2022qupitzx,
poor2023travaganza}, are all pen-and-paper; mechanizations have so far
been confined to the syntactic verifications of the Bian--Selinger line,
the Agda developments of the
rig-categorical approach (next subsection), and the present work.  The rule sets themselves are
still being revised: Blake's recent PROP-theoretic treatment uniformly
transfers and simplifies the presentations of six near-Clifford circuit
fragments (among them the two-qubit Clifford+$T$, three-qubit
Clifford+CS, and qutrit Clifford presentations discussed above),
obtaining minimal rule sets for several of
them~\cite{blake2026simpler}.  Each such simplification claims that the
dropped rules are derivable from those kept, a claim this library can
check.

\subsection{Complete Equational Theories via Rig Categories}\label{sec:rel-rig}

A second line of work, originating in programming-language semantics,
obtains completeness from categorical structure.
Choudhury, Karwowski, and Sabry present the reversible language $\Pi$ as
the free symmetric rig groupoid and mechanize the syntax and most proofs
in univalent HoTT-Agda (about 7\,500 lines), using the formalization for
normalization-by-evaluation, verification, and synthesis of reversible
Boolean circuits~\cite{choudhury2022symmetries}.  This is the
mechanization closest to ours.  Their completeness theorem is \emph{full
abstraction} of $\Pi$'s syntactic
groupoid against a univalent subuniverse of finite types, over the
classical (permutation) fragment, obtained with higher
inductive types and univalence.  Ours is completeness of concrete
finite relation sets against independently constructed groups, obtained
by coset normal forms in axiom-free, setoid-based, standard-library Agda.
The two overlap at one point: their development encounters the word
problem for $S_n$, which our factorial-carrier normal
form decides (\S\ref{sec:permutations}).

On the quantum side of this line of work, Carette, Heunen, Kaarsgaard, and
Sabry extend rig groupoids with square-root maps and prove the resulting
equational theory sound and complete for arbitrary Clifford circuits,
Clifford+$T$ circuits on at most two qubits, and the computationally
universal Gaussian Clifford+$T$ gate set (their Section~6, which builds
on the presentations of Selinger and of Bian and
Selinger)~\cite{carette2024squareroots}.  This development, too, is
partially mechanized: an Agda library extending agda-categories
formalizes the equational model, all of their \S6.1 (whose completeness
theorem for the two-qubit Clifford fragment imports a result of
Selinger's), and lemmas of two of the three
larger completeness proofs.  The authors report that no errors surfaced
in the paper proofs, that a missing rig-category coherence was found in
agda-categories itself, and that the weak setting of
agda-categories requires explicit unit and associativity bookkeeping;
our \aid{by-assoc} solver addresses the same bookkeeping for words
(\S\ref{sec:solvers}).  The two most recent results in this line are
still pen-and-paper: Fang, Heunen, and Kaarsgaard's
Hadamard-$\Pi$ adjoins the Hadamard gate to $\Pi$ and derives
completeness from a new finite presentation, with a synthesis algorithm,
of the orthogonal groups over
$\mathbb{Z}[1/\sqrt2]$~\cite{fang2026hadamard} (matrix groups of
the family whose presentations Li, Ross, and Selinger
studied~\cite{li2021on}); and Heunen, Kaarsgaard, and Lemonnier derive
sound and complete axiomatizations of controlled circuits uniformly, for
quantum and Toffoli-style gate sets at once, from coherence for
semisimple rig categories~\cite{heunen2026onerig}.  These results all
depend on finite presentations of concrete matrix
groups and on normal forms for them, which are the objects this
library mechanizes.  For example, a verified presentation of
$O_n(\mathbb{Z}[1/\sqrt2])$ in the style of \S\ref{sec:examples} would
give a machine-checked basis for the completeness of Hadamard-$\Pi$.

\subsection{Group Theory in Proof Assistants: Abstract Theory Versus Concrete Presentations}\label{sec:rel-gt}

Lean's mathlib~\cite{mathlib2020} defines \aid{FreeGroup} and
\aid{PresentedGroup} (the quotient of a free group by the normal
closure of a relator set) and develops Coxeter groups atop
them~\cite{mathlib-doc-presented}, together with a groupoid-based proof
of the Nielsen--Schreier theorem.  It also provides reduced-word normal
forms for free groups and transversal-based normal forms for amalgamated
products of groups.  These are general constructions; our focus is
verified coset tables and normal-form transports for concrete circuit
presentations.  Conversely,
the deep results of the Rocq/MathComp community (most famously the Odd
Order theorem~\cite{gonthier2013oddorder}) concern finite groups
represented as finite types; MathComp does have a presentation notation:
$G \mathrel{\backslash\mathtt{isog}}\ \mathtt{Grp}(x_1 \ldots x_n : s_1 =
t_1, \ldots)$ states that $G$ is isomorphic to the largest finite
quotient of the presented group, and MathComp uses it to characterize
the dihedral, semidihedral, and quaternion groups; but the proofs go
through the finite group theory of the target (orders and counting), and
no coset enumeration or normal form is involved.  In Isabelle, the AFP has free
groups with the ping-pong lemma~\cite{breitner2010freegroups} and a
recent combinatorial Nielsen--Schreier proof with a decision procedure
for the conjugacy problem in free groups~\cite{kharim2023freegroups}:
normal-form reasoning in the free group, adjacent in spirit to our
\aid{NormalFormInjective}, but again without presentations of concrete
target groups.  What this library adds is
verified coset tables for concrete presentations, of the kind
used in computational group theory~\cite{sims1994computation,
holt2005handbook}, with the certificate (tables, sections,
five equations) kept separate from the search that found it.

On the univalent side, cubical Agda~\cite{vezzosi2019cubical} and its
library~\cite{cubicallibrary} provide free groups and set quotients as
higher inductive types, including unique normal forms and decidable
equality for free groups on discrete generators;
agda-unimath develops finite group theory including the symmetric groups
and the sign homomorphism~\cite{agdaunimath,mangel2023sign}.  A HIT-based
presented group would turn equational reasoning into rewriting, but our
central devices are maps into the syntax (sections of normal
forms, Schreier transversals), which quotients complicate: a section
$\mathit{NF} \to \aid{Word}\;X$ that picks canonical representatives is
easy to define on a setoid and not, in general, definable on a quotient.
Since the library also targets the non-cubical standard library, we chose
setoids; \S\ref{sec:setoids} discusses the choice further.

\subsection{Formalized Rewriting and Completion}\label{sec:rel-rw}

The rewriting community has formalized its own meta-theory: IsaFoR/CeTA
contains Knuth--Bendix orders and completion~\cite{sternagel2013kb} and
abstract completion~\cite{hirokawa2019abstract}, certifying completion
procedures that, when they succeed, decide word problems via
convergent rewriting.  Our approach differs: we never
orient the relations or prove confluence; the normal form is an
arbitrary verified function, and uniqueness is proved against the
semantics.  This sidesteps the classical obstruction that finitely
presented monoids with decidable word problem need not admit finite
convergent presentations (Squier~\cite{squier1987word}; Squier, Otto,
and Kobayashi~\cite{squier1994finiteness}): a normal-form function is
a weaker artifact
than a convergent rewrite system, and correspondingly more often
available.  The polygraphic approach of Guiraud--Malbos and
collaborators~\cite{guiraud2018polygraphs,ara2025polygraphs} studies
presentations and their coherence in higher categories, with Lafont's
algebraic theory of Boolean circuits~\cite{lafont2003boolean} as an early
landmark directly ancestral to circuit presentations like ours; those
developments are pen-and-paper, and mechanizing polygraphic coherence
along the lines of this library is a possible direction for future work.

\subsection{Related Tools}\label{sec:tools}

Two unverified web tools compute normal forms of the kind this
library is built to prove correct and complete.  A
normalizer to \emph{Lafont normal form} for linear permutation circuits
implements the normalization lemmas of Lafont's algebraic theory of Boolean
circuits~\cite{lafont2003boolean,lafontlp-tool}; and a normalizer for the
symplectic groups $Sp(2n,\mathbb{Z}_p)$ computes Selinger-style staircase
normal forms~\cite{qupit-tool}.

\section{Discussion}\label{sec:discussion}

\subsection{Mechanization Statistics}\label{sec:stats}

The repository contains about 22\,000 lines of Agda in 74 files
(October 2026; \texttt{wc -l}, comments included).  The verified index
module reaches 61 of these; the remainder includes auxiliary examples
and unfinished experiments.
The reusable core (words, presentations, normal forms, the engine, the
constructions, the circuit layer) is under ten thousand lines, with the
\texttt{ForStdlib} additions and a support library of modular arithmetic
beside it; the verified examples add
about the same again, and the largest single objects are two of the three
amalgamation case studies, the qutrit Clifford+$T$ and $U_3$ studies,
about 4\,260 lines together.  Of
the 7\,000 lines under \texttt{Presentation}, about 1\,000 are the core
proper; 4\,400 are the five product constructions with their normal-form
and presentation theorems, the amalgamation alone 1\,800; 600 are the
solvers; and the rest is the join primitive and the $S_n$ and wreath
presentations over a plain alphabet that the $U_3$ study uses.

Typechecking cost is modest.  Checking the index module's dependency closure from
scratch with Agda~2.8.0 (61 modules, the standard library's interfaces
cached) takes 1\,min\,42\,s and peaks at 3.5\,GB of memory; a third of
the time is the amalgamated-product construction (31\,s), checked once
and for all; the three amalgamation case studies take 19\,s together,
and every other module is under 4\,s.

\subsection{Setoids Versus Quotients, Revisited}\label{sec:setoids}

We now assess the decision of \S\ref{sec:presentations} to present
monoids as setoids over inductive congruences rather than as quotient types.
The cost is that every function out of the presented monoid carries a
congruence proof, and records like \aid{Injection} bundle that bookkeeping.
The benefits are: (i) normal-form maps, coset tables, and sections are
plain structurally recursive programs, so the hundreds of table
well-definedness obligations in \S\ref{sec:examples} reduce to \aid{refl}
by evaluation;
(ii) sections, that is, maps into the syntax that
choose representatives (the essential ingredient of
\aid{NormalForm}, Schreier transversals, and amalgamation tails), are
directly definable for setoids but not, in general, for quotients;
(iii) the library remains inside plain Agda and the
standard library's algebra hierarchy, a requirement for upstreaming;
and, secondarily, (iv) derivations of $\approx$ are data, on which the
congruence-lifting lemmas of \S\ref{sec:constructions} are inductions.
The univalent alternative \cite{vezzosi2019cubical,cubicallibrary} would
turn equational reasoning into rewriting, but its eliminator still
demands a congruence proof for every function out of the quotient, and
(ii) and (iii) would be forfeited.  Since this domain consists largely of
computing canonical representatives, we found setoids the better choice;
with the standard library's function bundles, the witnesses of
\S\ref{sec:nfproperty} needed no new theory.

\subsection{What the Standard Library Lacked}

The \texttt{ForStdlib} layer (\S\ref{sec:forstdlib}) records what we
needed from textbook group theory and did not find in the standard
library: semidirect products,
amalgamated products or pushouts of monoids, extensions and exact
sequences, epimorphism structures, and the permutations of
$\mathsf{Fin}\;n$ bundled as a group.  None of these is hard (the
seven modules total 865 lines), but each was needed for the
semantic side of some presentation theorem, and we expect them to be
useful independently.  More generally, the standard
library's algebra hierarchy has many structures and comparatively few
constructions, and presentation theorems need the constructions.

\subsection{Proof Engineering Lessons}

We draw four lessons.  \emph{Choose indices for
the unifier.}  With the circuit layer's successor-only indexing
(\S\ref{sec:circuits}), coverage checking succeeds by injectivity of
constructors; with general arithmetic indices, every table would carry
stuck side conditions.  \emph{Make certificates finite.}  The five coset-table
hypotheses quantify over generators, cosets, and axiom instances only,
so in our finite-index case studies obligations split into finitely many closed goals, the
precondition for \aid{refl}-discharge; hypotheses phrased over all words
would have demanded inductions in every instance.  \emph{Let evaluation
prove what it can.}  The most useful tool in the development
is \aid{by-equal-nf Eq.refl}: once a level has a normal form, any true
closed equation at that level is proved by evaluation.  The case analyses
remain (the soundness law and
the uniqueness proof of each level are still proved by hand, case by
case), but the effort moves to constructing the next normal form,
and the purely computational obligations need no manual proof.
\emph{Separate search from checking.}  Coset tables were found by
ordinary computation outside Agda, and only their verification is
formal.  The formal development does not depend on how a table was
produced, so the search code is outside the trusted base and could be
changed freely.

\subsection{Limitations}\label{sec:threats}

We note two limitations.  First, our completeness theorems are stated against
semantic groups built compositionally ($\mathbb{Z}_N \wr S_n$, product
groups, extension groups) or against permutations; for the amalgamation
case studies the final theorems are isomorphisms between two
presentations, whose semantic reading as the concrete matrix
monoids (e.g.\ Clifford+$T$~\cite{giles2013remarks,glaudell2019qutrit},
or the unitaries over
$\mathbb{Z}[\tfrac12,i]$~\cite{amy2020numbertheoretic,bian2021un})
additionally uses the paper-mathematics identification of those groups
with the amalgamated products, which we have not formalized.
Second, statistics like ``every well-definedness obligation closed by
\aid{refl}'' describe our case studies, not a theorem about the method;
gate sets whose coset structure is not finitely enumerable per level
would need a different approach.

\section{Conclusion and Future Work}\label{sec:conclusion}

We presented an Agda library for presentations of monoids and groups,
normal forms, and completeness of finite relation sets, organized around
inductively defined circuits and a verified, monoid-level
Reidemeister--Schreier engine.  Completeness is reduced, by one generic
lemma, to soundness and a unique-normal-form witness.  Normal forms are built
by verified coset tables, one mixed-radix digit per level of a subgroup
tower, or by amalgamation where the tower branches.  The catalogue of
verified instances spans the symmetric groups as circuits (complete for
two semantics), cyclic groups, signed permutations and general wreath
products $\mathbb{Z}_N \wr S_n$, Pauli groups modulo phases in all prime
dimensions, and the presentations of the single-qubit and single-qutrit
Clifford+$T$ monoids and of $U_3(\mathbb{Z}[\tfrac12,i])$ as amalgamated
free products.  In the Clifford+$T$ case, the amalgam's alternating normal
form reproduces the syllable structure of the Matsumoto--Amano normal
form.  Every headline theorem is collected, fully
typed, in one index module, and the whole verified development is safe,
axiom-free Agda.

The work continues in three directions.  \emph{Clifford
completeness.}  The multi-qupit Clifford groups sit above the Pauli
groups (modulo phases) formalized here as extensions by the symplectic
groups $Sp(2n,\mathbb{Z}_p)$; with the Pauli presentations and the
extension-presentation theorem verified, the next target is
machine-checked completeness for multi-qubit and multi-qupit
Clifford circuits: formal counterparts of Selinger's qubit
presentation~\cite{selinger2015clifford}, the qutrit rule set
of~\cite{li2025qutrit}, and its generalization to all odd primes (work in
progress).  Their pen-and-paper proofs are long, regular case analyses of
the kind this framework handles.  \emph{More groups.}  The
constructions should extend to further families: the lamplighter groups
$\mathbb{Z}_N \wr \mathbb{Z}$ as a first infinite-index stress test of
the wreath machinery, HNN-style extensions to complement amalgamations,
and the remaining gate groups over rings such as $\mathbb{Z}[\tfrac12,i]$
and $\mathbb{Z}[\tfrac12]$~\cite{bian2021un,li2021on}, including
the orthogonal groups over $\mathbb{Z}[1/\sqrt2]$ whose finite
presentation underpins the completeness of the quantum programming
language Hadamard-$\Pi$~\cite{fang2026hadamard}.
\emph{Upstreaming.}  The \texttt{ForStdlib} constructions (semidirect
products, monoid amalgamations, extensions, epimorphisms, the permutation
group) are being prepared for the Agda standard library, and the
presentation framework itself is written to its conventions with that
destination in mind.

\section*{Disclosure of Use of AI in Research}
The authors wrote all of the code except for one major module, which
combines the factor presentations of a non-split group extension, and
some auxiliary code concerning semantics, which were written with Claude
Opus 4.8.  We also used Claude Opus 4.8 to assist with coding,
literature review, and the writing and editing of the paper.
Claude Fable 5.1, Claude Opus 5.5, and ChatGPT Astra 6 assisted with
subsequent editing and revision of the manuscript.

\section*{Data Availability Statement}
The artifact, version \texttt{popl-revision2a}, is available on
Zenodo~\citep{bian2026artifact}.  It contains the Agda
sources and a hyperlinked HTML rendering of
\texttt{MainTheorems.agda} and its imports (\texttt{html/index.html}) for
browsing without Agda.

\begin{acks}
This research was partially supported by the Quantum Science
and Technology-National Science and Technology Major Project under Grant
No.~2024ZD0300500, the National Natural Science Foundation of China
under Grant No.~92465202, and the Beijing Science and Technology Planning Project under Grant No.~Z25110100810000.
\end{acks}

\bibliographystyle{ACM-Reference-Format}
\bibliography{refs}

\end{document}